\documentclass[10pt,amsmath,amssymb,twocolumn,superscriptaddress,nofootinbib,aps,prd,twocolumn,floatfix]{revtex4-1}

\usepackage{graphicx}  % needed for figures
\graphicspath{{./plots/}}
\usepackage{bm}        % for math
\usepackage{enumitem}
\usepackage{etoolbox}

\usepackage[colorlinks=true]{hyperref}

\providecommand{\texorpdfstring}[2]{#1}

\newcommand{\lr}[1]{ \left( #1 \right) }
\newcommand{\lrs}[1]{ \left[ #1 \right] }

\newcommand{\vev}[1]{ \langle \, #1 \, \rangle }

\newcommand{\tr}{ {\rm Tr} \, }
\newcommand{\re}{ {\rm Re} \, }

\renewcommand{\Re}{ {\rm Re} \, }

\newcommand{\MeV}{ \, {\rm MeV} }
\newcommand{\GeV}{ \, {\rm GeV} }

\newtoggle{twocolumn}
\toggletrue{twocolumn}

\newcommand{\red}[1]{#1}

\begin{document}
\sloppy

\title{Electric conductivity in finite-density \texorpdfstring{$SU(2)$}{SU(2)} lattice gauge theory with dynamical fermions}

\author{P.~V.~Buividovich}
\email{pavel.buividovich@liverpool.ac.uk}
\affiliation{Department of Mathematical Sciences, University of Liverpool, Liverpool, L69 7ZL, UK}
%\affiliation{\UL}

\author{D.~Smith}
\email{d.smith@gsi.de}
\affiliation{Institut f\"ur Theoretische Physik, Justus-Liebig-Universit\"at, 35392 Giessen, Germany}
\affiliation{Helmholtz Research Academy Hesse for FAIR (HFHF), Campus Giessen, 35392 Giessen, Germany}
\affiliation{Facility for Antiproton and Ion Research in Europe GmbH (FAIR GmbH), 64291 Darmstadt, Germany}

\author{L.~{von Smekal}}
\email{lorenz.smekal@physik.uni-giessen.de}
\affiliation{Institut f\"ur Theoretische Physik, Justus-Liebig-Universit\"at, 35392 Giessen, Germany}
\affiliation{Helmholtz Research Academy Hesse for FAIR (HFHF), Campus Giessen, 35392 Giessen, Germany}

\date{October 20th, 2020}

\begin{abstract}
We study the dependence of the electric conductivity on chemical potential in finite-density $SU\lr{2}$ gauge theory with $N_f = 2$ flavours of rooted staggered sea quarks, in combination with Wilson-Dirac and Domain Wall valence quarks. The pion mass is reasonably small with $m_{\pi}/m_{\rho} \approx 0.4$. We concentrate in particular on the vicinity of the chiral crossover, where we find the low-frequency electric conductivity to be most sensitive to small changes in fermion density. Working in the low-density QCD-like regime with spontaneously broken chiral symmetry, we obtain an estimate of the first nontrivial coefficient $c\lr{T}$ of the expansion of conductivity $\sigma\lr{T,\mu} = \sigma\lr{T,0} \lr{1 + c\lr{T} \lr{\mu/T}^2 + O\lr{\mu^4}}$ in powers of $\mu$, which has rather weak temperature dependence and takes its maximal value $c\lr{T} \approx 0.10 \pm 0.07$ around the critical temperature. At larger densities and lower temperatures, the conductivity quickly grows towards the diquark condensation phase, and also becomes closer to the free quark result. As a by-product of our study we confirm the conclusions of previous studies with heavier pion that for $SU\lr{2}$ gauge theory the ratio of crossover temperature to pion mass $T_c/m_{\pi} \approx 0.4$ at $\mu=0$ is significantly smaller than in real QCD.
\end{abstract}

\maketitle

\section{Introduction}
\label{sec:intro}

Since quarks in QCD have finite electric charge, a hot QCD medium is characterized by some finite electric conductivity. It can be directly accessed in heavy-ion collision experiments via the dilepton emission rate \cite{Campbell:1704.06307,McLerran:85:1}, and is also of direct importance for the lifetime of strong magnetic fields generated in off-central heavy-ion collisions \cite{Skokov:1305.0774,Gursoy:2009.09727}.

The temperature dependence of the electric conductivity in QCD and QCD-like theories has been extensively studied by now. A lot of first-principle results are available from lattice gauge theory simulations \cite{Brandt:1710.07050,Kaczmarek:1604.06712,Kaczmarek:1604.07544,Meyer:1512.07249,Aarts:1412.6411,Aarts:1307.6763,Buividovich:10:1,Kaczmarek:1012.4963} (see also \cite{Nikolaev:2008.12326} for a recent summary of the lattice studies of electric conductivity). The electric conductivity was also calculated using a variety of approximation methods which complement lattice simulations, for instance, based on Boltzmann or Schwinger-Dyson equations \cite{Greiner:1408.7049,Puglisi:1408.7043,Qin:1307.4587}, or hadron gas models \cite{Greiner:1602.05085,Fernandez:hep-ph/0512283}.

However, there are practically no first-principle results regarding the dependence of the electric conductivity on baryon chemical potential, apart from AdS/CFT calculations \cite{Karch:0705.3870,Kim:0803.0318} which are not directly applicable to non-supersymmetric QCD. Symmetries of the QCD action suggest that the electric conductivity should be an even function of the chemical potential $\mu$, and thus can be expanded in powers of $\mu$ as
\begin{eqnarray}
\label{sigma_vs_mu_parameterization}
 \frac{\sigma\lr{T, \mu}}{T} = \frac{\sigma\lr{T, 0}}{T} \, \lr{ 1 + c\lr{T} \, \lr{\frac{\mu}{T}}^2 + O\lr{\mu^4}} .
\end{eqnarray}

In a calculation based on the off-shell Parton-Hadron-String Dynamics (PHSD) transport approach \cite{Cassing:1312.3189} the coefficient $c\lr{T}$ in (\ref{sigma_vs_mu_parameterization}) was estimated as $c\lr{T} \approx 0.46$ for $T$ near the deconfinement transition \cite{Cassing:1312.3189}. A study based on the Boltzmann equation within the quasiparticle approach also gives a result consistent with this estimate \cite{Srivastava:1501.03576}, although only for a single non-zero value of $\mu$. Within the dynamical quasiparticle model the dependence of the electric conductivity on the baryon chemical potential was found to be rather weak \cite{Bratkovskaya:1911.08547}, which is consistent with results obtained using the Functional Renormalization Group \cite{Smekal:1807.04952}. On the other hand, a kinetic theory calculation based on the hadron resonance gas model \cite{Kadam:1712.03805} suggests a strong dependence of $\sigma$ on $\mu$ in the low-temperature hadronic phase, with $\sigma/T$ changing almost by an order of magnitude as the chemical potential varies from $\mu = 0.1 \GeV$ to $\mu = 0.3 \GeV$. A detailed analysis of pion and nucleon loop contributions to $\sigma$ reveals even a non-monotonic dependence of electric conductivity on $\mu$ \cite{Ghosh:1607.01340}.

These estimates imply that a finite chemical potential can significantly change the electric conductivity in the physically interesting part of the QCD phase diagram with $\mu \gtrsim T$, where the QCD critical point is believed to be located. This region of the phase diagram is in the focus of ongoing heavy-ion collision experiments at RHIC and LHC. Planned experiments at NICA and FAIR facilities will achieve even larger baryon densities at lower temperatures, hence even larger values of the ratio $\mu/T$. Thus it is important to study the density dependence of $\sigma$ in non-Abelian gauge theory from first principles in order to correctly interpret the experimental data on dilepton emission rates.

\red{As is well known, due to the notorious fermionic sign problem, first-principle lattice QCD simulations can only be performed at zero chemical potential. Methods such as Taylor series expansion, reweighting or analytic continuation from imaginary chemical potential can be used to obtain more or less reliable results for small values of $\mu/T$. However, these methods often make the extraction of physical observables from lattice simulations much more technically challenging than for the case of $\mu = 0$.}

If one is interested in obtaining qualitative estimates rather than high-precision results, it is often helpful to consider QCD-like theories which behave similarly to QCD in some regions of their phase diagram, but have no fermionic sign problem. Examples include gauge theories with $SU\lr{2}$ \cite{Kogut:hep-lat/0105026,Kogut:hep-ph/0001171} and $G_2$ \cite{Smekal:1203.5653} gauge groups, as well as QCD at finite isospin chemical potential \cite{Son:hep-ph/0005225,Kamikado:1207.0400,Endrodi:1712.08190}.

In this work we perform a numerical study of the dependence of the electric conductivity on the fermion chemical potential in $SU\lr{2}$ gauge theory with dynamical fermions. ``Baryons'' in $SU\lr{2}$ gauge theory are diquarks, bound states of two quarks, which have the same mass $m_{\pi}$ as the pion and are thus much lighter than baryons in real QCD. Diquarks hence condense for $\mu \gtrsim m_{\pi}/2$ \cite{Kogut:hep-lat/0105026,Kogut:hep-ph/0001171} (see also Fig.~\ref{fig:phase_diagram} below), significantly earlier than for real QCD where condensation of nucleons with mass $m_n \gg m_{\pi}$ happens at $\mu \approx m_n/3 \gg m_{\pi}/2$. At not very large values of the chemical potential outside of the diquark condensation phase, however, the properties of finite-density $SU\lr{2}$ gauge theory are expected to be similar to those of real QCD. In particular, this similarity makes our estimate of the coefficient $c\lr{T}$ in the expansion (\ref{sigma_vs_mu_parameterization}) relevant for real QCD, in a way analogous to orbifold equivalence, see, e.g.~\cite{Hanada:1103.5480}.

Our main finding from the analysis of the data in low-density QCD-like regime is that the density dependence of electric conductivity is rather weak in the temperature range $T/T_c = 0.7 \ldots 4.0$ which we've considered in our study. The conductivity is most sensitive to the quark density in the vicinity of the chiral crossover, where our estimate for the coefficient $c\lr{T}$ in (\ref{sigma_vs_mu_parameterization}) is $c\lr{T_c} \approx 0.10 \pm 0.07$, noticeably larger than the corresponding free-quark result.\footnote{Of course, for a free quark gas the exact zero-frequency limit of the electric conductivity is ill-defined. However, it gets some finite value within numerical analytic continuation methods used to extract conductivity from Euclidean correlators, such as the Backus-Gilbert method used in this work.} This estimate implies that the chemical potential should be at least several times larger than the temperature in order to significantly affect the electric conductivity. The temperature dependence of the coefficient $c\lr{T}$ also appears to be rather weak.

Another part of the phase diagram where finite-density $SU\lr{2}$ gauge theory is expected to behave similarly to QCD is the conjectured quarkyonic phase at very low temperatures and high densities $\mu \gg m_n$ \cite{Pisarski:0706.2191,Braguta:1605.04090}. Calculation of the electric conductivity in this part of the phase diagram could shed more light on the properties of the quarkyonic/color-superconducting phase. As we will see, however, measurements of the electric conductivity in this low-$T$, large-$\mu$ regime are numerically very challenging, and we will leave a detailed study of this for further work.

We also present data on the phase diagram of finite-density $SU\lr{2}$ gauge theory with $N_f=2$ fermion flavours which complements previous results \cite{Hands:hep-lat/0604004,Hands:1001.1682,Hands:1210.4496,Hands:1502.01219,Braguta:1605.04090,Braguta:1711.01869,Holicki:1701.04664,Etou:1910.07872,Huber:1909.12796,Hands:1912.10975} obtained either on smaller and coarser lattices, or for smaller temperatures and larger densities, or with different lattice actions. We confirm the findings of \cite{Hands:1210.4496} that the chiral crossover temperature in $SU\lr{2}$ gauge theory is $3$ to $5$ times smaller than the pion mass, depending on the chemical potential, in contrast to real QCD where $T_c \gtrsim m_{\pi}$ at $\mu = 0$.

The outline of the paper is the following: in Section~\ref{sec:lattice_setup} we present the details of our lattice setup and discuss the mixed fermionic action used to calculate the electric conductivity. In Section~\ref{sec:phase_diagram} we study the phase diagram of $SU\lr{2}$ gauge theory with $N_f = 2$ rooted staggered fermion flavours in the $\mu-T$ plane. In Section~\ref{sec:conductivity_preliminary} we discuss our numerical approach to extract the electric conductivity from current-current correlators. In Section~\ref{sec:conductivity_results} we present numerical results for the electric conductivity, estimated using both the simple ``correlator midpoint'' estimate as well as using the more advanced Backus-Gilbert method. We briefly summarize our findings in the concluding Section~\ref{sec:conclusions}. Some technical details of our calculations and analytic expressions for electric conductivities of a free quark gas and a free pion gas at finite density are relegated to Appendices.

\section{Lattice setup}
\label{sec:lattice_setup}

Gauge field configurations were generated using the standard Hybrid Monte-Carlo algorithm with $N_f=2$ mass-degenerate rooted staggered fermions and a tree-level improved Symanzik gauge action. We accelerate both the HMC algorithm and the measurements of current-current correlators on GPUs. HMC is implemented with single-precision arithmetics within the \texttt{CUDA} framework, and measurements use double precision and are implemented using \texttt{OpenCL}. The same algorithmic and lattice setup has been also used recently in \cite{Smith:1910.04495}.

We use lattices with spatial sizes $L_s = 24$ and $L_s = 30$ and temporal sizes $L_t = 4 \ldots 30$, changing in steps of two. In this paper we use a fixed-scale approach, choosing a single value $\beta = 1.7$ of the inverse gauge coupling which compromises between being sufficiently far from the artificial strong-coupling bulk phase \cite{Scheffler:1311.4324} and still having a reasonably large lattice spacing and lattice volume. The chemical potential takes values $a \mu = 0.0, \, 0.05, \, 0.1, \, 0.2, \, 0.5$ in units of inverse lattice spacing $a$ for $L_s = 24$, and $a \mu = 0.0, \, 0.05, \, 0.20$ for $L_s = 30$. Our largest value of the chemical potential, $a \mu = 0.5$, thereby represents a kind of compromise between approaching the diquark condensation phase (see Fig.~\ref{fig:phase_diagram}) while still staying reasonably well below half-filling and eventual saturation of the quark density as obvious lattice artifacts. As discussed below, to facilitate diquark condensation in a finite volume, for $L_t \geq 10$ we also generate gauge field configurations with a small diquark source term $a \lambda = 5 \cdot 10^{-4}$.

The numbers of gauge field configurations used in this work are summarized in Table~\ref{tab:num_configs}. To obtain these ensembles, we have saved gauge field configurations after every 3rd full Hybrid Monte-Carlo update, which appears to be enough to have reasonably small autocorrelations in the data for current-current correlators.

\begin{figure*}[h!tpb]
  \centering
  \includegraphics[angle=-90,width=0.45\textwidth]{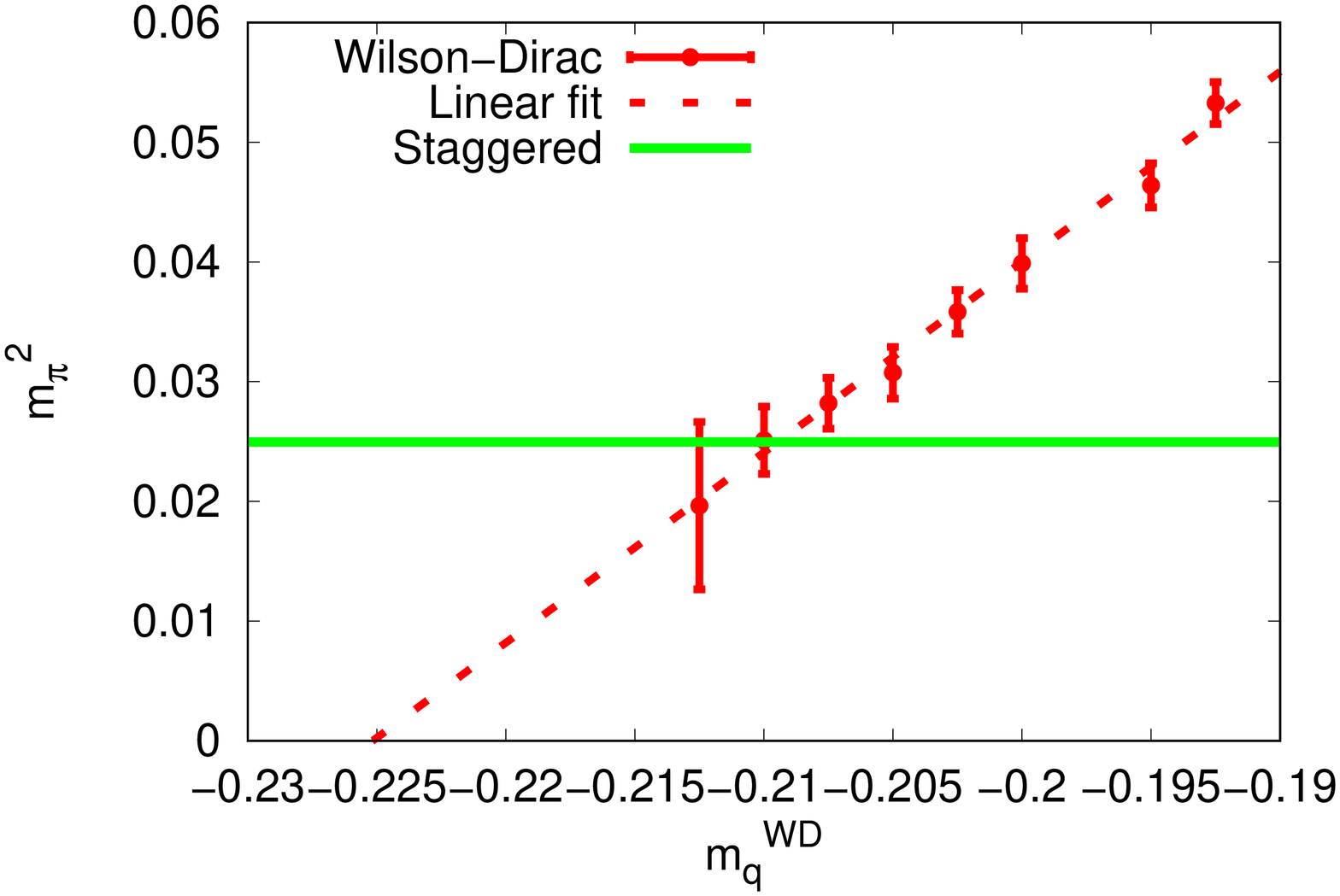}
  \includegraphics[angle=-90,width=0.45\textwidth]{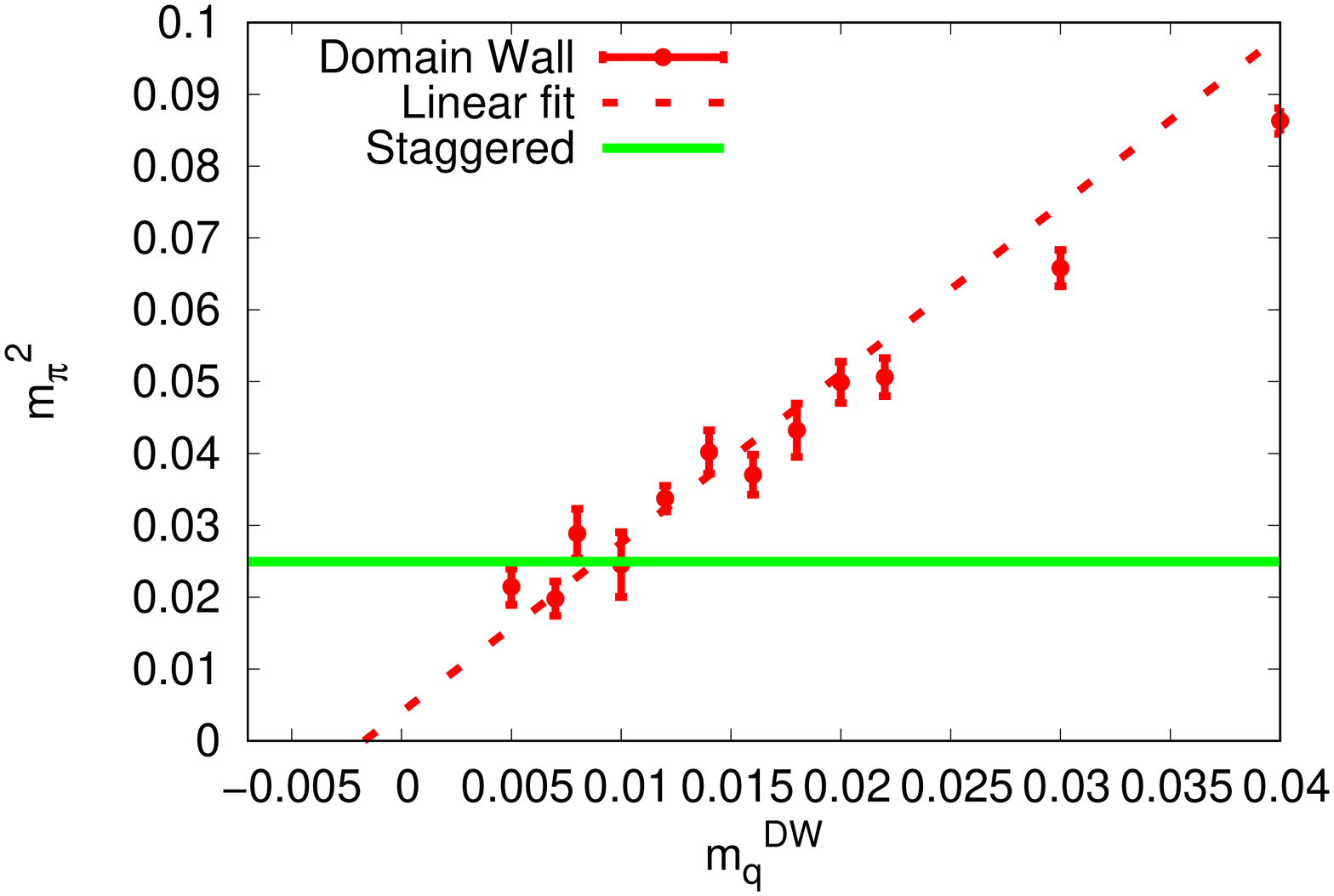}\\
  \caption{Squared pion mass $m_{\pi}^2$ calculated with Wilson-Dirac (on the left) and Domain Wall (on the right) valence fermions as a function of bare valence quark mass $m_q$.}
  \label{fig:mpi_vs_mq}
\end{figure*}

The measurements of current-current correlators are performed mainly using Wilson-Dirac (WD) fermions. One of the technical advantage of using WD fermions for measuring the electric conductivity is that all data points in the current-current correlator in Green-Kubo relations (\ref{GreenKubo_conductivity}) can be treated uniformly, whereas for staggered fermions even and odd time slices are typically treated separately \cite{Aarts:hep-lat/0703008} in order to filter out the contributions from non-taste-singlet states, which effectively decreases the signal-to-noise ratio.

In addition, some of the measurements are also made with Domain-Wall (DW) fermions \cite{Furman:hep-lat/9405004}, providing a cross-check for the Wilson-Dirac data. We use the distance between domain walls (lattice size along the fifth dimension) $L_5 = 16$, which is typically sufficient to suppress additive mass renormalization \cite{Edwards:hep-lat/0510062,Berkowitz:1704.01114} for DW fermions. Such a mixed lattice action with staggered sea fermions and DW valence fermions has already been used in a number of studies of the nucleon axial charge \cite{Renner:hep-lat/0409130,Edwards:hep-lat/0510062,Berkowitz:1704.01114}. However, our primary motivation for using DW valence quarks is that we re-use fermion propagators entering the current-current correlators (\ref{current_current_connected}) to calculate also correlators of axial and vector currents. Those correlators are related to so-called anomalous transport coefficients \cite{Landsteiner:1102.4577,Buividovich:13:8}, which will be the subject of another forthcoming work. Since the axial anomaly is very subtle for staggered fermions, the use of DW valence fermions with good chiral properties is a big advantage for this kind of calculations.

To improve the chiral properties of DW and WD fermions without using much finer and larger lattices, we follow \cite{Edwards:hep-lat/0510062} and use HYP smearing \cite{Hasenfratz:hep-lat/0103029} for gauge links in the DW and WD Dirac operators.

\begin{table}
  \centering
  \begin{tabular}{||c||c|c|c|c||c|c|c|c|c||}
   \hline
   & \multicolumn{4}{|c||}{ $\lambda = 0$} & \multicolumn{5}{c||}{ $a \lambda = 5 \cdot 10^{-4}$ } \\ \hline
$L_t$ \textbackslash \, $a \mu$ & 0.0 & 0.1 & 0.2 & 0.5 & 0.0 & 0.05 & 0.1 & 0.2 & 0.5 \\
     \hline
   	4   & 402  & 202 & 202  & 202         &   0 &   0 &   0 &    0 &   0 \\
	6   & 202  & 202 & 202  & 202         &   0 &   0 &   0 &    0 &   0 \\
	8   & 402  & 202 & 802  & 202         &   0 &   0 &   0 &    0 &   0 \\
	10  & 402  & 202 & 1661 & 202         & 503 & 994 & 202 & 1188 & 233 \\
	12  & 791  & 202 & 1368 & 202         & 339 & 639 & 202 &  802 & 249 \\
	14  & 1661 & 202 & 946  & 102         & 267 & 420 & 395 &  724 & 184 \\
	16  & 1661 & 202 & 635  & 50          & 397 & 692 & 292 &  731 & 267 \\
	18  & 1545 & 469 & 549  & 62          & 321 & 230 & 314 &  331 & 179 \\
    20  & 1074 & 0   & 454  & 0           & 268 & 418 & 230 &  451 & 132 \\
    22  & 0    & 0   & 0    & 0           & 227 & 324 & 191 &  274 &  0  \\
    24  & 0    & 0   & 0    & 0           & 95  & 132 & 195 &  161 &  0  \\
    26  & 0    & 0   & 0    & 0           & 85  & 104 & 203 &    0 &  0  \\
    \hline
  \end{tabular}
  \caption{Numbers of gauge field configurations with spatial size $L_s = 24$ used in this work.}
  \label{tab:num_configs}
\end{table}

\begin{table}
   \centering
   \begin{tabular}{||c||c|c|c||c|c|c||}
    \hline
& \multicolumn{3}{|c||}{ $\lambda = 0$} & \multicolumn{3}{c||}{ $a \lambda = 5 \cdot 10^{-4}$ } \\
    \hline
$L_t$ \textbackslash \, $a \mu$ & 0.0 & 0.05 & 0.2 & 0.0 & 0.05 & 0.2 \\
     \hline
   	4   & 1602  & 1452 & 202    & 0 & 0 & 0 \\
	6   & 1602  & 678 & 202    & 0 & 0 & 0 \\
	8   & 1495  & 348 & 202    & 0 & 0 & 0 \\
	10  & 880  & 211 & 202    & 0 & 0 & 0 \\
	12  & 580  & 262 & 123    & 0 & 0 & 420\\
	14 & 0 & 0 & 224   &  382    & 309   & 259   \\
	16 & 0 & 0 & 145   &  527    & 370 & 162   \\
	18 & 0 & 0 & 94    &  396    & 267 & 114  \\
    20 & 0 & 0 & 63    &  338    & 337   & 69   \\
    22 & 0 & 0 & 47    &  283    & 279   & 99     \\
    \hline
   \end{tabular}
  \caption{Numbers of gauge field configurations with spatial size $L_s = 30$ used in this work.}
  \label{tab:num_configs_Ns30}
\end{table}

As in \cite{Renner:hep-lat/0409130,Edwards:hep-lat/0510062,Berkowitz:1704.01114}, bare quark masses in the WD and DW Dirac operators are tuned to match the pion mass $m_{\pi}^{stag} = 0.158 \pm 0.002$ obtained with staggered valence quarks with $m_q^{stag} = 0.005$. The dependence of the squared pion mass $m_{\pi}^2$ on the bare quark masses of the WD and DW fermions on $24^3 \times 48$ lattice with $\beta = 1.7$ is illustrated in Fig.~\ref{fig:mpi_vs_mq}. One can see that with a good precision $m_{\pi}^2$ depends on $m_q$ as $m_{\pi}^2 \sim m_q + \Delta m$ in accordance with the Gell-Mann-Oakes-Renner relation, where $\Delta m$ accounts for additive mass renormalization. From these data we have estimated that the bare quark mass should be $m_q^{WD} = -0.21$ for WD fermions and $m_q^{DW} = 0.01$ for DW fermions in order to match $m_{\pi}^{stag} = 0.158 \pm 0.002$. To illustrate the effect of HYP smearing on additive mass renormalization (and hence on the chiral properties of the lattice fermions), let us note that for WD fermions without HYP smearing the bare quark mass should be as large as $m_q = -0.685$ to obtain $m_{\pi} = 0.158$.

The mass of the $\rho$-meson obtained from the same ensemble on $24^3 \times 48$ lattice with $\beta=1.7$ is $m_{\rho} = 0.36 \pm 0.07$ for WD fermions and $m_{\rho} = 0.44 \pm 0.05$ for DW fermions, thus the ratio of pion and $\rho$-meson masses $m_{\pi}/m_{\rho} \approx 0.4$ is reasonably small. While not yet physical, it is smaller than in the previous studies of $SU\lr{2}$ gauge theory. The pion Compton wavelength is almost four times smaller than the lattice size, $m_{\pi} L_s \approx 3.7$ for $L_s = 24$ and $m_{\pi} L_s = 4.74$ for $L_s = 30$, hence we expect finite-size artifacts to be reasonably small.

\section{Phase diagram of finite-density \texorpdfstring{$SU\lr{2}$}{SU(2)} gauge theory}
\label{sec:phase_diagram}

In addition to the chiral condensate $\vev{\bar{\psi}\psi}$ and its susceptibility which are conventionally used to map out the chiral crossover on the QCD phase diagram, the order parameters of $SU\lr{2}$ gauge theory also include the diquark condensate $\vev{\psi\psi}$. In this Section we study these order parameters within our lattice setup and map out the boundaries of regimes which favour chiral or diquark condensates.

\begin{figure}[h!tpb]
  \centering
  \includegraphics[width=0.45\textwidth]{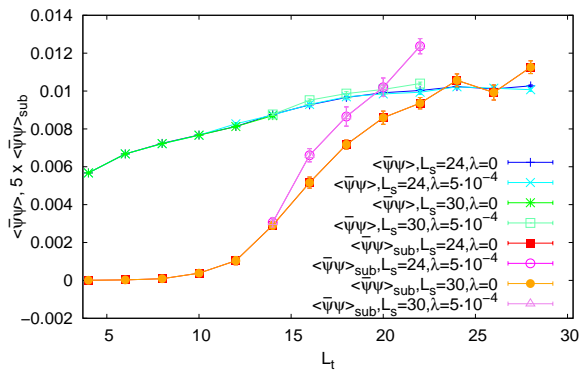}
  \caption{The effect of subtraction (\ref{chiral_condensate_subtraction}) on the temperature-dependent chiral condensate at zero chemical potential and at different lattice sizes and values of the diquark source.}
  \label{fig:condensates_subtraction_mu0.0}
\end{figure}

We first discuss the chiral and diquark condensates and the corresponding susceptibilities. There are two subtleties which have to be taken into account when interpreting the raw lattice data for these observables. First, the chiral condensate contains a UV divergent additive part which might also depend on temperature and chemical potential. As discussed in \cite{WolfgangUngerPhDThesis,Smith:1910.04495}, this UV divergent part can be removed by subtracting the first-order term of the Taylor expansion of the chiral condensate in powers of the bare quark mass $m_q$:
\begin{eqnarray}
\label{chiral_condensate_subtraction}
 \vev{\bar{\psi} \psi}_{sub} = \vev{\bar{\psi} \psi} -  \frac{\partial \vev{\bar{\psi} \psi}}{\partial m_q} \, m_q .
\end{eqnarray}
The effect of this subtraction on the $L_t$ dependence of the chiral condensate at $\mu = 0$ is illustrated in Fig.~\ref{fig:condensates_subtraction_mu0.0} for different lattice volumes and values of the diquark source. One can see that only after subtraction one can observe an expected temperature dependence of the chiral condensate and identify an inflection point which indicates a crossover between the high- and low-temperature regimes with (approximately) restored and spontaneously broken chiral symmetry.

\begin{figure*}[h!tpb]
  \centering
  \includegraphics[width=0.45\textwidth]{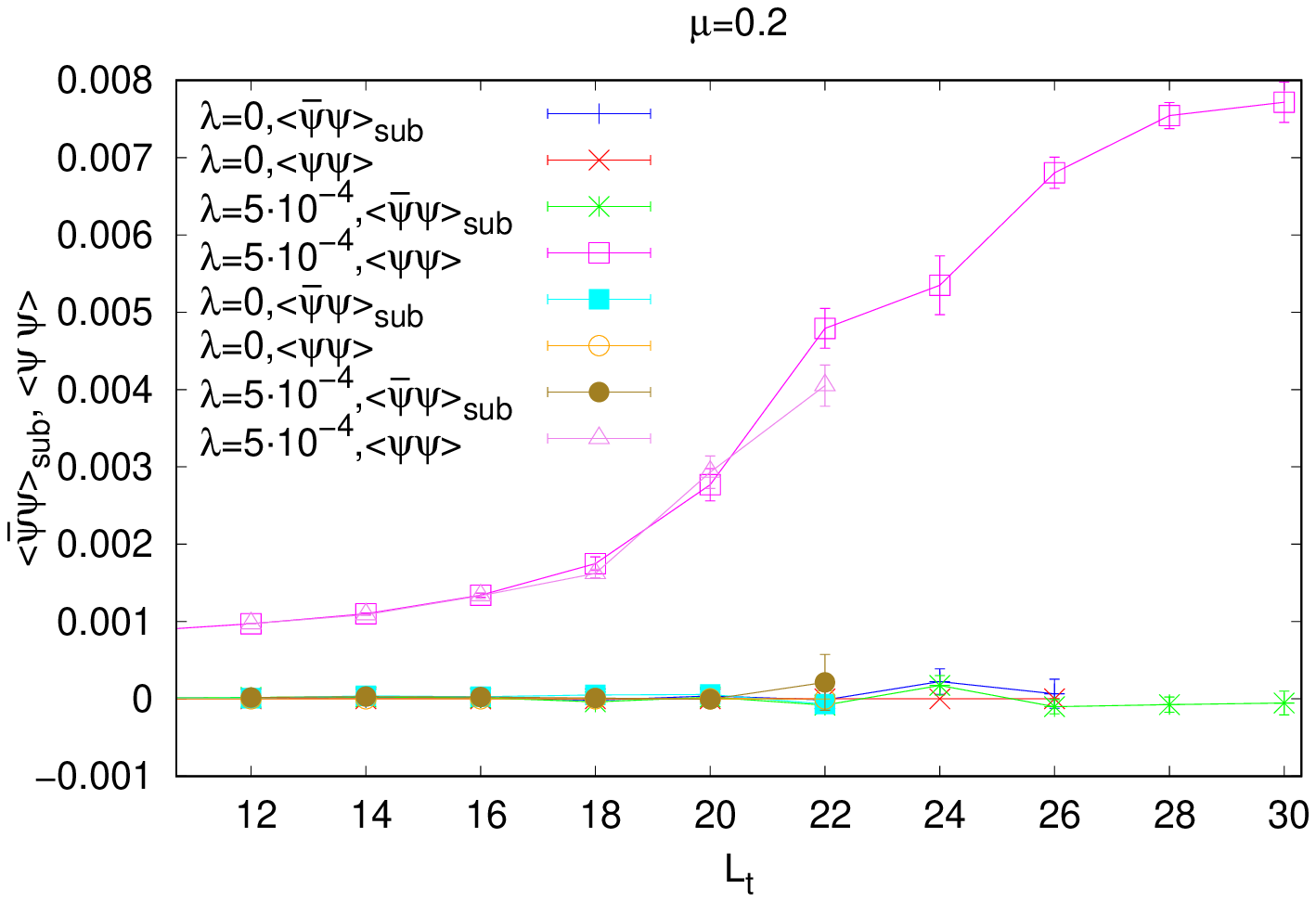}
  \includegraphics[width=0.45\textwidth]{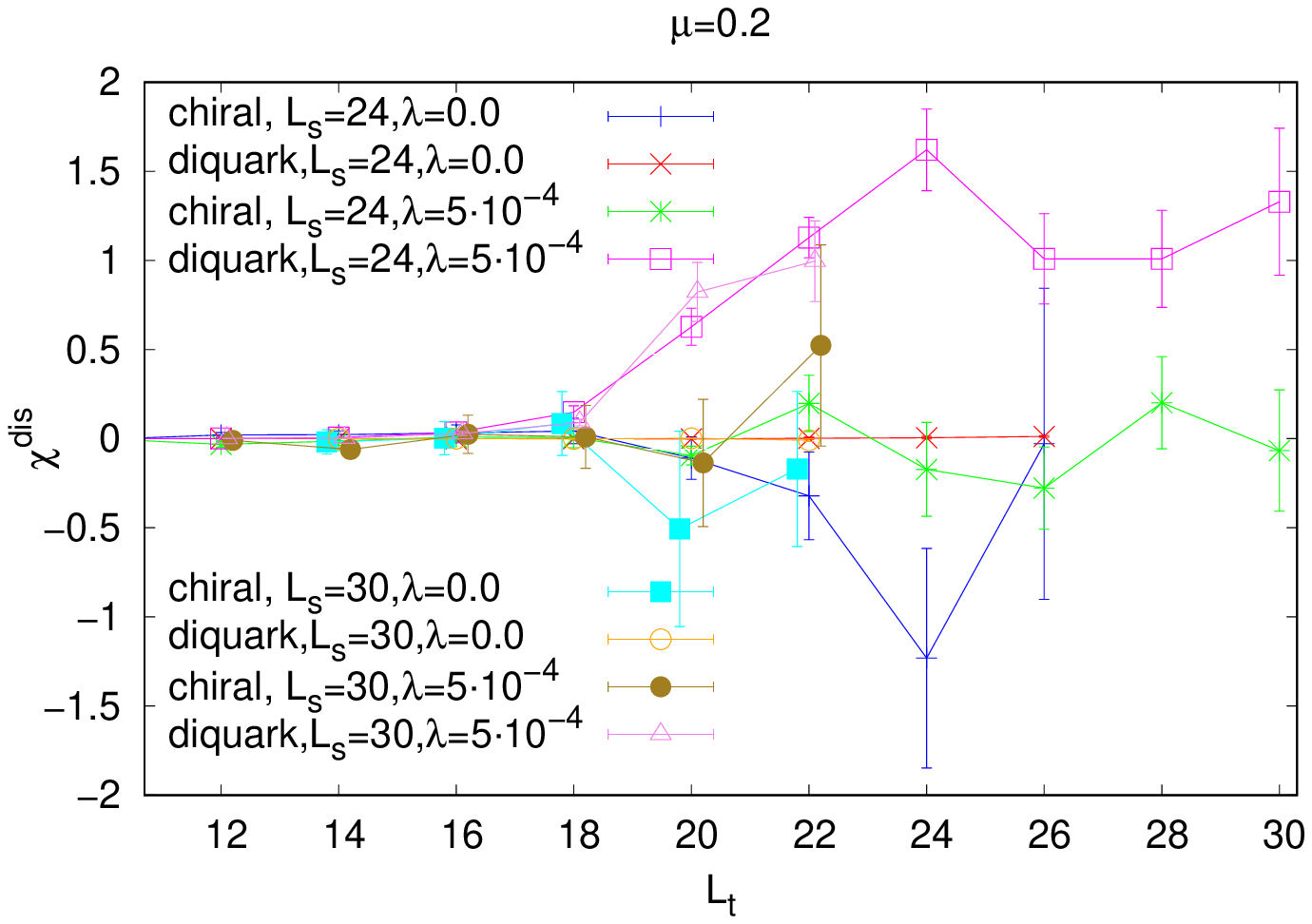}\\
  \caption{Chiral and diquark condensates (left) and susceptibilities (right) at $a \mu=0.2$ as functions of the temporal lattice size $L_t$ at $L_s=24$ and $L_s=30$ with and without a small diquark source.}
  \label{fig:effect_of_lambda}
\end{figure*}

The second subtlety is that in the chiral limit and at $\mu = 0$ the ground states with nonzero chiral and diquark condensates have equal energies. An introduction of a Dirac mass term, which is inevitable in Hybrid Monte-Carlo (HMC) simulations, breaks this degeneracy and biases the system towards the phase with nonzero chiral condensate, which makes it difficult to observe the signatures of the diquark condensation phase. In order to counteract this bias, for simulations at sufficiently low temperatures ($L_t > 10$) we introduce a small diquark source term in the action of the form $\lambda \psi \psi$ with $a \lambda = 5 \cdot 10^{-4}$ (in lattice units) which makes the diquark condensation more energetically favourable. As illustrated in Fig.~\ref{fig:effect_of_lambda}, the presence of this small source term has little effect outside of the diquark condensation phase. As illustrated in Fig.~\ref{fig:conductivity_MP_summary} below, it also has practically no effect on the electric conductivity. On the other hand, in the diquark condensation phase it acts to rotate the chiral condensate into a diquark condensate and also produces a clear wide peak in the diquark susceptibility typical for a crossover.

\begin{figure*}[h!tpb]
  \centering
  \includegraphics[width=0.45\textwidth]{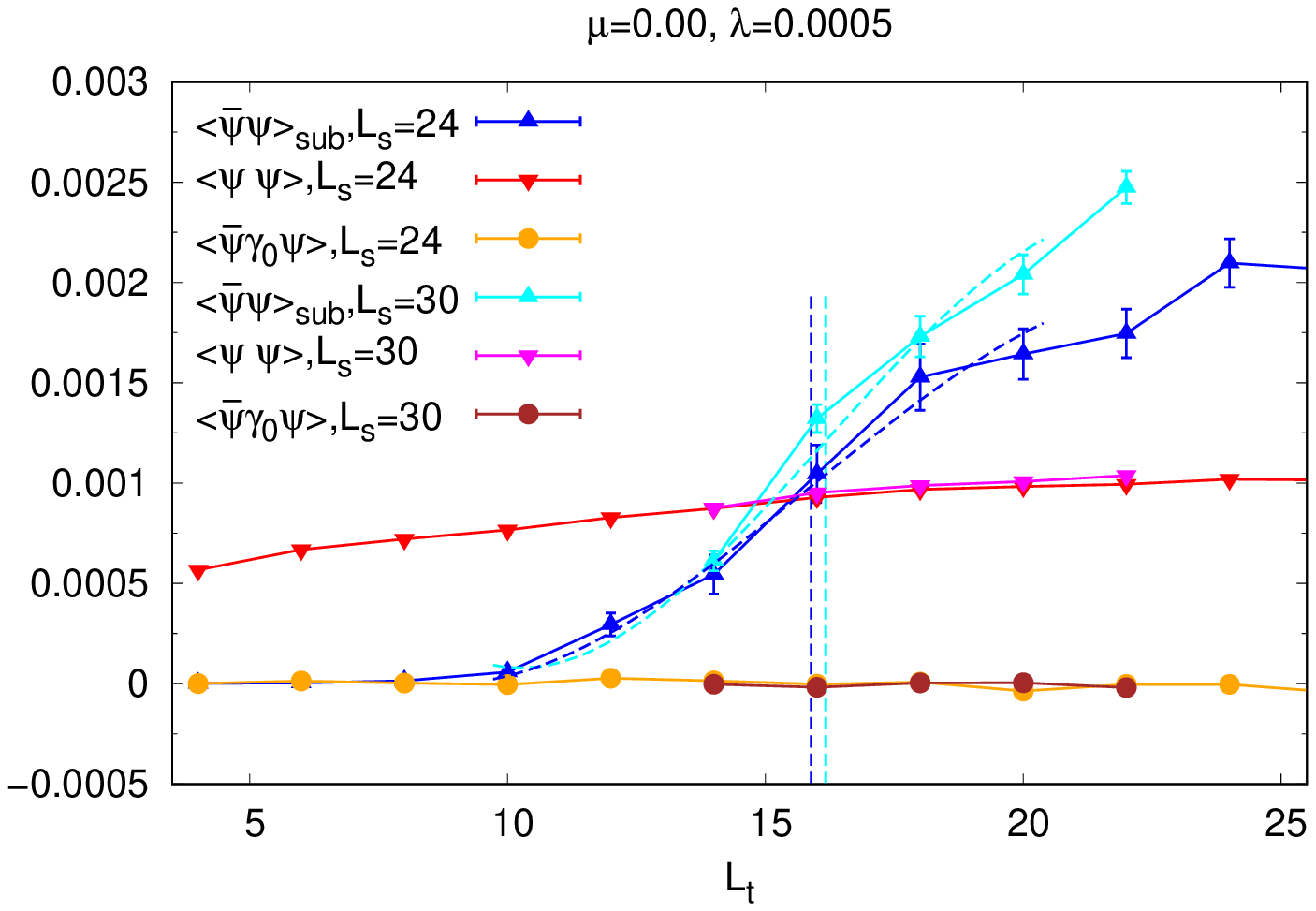}
  \includegraphics[width=0.45\textwidth]{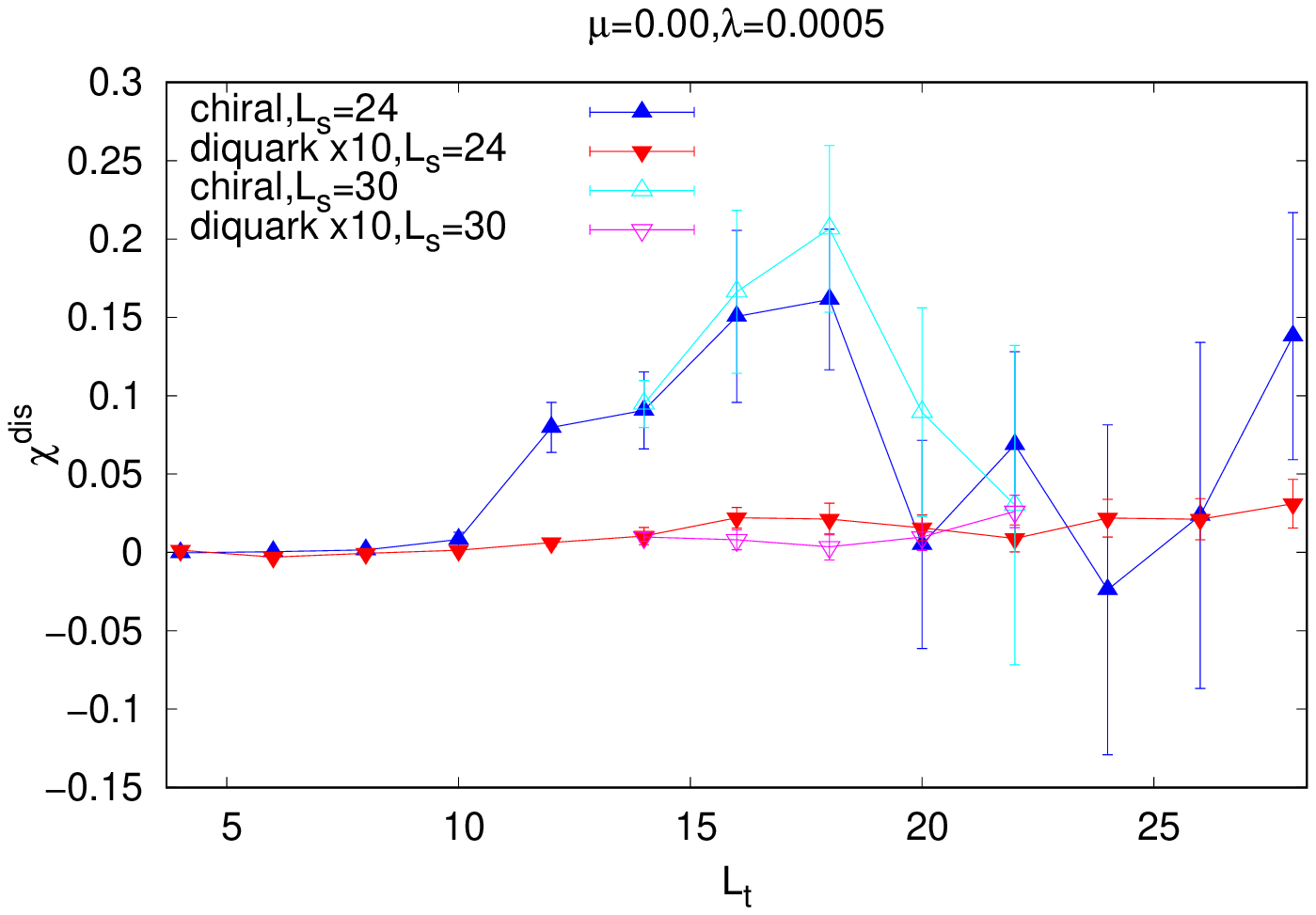}\\
  \includegraphics[width=0.45\textwidth]{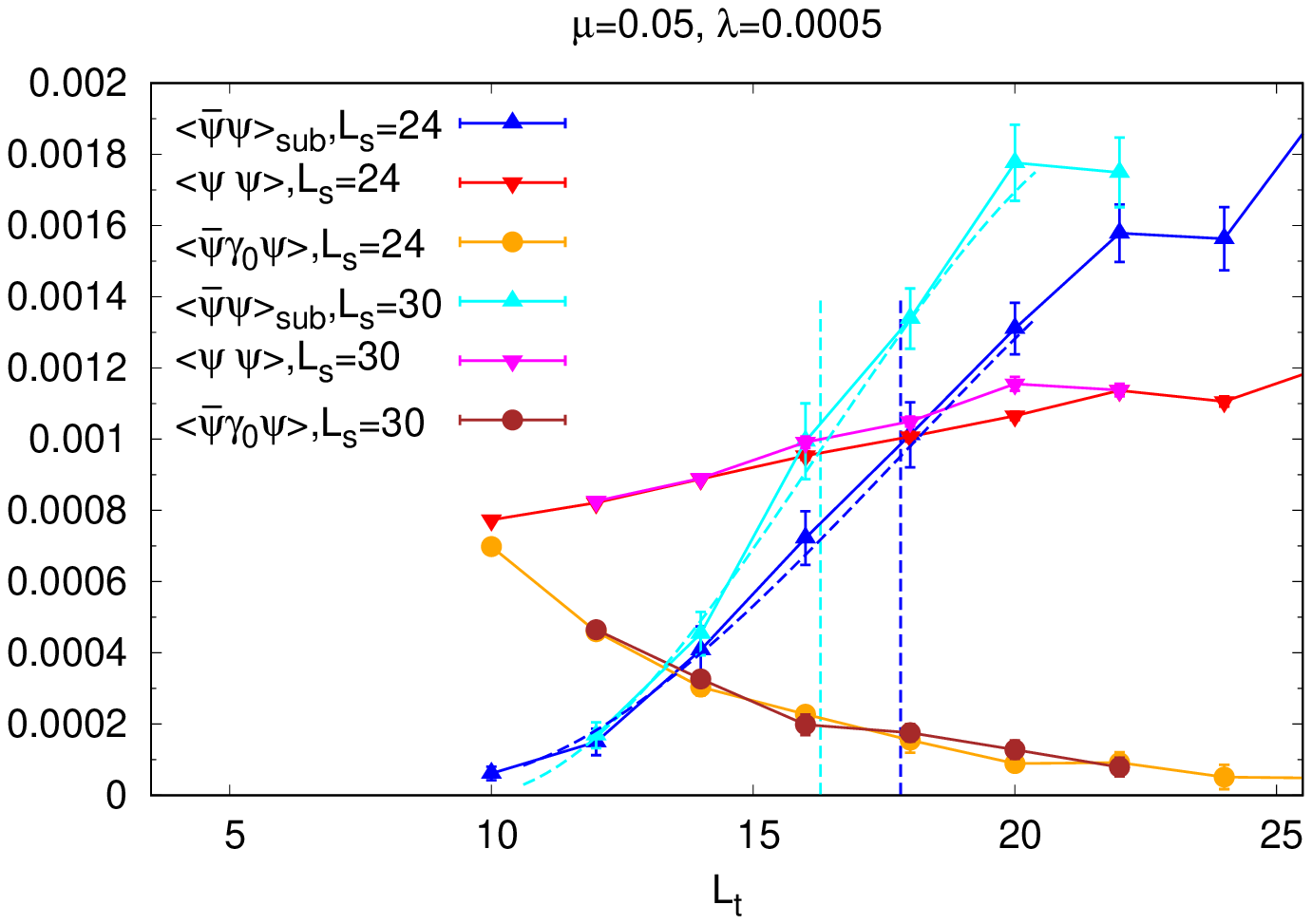}
  \includegraphics[width=0.45\textwidth]{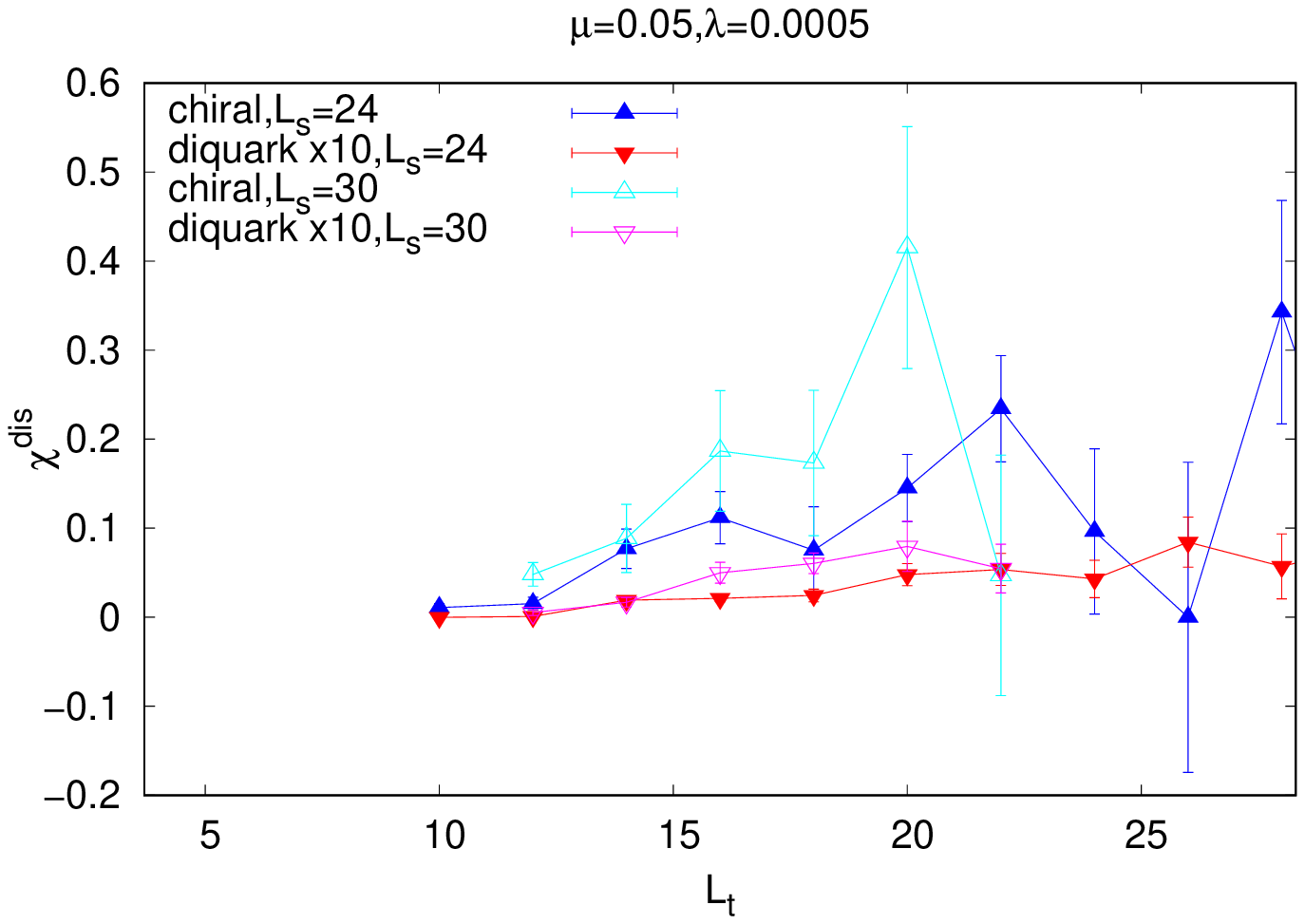}\\
  \includegraphics[width=0.45\textwidth]{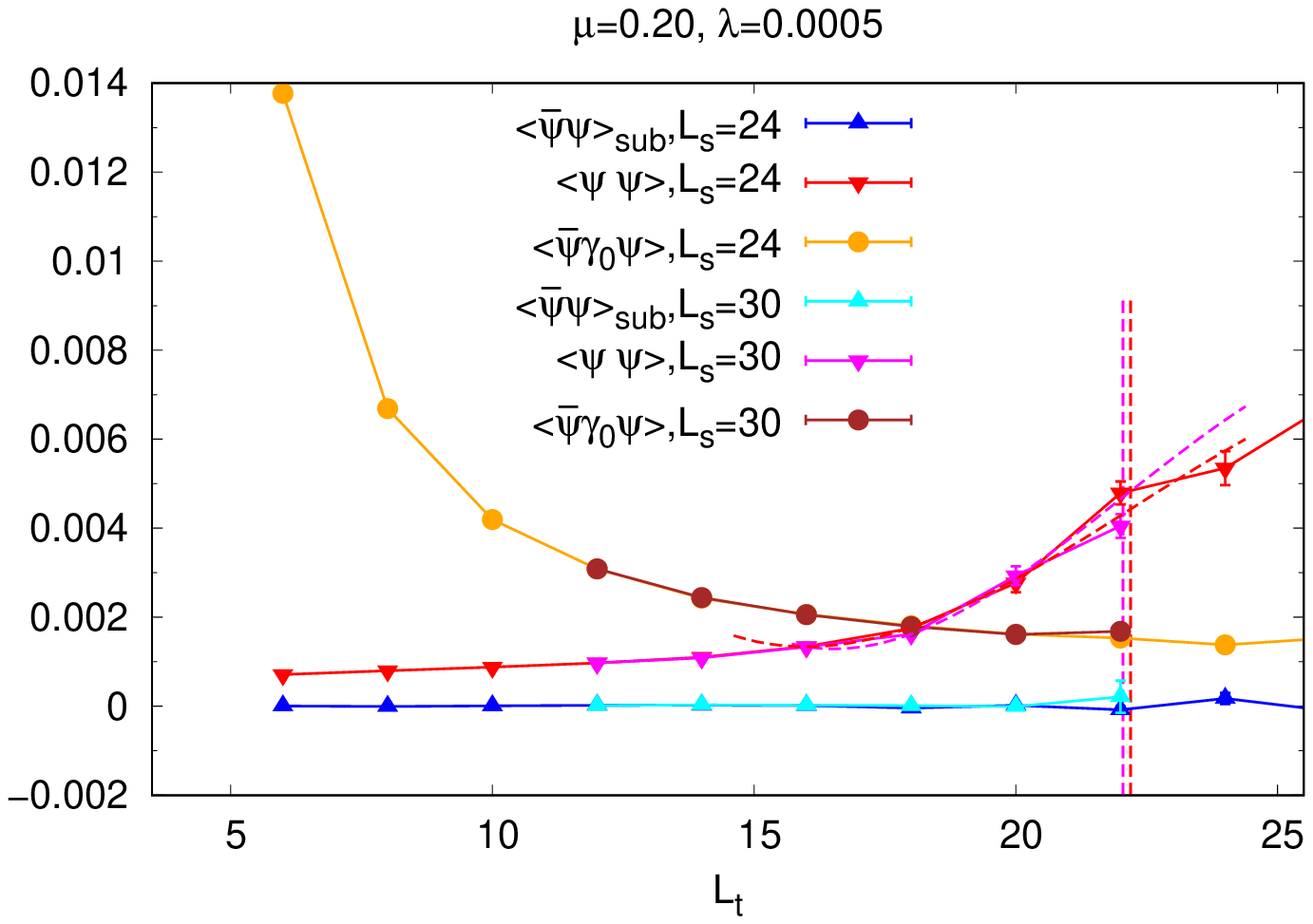}
  \includegraphics[width=0.45\textwidth]{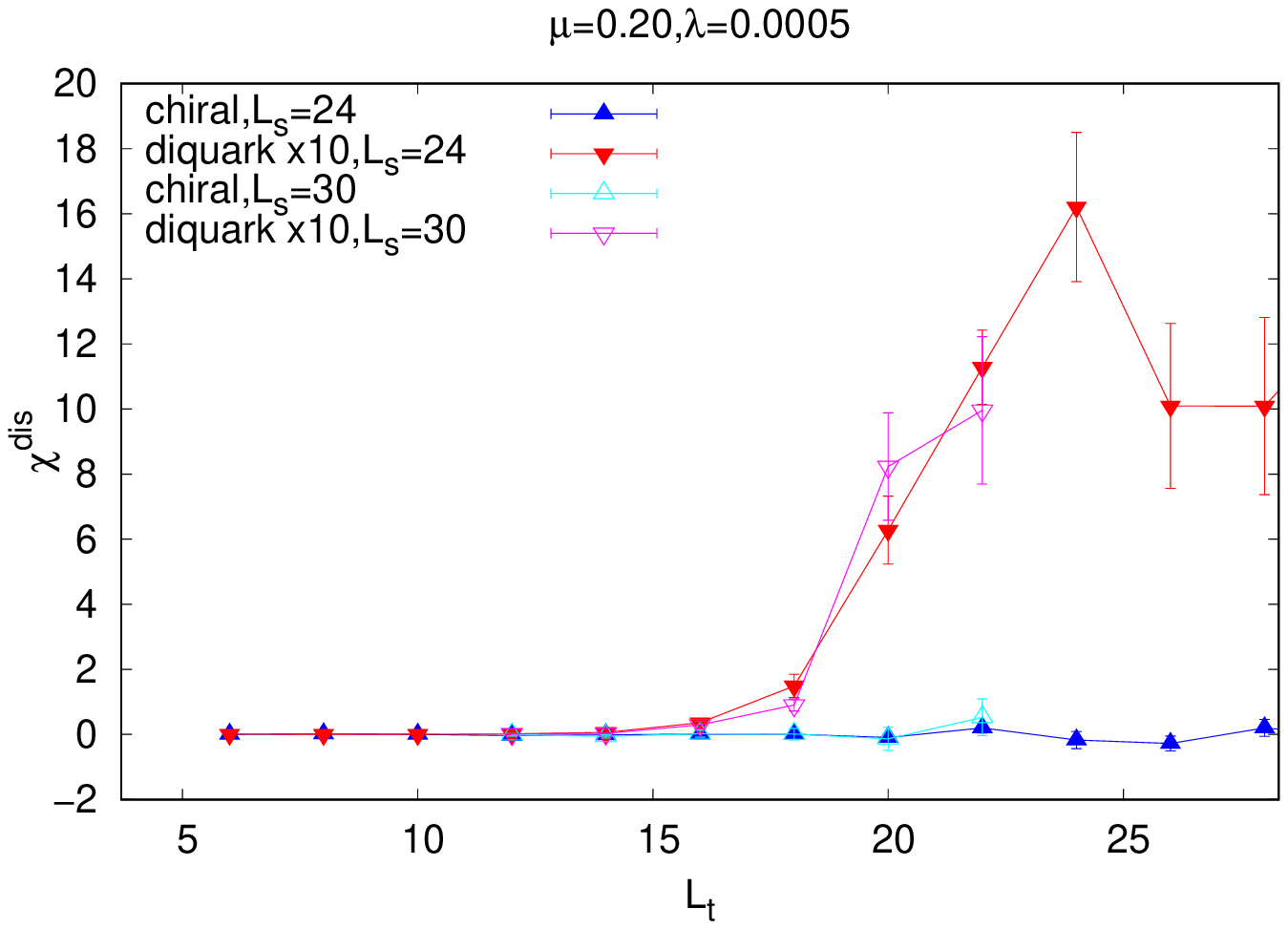}\\
  \caption{On the left: Quark density, subtracted chiral condensate and diquark condensate as functions of the temporal lattice size $L_t$ on lattices with $L_s=24$ and $L_s=30$. On the right: chiral susceptibility and diquark susceptibility as functions of $L_t$.}
  \label{fig:condensates_vs_t}
\end{figure*}

In Fig.~\ref{fig:condensates_vs_t} we illustrate how the quark density, the subtracted chiral condensate, the diquark condensate and their corresponding disconnected susceptibilities depend on the temporal lattice size $L_t$ at three different values of the chemical potential $a\mu = 0$, $a\mu = 0.05$, $a\mu = 0.2$, and at different spatial lattice sizes $L_s = 24$ and $L_s = 30$.

For $a\mu = 0$ and $a \mu = 0.05 < a m_{\pi}/2$ we expect the conventional chiral symmetry breaking pattern, and for $a\mu = 0.2$ and sufficiently low temperatures we approach the diquark condensation phase. Using this data we can locate chiral crossover and diquark condensation using the inflection points of either the chiral or the diquark condensates, considered as functions of $L_t$. For $a \mu = 0.0$ and $a \mu = 0.05$ (i.e. at $\mu < m_{\pi}/2$) we use the chiral condensate, for $a \mu \geq 0.1$ the diquark condensate is used. To identify the inflection point, we fit the chiral/diquark condensate data points with a third-order polynomial. The inflection point of the fitting polynomial is used as an estimate of the crossover position. These fits and the positions of the corresponding inflection points are shown on Fig.~\ref{fig:condensates_vs_t} with dashed lines. The estimates of critical $L_t$ obtained in this way are also in good agreement with the observed peaks in the corresponding susceptibilities. 

Note that the thermodynamic singularity in the chiral susceptibility is known to be associated with disconnected fermionic diagrams and the contribution of connected fermionic diagrams has only a mild temperature dependence (see e.g.~\cite{Bhattacharya:1402.5175}). \red{We have measured both the connected and the disconnected contributions to susceptibility, and our numerical results confirm this well-known feature of the chiral crossover.} For this reason on Fig.~\ref{fig:condensates_vs_t} we only plot the disconnected contribution. Since susceptibilities are much noisier observables than the condensates, we have decided to use the condensates instead of susceptibility peaks to identify the crossover position.

\begin{figure}[h!tpb]
  \centering
  \includegraphics[angle=-90,width=0.4\textwidth]{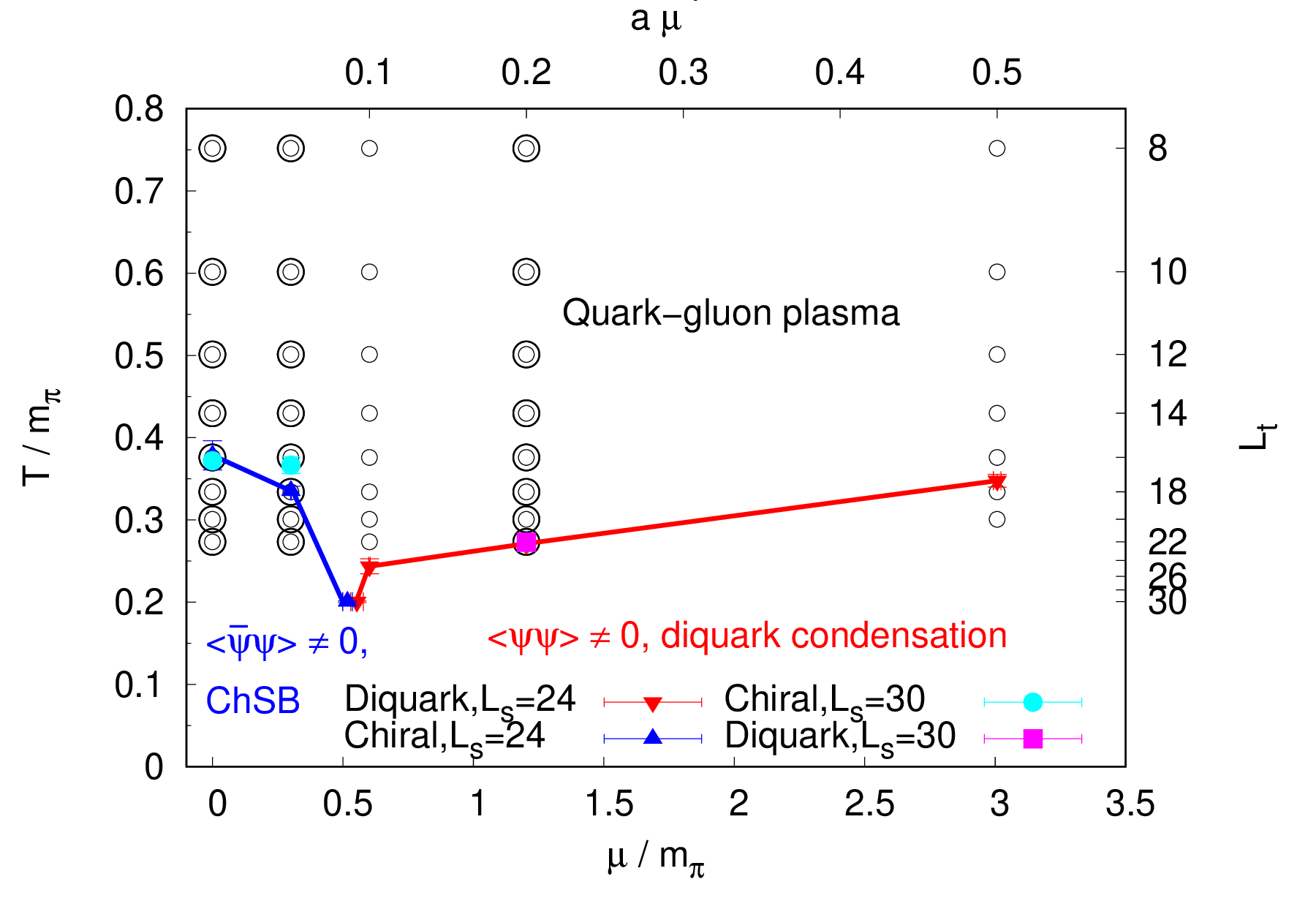}
  \caption{Numerical estimate of the phase diagram of finite-density $SU\lr{2}$ gauge theory with $N_f = 2$ rooted staggered fermions. Blue and red points correspond to inflection points of $L_t$ dependence of the chiral and diquark condensates, respectively, with lattice size $L_s=24$. Cyan and magenta points are similar estimates obtained on lattices with $L_s=30$. Circles and double circles correspond to parameter values for which the electric conductivity was analyzed with $L_s=24$ and $L_s=30$, respectively.}
  \label{fig:phase_diagram}
\end{figure}

The resulting estimates for the boundaries of the chirally broken phase and the diquark condensation phase are shown in Fig.~\ref{fig:phase_diagram}. Blue and red points correspond to inflection points of $L_t$ dependence of the chiral and diquark condensates, respectively. In this plot, the two points with the lowest temperature (corresponding to $L_t = 30$) were obtained in a different way: here, we have fixed $L_t = 30$ and looked for an inflection point in the $\mu$ dependence of the both the chiral (blue point) and the diquark (red point) condensates. The results coincide within statistical and fitting uncertainties, which suggests that the two phases coexist in this region of the phase diagram.

Our findings for the phase diagram agree well with previous results obtained in lattice simulations on sufficiently fine lattices \cite{Hands:1210.4496,Hands:1502.01219,Hands:1912.10975}, as well as within the functional renormalization group approach in effective low-energy theories \cite{Smekal:1112.5401,Strodthoff:1306.2897}. The chiral crossover moves towards lower temperatures as $\mu$ is increased towards the diquark condensation threshold. In particular, above this threshold, the critical temperature of the superfluid diquark condensation phase only rather weakly depends on $\mu$ as observed previously in two-color QCD \cite{Hands:1210.4496,Braguta:1605.04090,Braguta:1808.06466} and analogously for the pion condensation phase in QCD at finite isospin density as well \cite{Endrodi:1712.08190}.

\red{We note that in the absence of any order parameter, we expect the finite-temperature confinement-deconfinement phase transition at small densities $\mu \lesssim m_{\pi}/2$ to be either a crossover or a weak first-order phase transition. The fact that temperature dependence of both the chiral condensate and chiral susceptibility becomes somewhat more pronounced at a larger lattice size $L_s=30$ still leaves a possibility of a finite-order phase transition, but can be also consistent with a crossover scenario in which the chiral susceptibility is bounded by a large but finite value related to finite quark mass. In view of the fact that in the heavy quark mass limit the deconfinement transition in $SU\lr{2}$ gauge theory is a second-order phase transition \cite{Engels:hep-lat/9509091}, the scenario of first-order phase transition seems unlikely. However, the precise nature of the deconfinement transition can only be determined in a detailed study of volume dependence, which is out of the scope of this paper.}

An interesting feature of the phase diagrams obtained from lattice simulations both in this work and in \cite{Hands:1210.4496,Hands:1912.10975} is that the chiral crossover happens at temperatures which are several times lower than the pion mass. From our data we estimate $T_c/m_{\pi} \approx 0.37$ at $\mu = 0$, and $T_c/m_{\pi} \approx 0.2$ at $a \mu = 0.1$ near the diquark condensation threshold. Although the results of \cite{Hands:1210.4496,Hands:1912.10975} were obtained for considerably larger pion masses (larger values of $m_{\pi}/m_{\rho}$), the ratios $T_c/m_{\pi}$ at both $\mu = 0$ and $\mu = m_{\pi}/2$ obtained in these works are consistent with our estimates. Such small values of $T_c/m_{\pi}$ are in sharp contrast with real QCD, where $T_c \approx 155 \MeV$ \cite{Bazavov:1111.1710}, $m_{\pi} \approx 135 \MeV$ and hence $T_c/m_{\pi} = 1.15 > 1$. This difference might be explained by the fact that there are 5 Goldstone bosons in $SU\lr{2}$ gauge theory with $N_f = 2$ flavours \cite{Kogut:hep-ph/0001171}, in contrast to the 3 pions in $N_f = 2$ QCD.

\section{Numerical measurements of electric conductivity}
\label{sec:conductivity_preliminary}

By virtue of Green-Kubo relations \cite{Meyer:1104.3708}, within the linear response approximation the electric conductivity $\sigma\lr{\omega}$ is related to correlators of same-direction vector currents:
\begin{eqnarray}
\label{GreenKubo_conductivity}
 \frac{1}{V} \sum\limits_{\vec{x}}\vev{j_i\lr{\tau, \vec{x}} j_i\lr{0, \vec{0}}}
 \equiv
 G\lr{\tau}
 = \nonumber \\ =
 \int\limits_{0}^{\infty} d\omega  \, K\lr{\tau, \omega} \, \sigma\lr{\omega} ,
 \nonumber \\
 K\lr{\tau, \omega}
 =
 \frac{\omega}{\pi} \,
 \frac{\cosh\lr{\omega \lr{ \tau - \frac{1}{2T} }}}{\sinh\lr{\frac{\omega}{2 T}}}
\end{eqnarray}
where $j_i\lr{\tau, \vec{x}}$ is the vector current density in some fixed spatial direction $i = 1,2,3$, $\sum\limits_{\vec{x}}$ denotes summation over spatial lattice coordinates, $V = a^3 L_s^3$ is the spatial lattice volume and $\tau \in \lrs{0 \ldots a L_t}$.

While the inversion of the relation (\ref{GreenKubo_conductivity}) is a numerically ill-posed problem, a number of practical inversion methods have been developed which either take into account some prior knowledge of $\sigma\lr{\omega}$ or return $\sigma\lr{\omega}$ smeared over a certain frequency range of order of temperature \cite{Meyer:1104.3708}. In this work we use the Backus-Gilbert method \cite{Meyer:1104.3708} with Tikhonov regularization \cite{Ulybyshev:1707.04212,Tripolt:1801.10348} as implemented in \cite{UlybyshevGreenKuboGitHub}.

Within the Backus-Gilbert method we construct the linear estimator of the conductivity based on the Euclidean current-current correlator (\ref{GreenKubo_conductivity}):
\begin{eqnarray}
\label{BG_estimator_def}
 \sigma_{BG}\lr{\omega}
 =
 \sum\limits_{\tau} q_{\tau}\lr{\omega} G\lr{\tau}
 = \nonumber \\ =
 \int\limits_{0}^{+\infty} \delta_{BG}\lr{\omega, \omega'} \sigma\lr{\omega'} ,
 \nonumber \\
 \delta_{BG}\lr{\omega, \omega'} = \sum\limits_{\tau} q_{\tau}\lr{\omega} K\lr{\tau, \omega'} .
\end{eqnarray}
where the resolution functions $q_{\tau}\lr{\omega}$ are chosen in such a way that, combined with the Green-Kubo kernel $K\lr{\tau, \omega}$ in (\ref{GreenKubo_conductivity}), they yield a smearing function $\delta_{BG}\lr{\omega, \omega'}$ which approximates the $\delta$-function as closely as possible. In the Backus-Gilbert method, we minimize the ``dispersion'' $\int\limits_0^{\infty} d\omega' \delta_{BG}^2\lr{\omega, \omega'} \lr{\omega - \omega'}^2$. This minimization requires an inversion of a certain ill-conditioned matrix constructed from the kernel $K\lr{\tau, \omega}$. With Tikhonov regularization this inversion is regularized by replacing the inverse singular values $1/x_i$ of this matrix by $x_i/\lr{x_i^2 + \Delta^2}$ with some small $\Delta$. This effectively cuts off the singular values $x_i$ which are smaller than $\Delta$ and thus makes the matrix inversion well-defined.

In contrast to other regularization schemes which use the covariance matrix for the Euclidean correlator in (\ref{GreenKubo_conductivity}), with Tikhonov regularization the resolution functions do not depend on the data and thus neither on the chemical potential $\mu$, which allows for a more meaningful comparison of data obtained at different values of $\mu$, and with the error-free data for free quarks as well. We calculate statistical errors for the smeared conductivity using data binning \cite{Ulybyshev:1707.04212}.

\begin{figure}[h!tpb]
  \centering
  \includegraphics[angle=-90,width=0.45\textwidth]{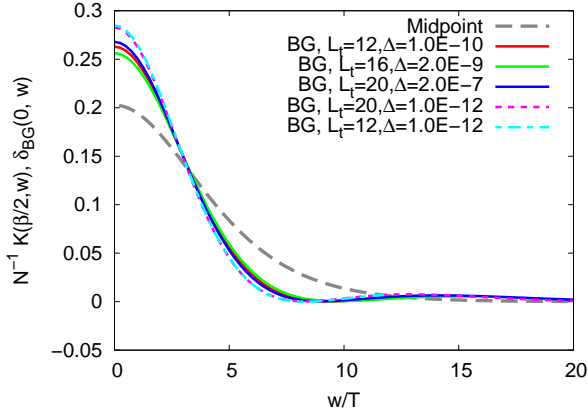}
  \caption{Comparison of resolution functions $\delta_{BG}\lr{0, \omega}$ and $\lr{\pi T^2}^{-1} K\lr{a L_t/2, \omega}$ for Backus-Gilbert and midpoint estimates of the low-frequency limit of the electric conductivity.}
  \label{fig:resolution_functions}
\end{figure}

Since the smeared conductivity $\sigma_{BG}\lr{\omega}$ in practice quite strongly depends on the regularization of the matrix inversion and the value of regularization parameters, the Backus-Gilbert method to some extent still suffers from the inherent ambiguity which is typical for numerically ill-defined analytic continuation problems \cite{Meyer:1104.3708}. To assess any residual ambiguity, in addition, we also consider an alternative simple estimator of the low-frequency conductivity \cite{Kaczmarek:1012.4963}. Namely, according to the Green-Kubo relation in (\ref{GreenKubo_conductivity}), the current-current correlator on the l.h.s. of Eq.~(\ref{GreenKubo_conductivity}), at the maximal Euclidean time separation $\tau = a L_t/2$, is related to the electric conductivity as
\begin{eqnarray}
\label{MP_estimate1}
 G\lr{a L_t/2} =
 \int\limits_{0}^{\infty} d \omega
 \, K\lr{a L_t/2, \omega} \, \sigma\lr{\omega} .
\end{eqnarray}
The function $K\lr{a L_t/2, \omega} = \omega/\pi \lr{\sinh\lr{\frac{\omega}{2T}}}^{-1}$ is localized within the region of small frequencies $\omega \sim T$ and can be also considered as a ``smeared'' $\delta$-function similar to the one used in the Backus-Gilbert method. The norm and width of this function are:
\begin{eqnarray}
\label{MP_norm_width}
 \mathcal{N} \equiv \int\limits_{0}^{\infty} d \omega
 \, K\lr{a L_t/2, \omega} = \pi T^2 ,
 \nonumber \\
 \Delta\omega = \sqrt{\mathcal{N}^{-1} \, \int\limits_{0}^{\infty} d\omega\, \omega^2 \, K\lr{a L_t/2, \omega} }
 = %\nonumber \\ =
 \sqrt{2} \pi T .
\end{eqnarray}
We can thus use the value of the Euclidean correlator at midpoint as an estimator $\sigma_{MP}$ of electric conductivity $\sigma\lr{\omega}$ smeared over frequencies in the range $\omega \lesssim \sqrt{2} \pi T \approx 4.4 \, T$:
\begin{eqnarray}
\label{MP_estimator_def}
 \sigma_{MP} = \frac{1}{\pi T^2}\, G\lr{a L_t/2} .
\end{eqnarray}
In Fig.~\ref{fig:resolution_functions} we compare the resolution function $\mathcal{N}^{-1} \, K\lr{a L_t/2, \omega}$ for the midpoint estimator (\ref{MP_estimator_def}) with resolution functions in the Backus-Gilbert method.

The contribution of connected fermionic diagrams to the current-current correlator in (\ref{GreenKubo_conductivity}) for a single gauge field configuration can be written as a single trace over fermionic indices (spin, color and lattice coordinates):
\begin{eqnarray}
\label{current_current_connected}
 \vev{j_{x,\mu} j_{y,\nu}}_{conn}
 = \nonumber \\ =
 C_{em} \, \left. \tr\lr{
  \frac{\partial D}{\partial \theta_{x,\mu}} D^{-1}
  \frac{\partial D}{\partial \theta_{y,\nu}} D^{-1}
 } \right|_{\theta=0} ,
\end{eqnarray}
where $x$, $y$ and $\mu$, $\nu$ label the sites and the directions on the four-dimensional (for WD fermions) or five-dimensional (for DW fermions) lattice, and $D$ is either the WD or DW Dirac operator in the background of the non-Abelian gauge fields and an Abelian lattice gauge field $\theta_{x,\mu}$, with link factors $e^{i \theta_{x,\mu}}$. The electric charge factor
\begin{eqnarray}
\label{charge_factor_def}
 C_{em} = \sum\limits_{f=u,d} q_f^2 = 5/9
\end{eqnarray}
is the sum of squared quark charges for the $u$- and $d$-quarks. In this work we follow most of the previous lattice QCD studies of the electric conductivity \cite{Brandt:1710.07050,Kaczmarek:1604.06712,Kaczmarek:1604.07544,Meyer:1512.07249,Aarts:1412.6411,Aarts:1307.6763,Buividovich:10:1,Kaczmarek:1012.4963} and present all the results with $C_{em}$ factored out.

At $x = y$ there is also an additional contact term contribution $\left. \tr\lr{\frac{\partial^2 D}{\partial \theta_{x,\mu}^2} D^{-1}}\right|_{\theta=0}$ to the correlator (\ref{current_current_connected}). This contact term affects only the high-frequency behavior of the electric conductivity, and we disregard it in the following. The time slice $\tau = 0$ in the current-current correlator (\ref{GreenKubo_conductivity}), for which this contact term is relevant, is discarded within the Backus-Gilbert method.

In the presence of a nonzero diquark source $\lambda$ the expression for the connected contribution (\ref{current_current_connected}) is somewhat more complicated, and is given in Appendix~\ref{apdx:current_diquarks}.

The contributions from disconnected fermionic diagrams
\begin{eqnarray}
\label{current_current_disconnected}
 \vev{j_{x,\mu} j_{y,\nu}}_{disc}
 =
 C_{disc}
 \times \nonumber \\ \times
 \left.
 \tr\lr{ \frac{\partial D}{\partial \theta_{x,\mu}} D^{-1} }
 \right|_{\theta=0} \,
 \left.
 \tr\lr{ \frac{\partial D}{\partial \theta_{y,\nu}} D^{-1} }
 \right|_{\theta=0} ,
\end{eqnarray}
to the current-current correlator (\ref{GreenKubo_conductivity}) are typically small and noisy, and in addition weighted by the charge factor $C_{disc} = \lr{\sum\limits_{f=u,d} q_f}^2 = 1/9$, which is five times smaller than the charge factor $C_{em} = 5/9$ for connected diagrams. Nevertheless, in Section~\ref{sec:conductivity_results} below we explicitly check the smallness of disconnected contribution at finite densities.

In order to obtain the four-dimensional conserved vector current for DW fermions, one should sum the correlator (\ref{current_current_connected}) over the fifth dimension, i.e. over $x_5 = 0 \ldots L_5-1$ and $y_5 = 0 \ldots L_5 - 1$ \cite{Furman:hep-lat/9405004}. This increases the number of Dirac operator inversions required to calculate (\ref{current_current_connected}) by a factor of $L_5$, thus making calculations with DW fermions significantly more expensive than with WD fermions. In Appendix~\ref{apdx:current_calculation} we discuss a small trick which allows to halve this numerical cost. Further, to obtain physical results with DW fermions it is crucial to subtract the contribution of five-dimensional bulk Dirac modes, which becomes quite significant at high temperatures and/or at small values of $\tau$ in (\ref{GreenKubo_conductivity}). As discussed in \cite{Furman:hep-lat/9405004}, this contribution can be compensated by the contribution of bosonic Pauli-Villars fields which live on the five-dimensional lattice with two times smaller size $L_5$ in the fifth dimension. This contribution is equal to minus twice the correlator (\ref{current_current_connected}) calculated with $L_5^{PV} = L_5/2$ and the bare mass $m^{PV} = 1$ in the DW Dirac operator.

\section{Numerical results}
\label{sec:conductivity_results}

\subsection{Euclidean correlators and midpoint conductivity estimates}
\label{subsec:numres_corrs}

\begin{figure*}[h!tpb]
  \centering
  \includegraphics[angle=-90,width=0.45\textwidth]{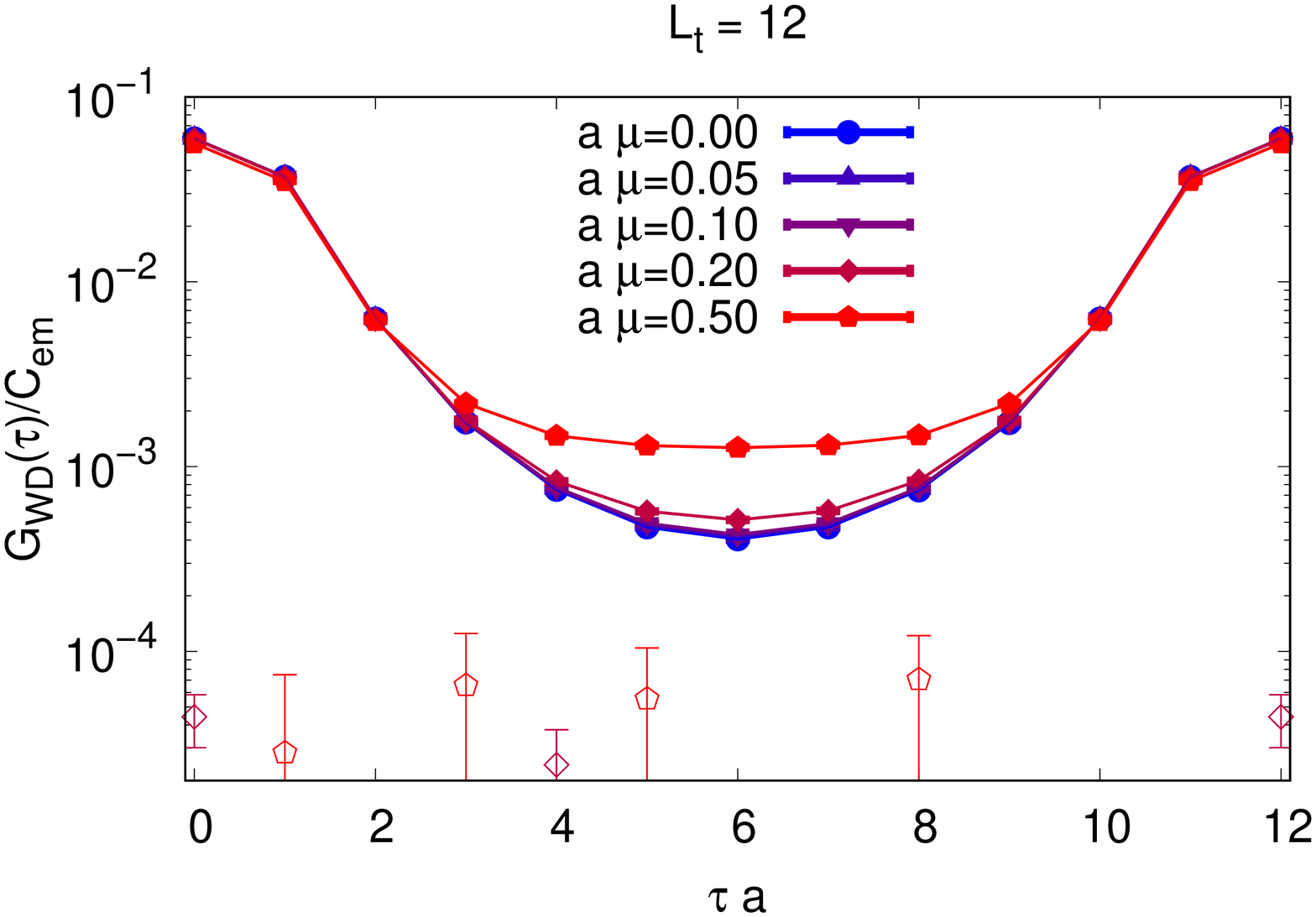}
  \includegraphics[angle=-90,width=0.45\textwidth]{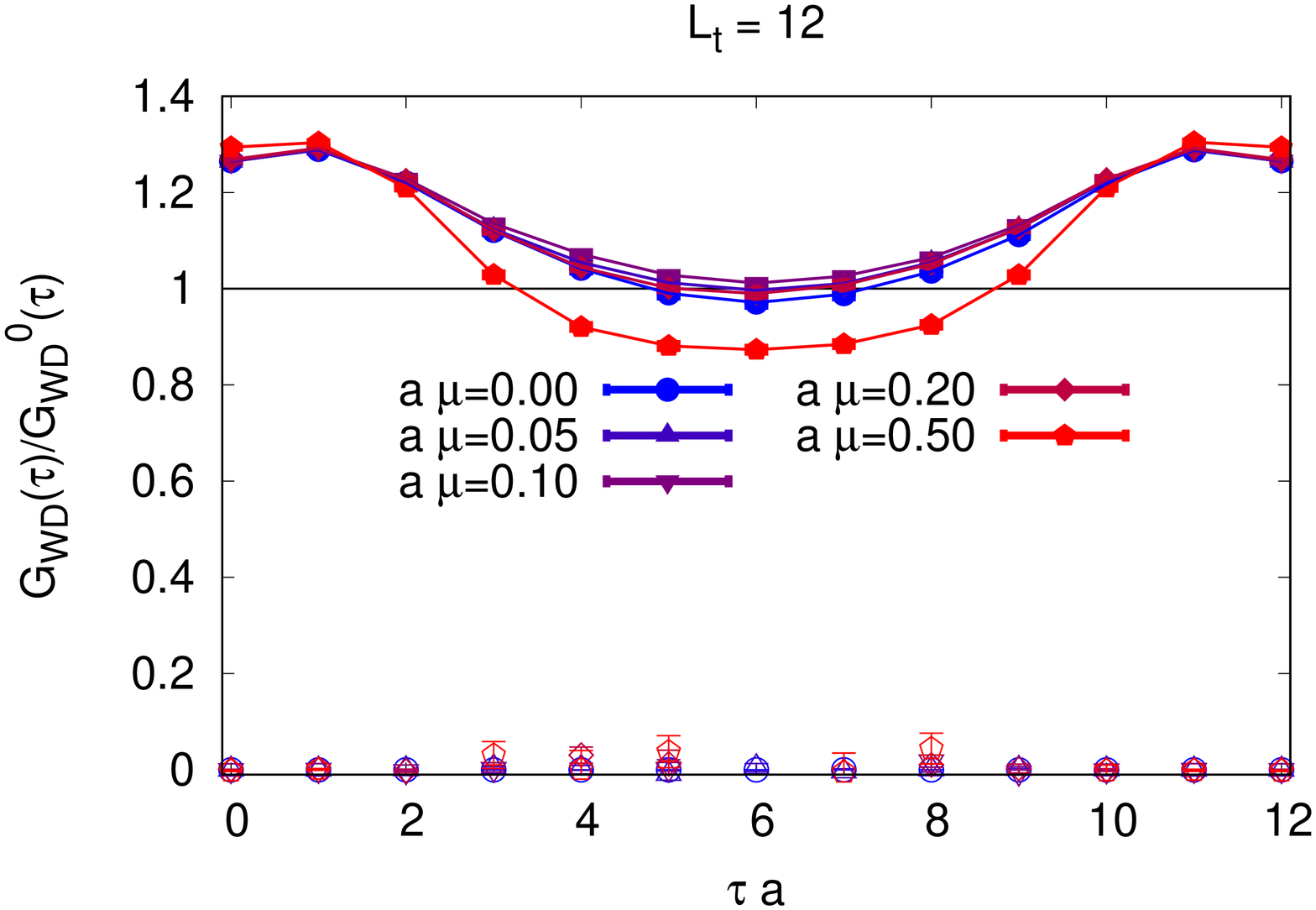}\\
  \includegraphics[angle=-90,width=0.45\textwidth]{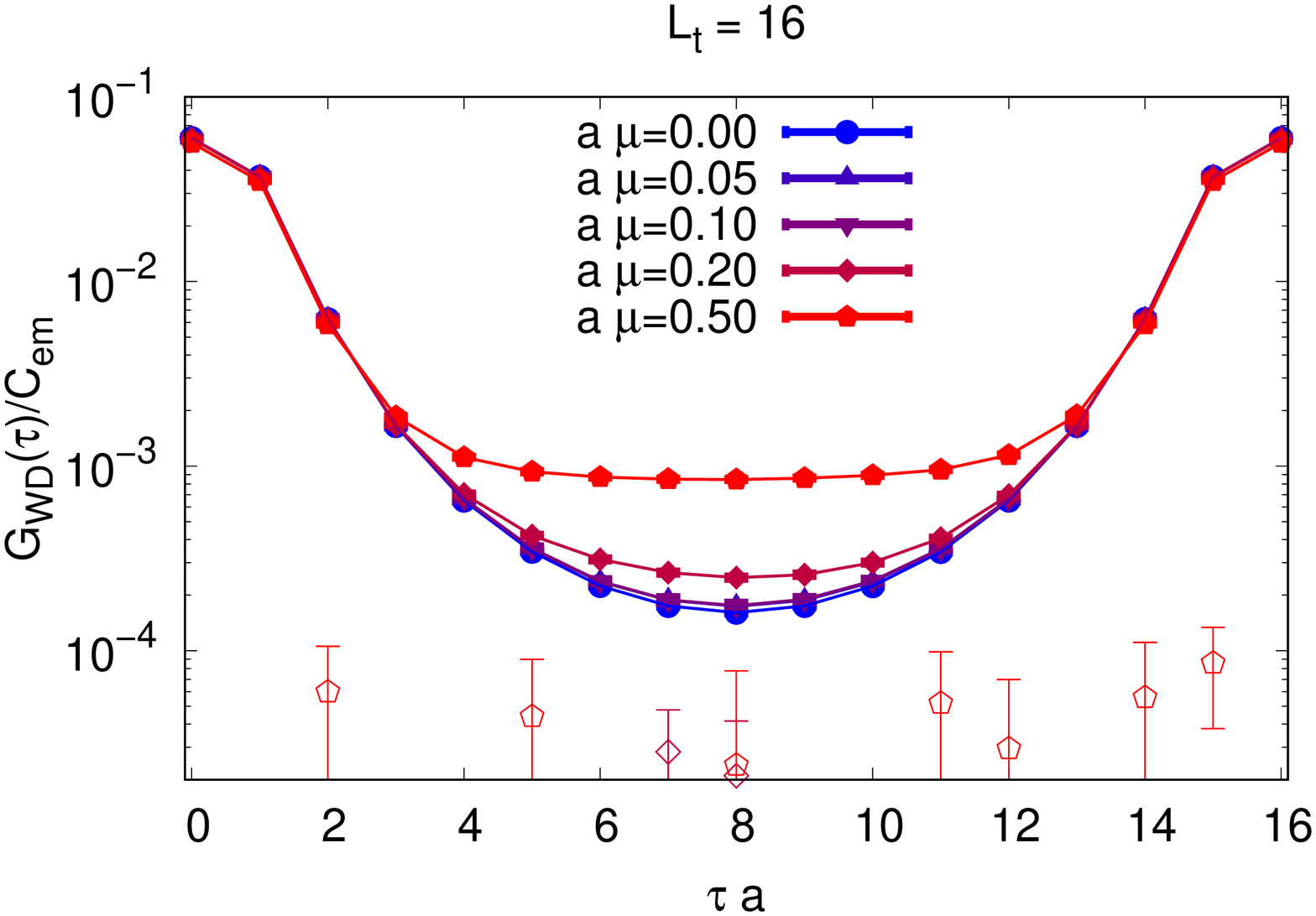}
  \includegraphics[angle=-90,width=0.45\textwidth]{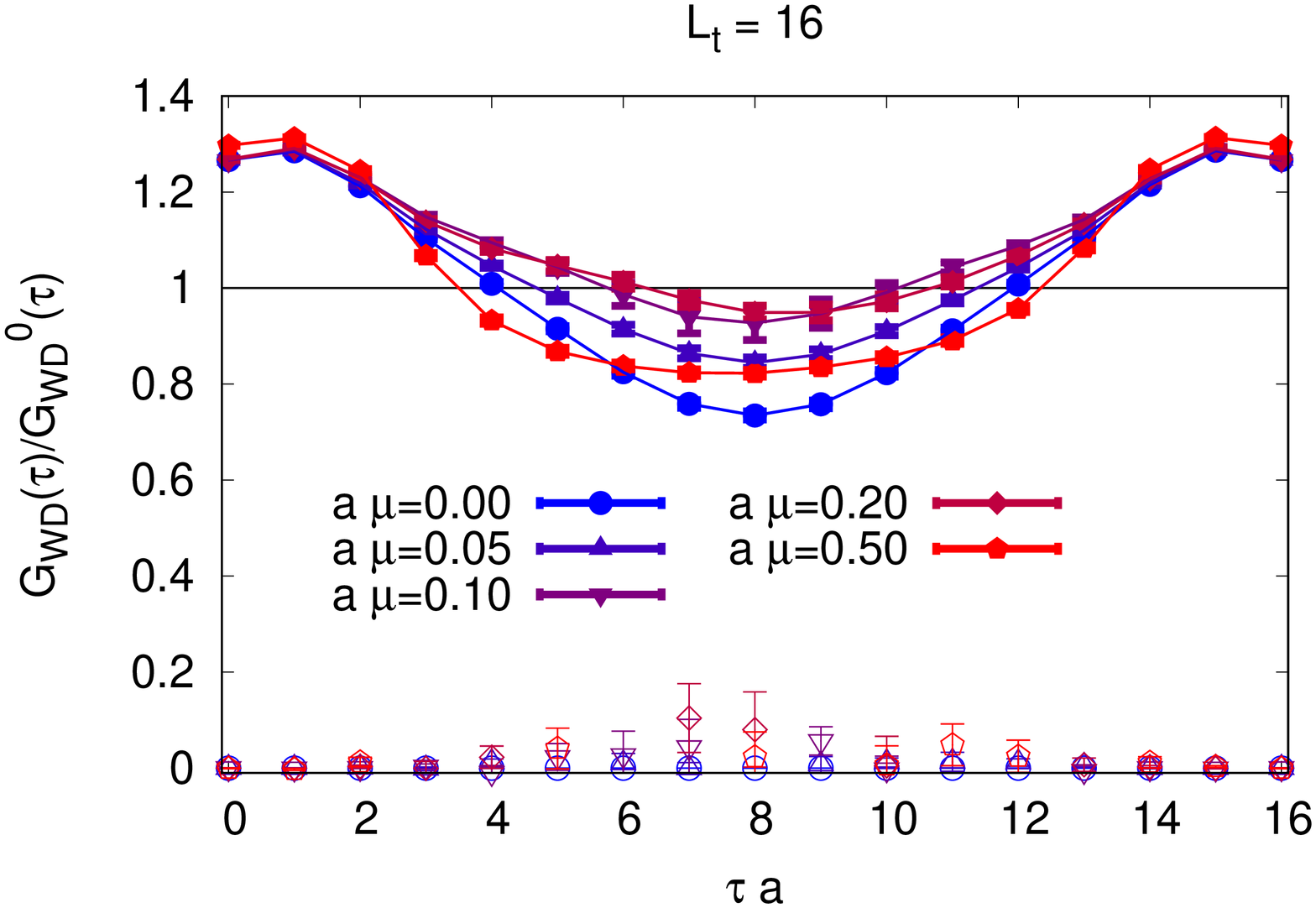}\\
  \includegraphics[angle=-90,width=0.45\textwidth]{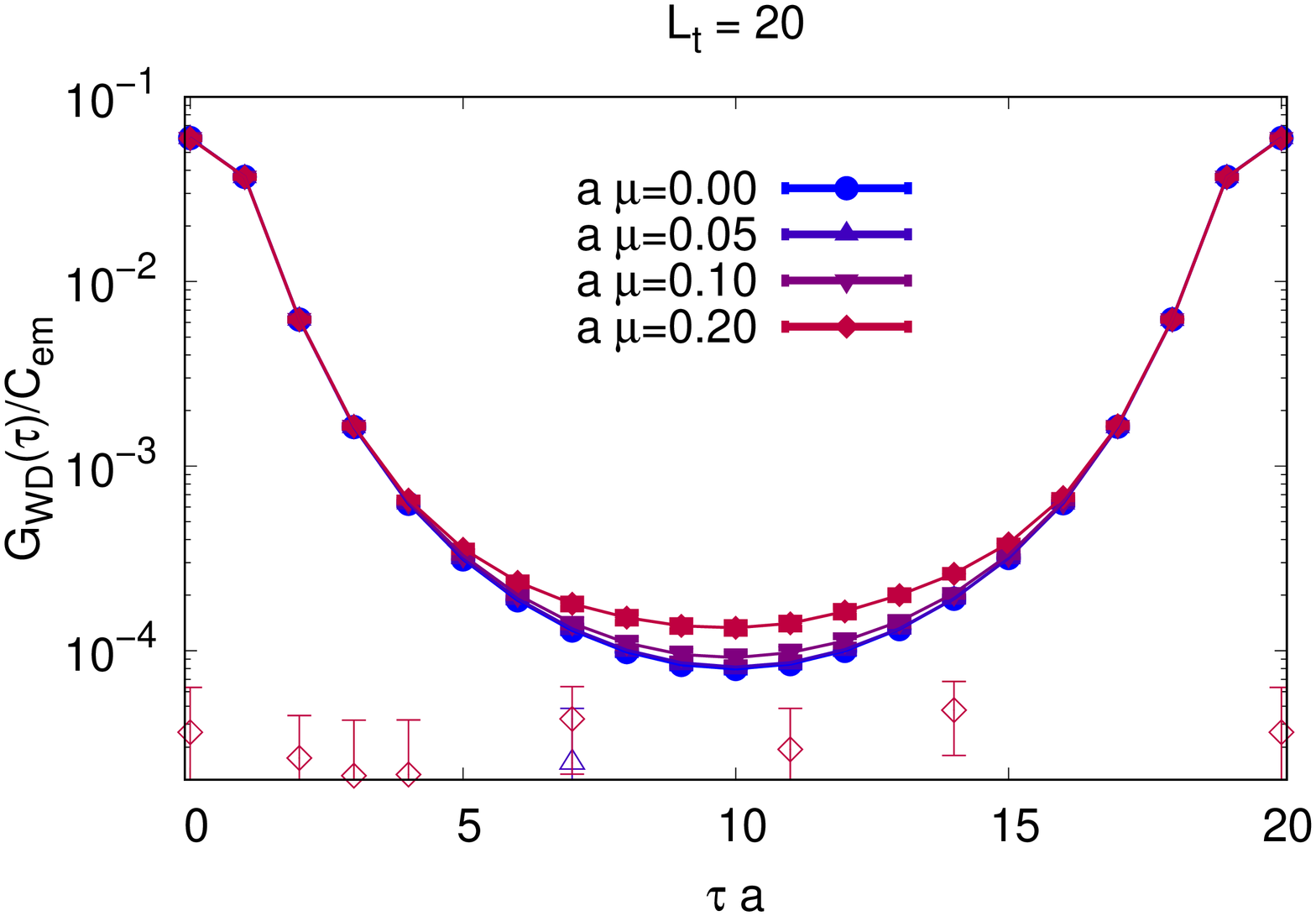}
  \includegraphics[angle=-90,width=0.45\textwidth]{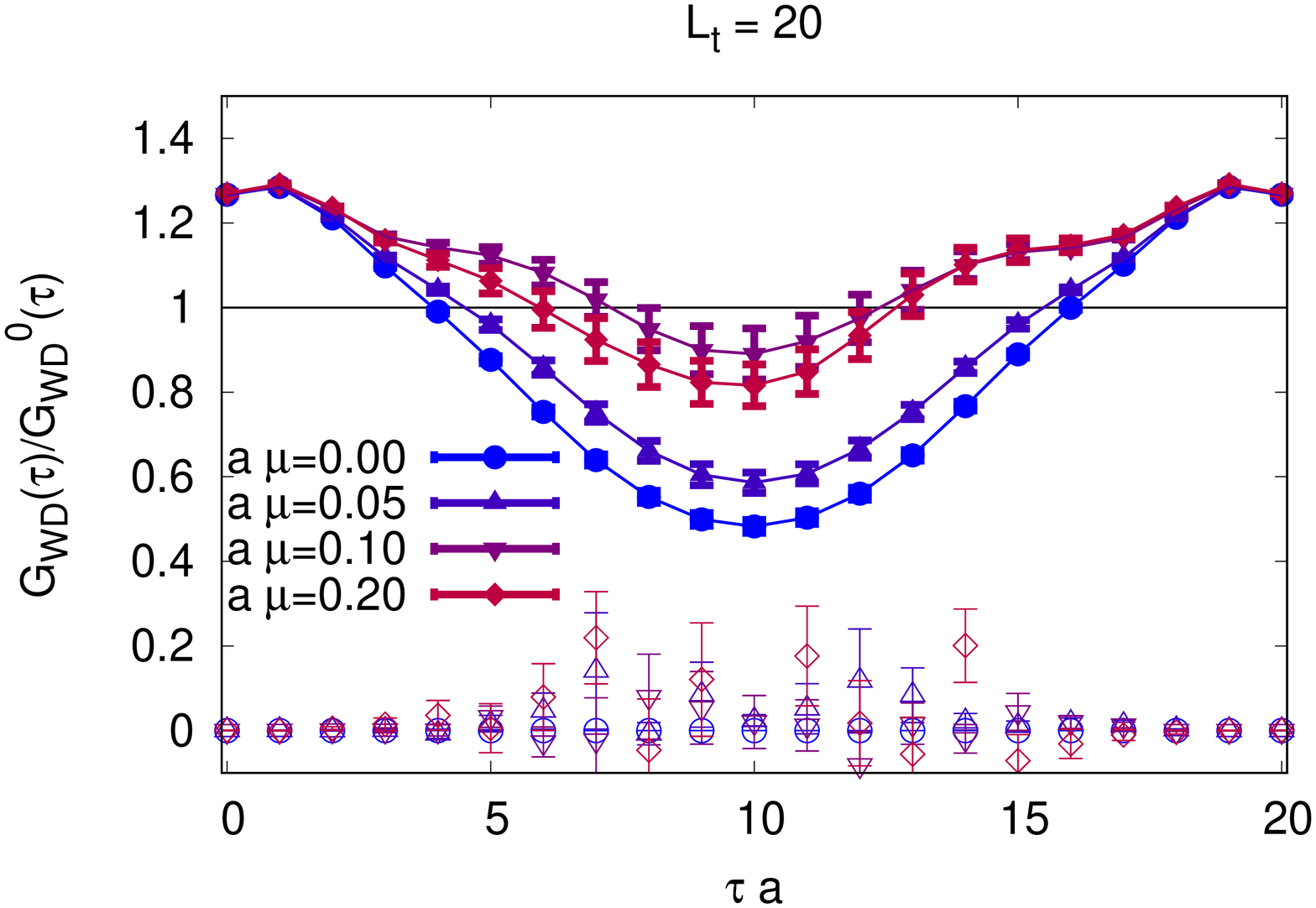}\\
  \caption{\textbf{On the left:} current-current correlators $G_{WD}\lr{\tau}$ obtained with Wilson-Dirac fermions on the lattice with $L_s = 24$. \textbf{On the right:} the ratio of current-current correlators $G_{WD}\lr{\tau}$ in $SU\lr{2}$ theory to the current-current correlators $G_{WD}^0\lr{\tau}$ calculated for non-interacting Wilson-Dirac quarks. On both sides, empty symbols show the corresponding disconnected contribution, multiplied by the ratio of charge factors $C_{disc}/C_{em} = 1/5$.}
  \label{fig:jVjV_WD}
\end{figure*}

We start the discussion of our numerical results by considering Euclidean current-current correlators which enter the Green-Kubo relations (\ref{GreenKubo_conductivity}). In Fig.~\ref{fig:jVjV_WD} we plot connected current-current correlators obtained with Wilson-Dirac fermions at three different temperatures corresponding to $L_t = 12, \, 16, \, 20$ and compare them with the corresponding disconnected contributions, as well as with the corresponding correlators for free quarks on the same lattice (plots on the right). Both connected and disconnected contributions were calculated for all configurations in the ensembles listed in Table~\ref{tab:num_configs}, and in addition averaged over $10 \ldots 30$ random source positions in order to reduce statistical errors.

One can see that as the chemical potential gradually increases, the connected current-current correlators (\ref{current_current_connected}) around mid-point become larger and more flat, which agrees qualitatively with the expected growth of the electric conductivity in a finite-density system. On the other hand, the behavior at small Euclidean time separations, which is most sensitive to the high-frequency part of the spectrum, is practically unaffected by finite density.

We've also invested a significant amount of CPU/GPU time into measuring disconnected contributions (\ref{current_current_disconnected}), and were not able to detect any statistically significant deviation from zero. For small values of the chemical potential we were able to reduce statistical errors of disconnected contributions such that they are at least $3-4$ orders of magnitude smaller than the connected contributions. However, for larger $\mu$ and smaller $T$ we were not able to reduce statistical errors of disconnected contributions below $10 \ldots 20 \%$ of the connected ones. Thus we cannot rule out that disconnected contributions might become important at very large densities and low temperatures, for instance, in the quarkyonic phase.

\begin{figure*}[h!tpb]
  \centering
  \includegraphics[angle=-90,width=0.45\textwidth]{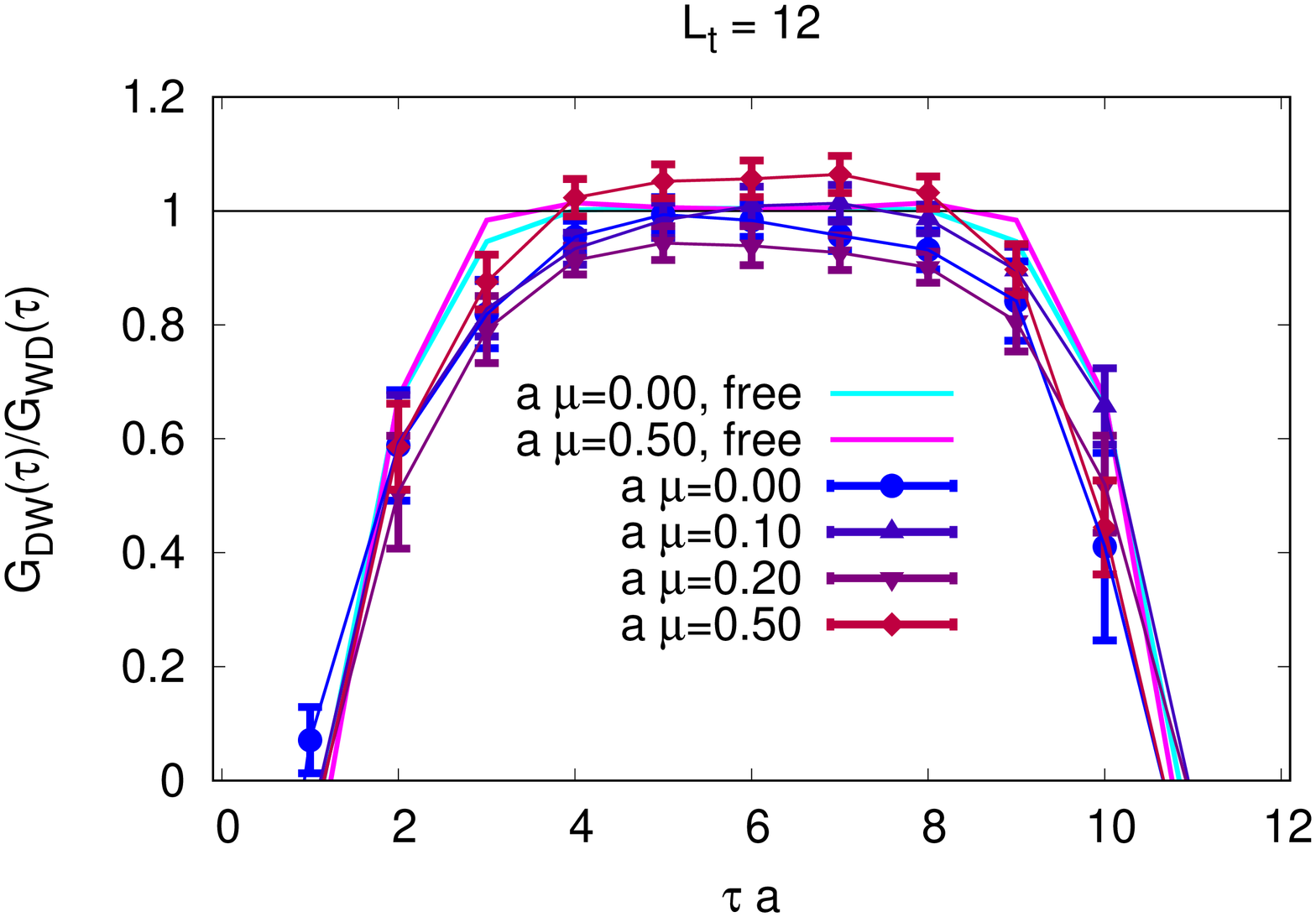}
  \includegraphics[angle=-90,width=0.45\textwidth]{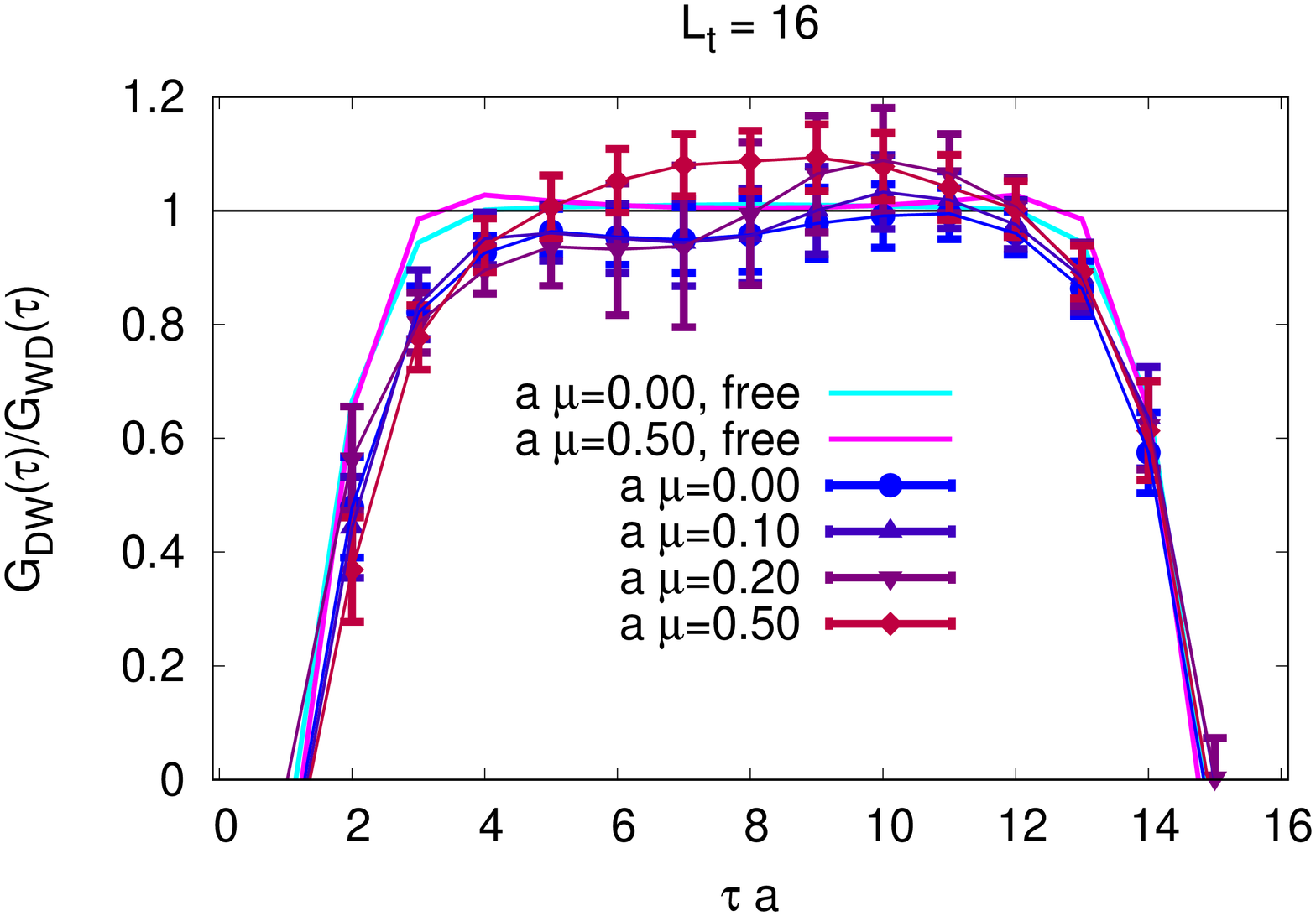}\\
  \caption{The ratio of current-current correlators $G_{WD}\lr{\tau}$ and $G_{DW}\lr{\tau}$ obtained with Wilson-Dirac and Domain Wall fermions, respectively, on the lattice with $L_s = 24$. For comparison, we also plot the corresponding ratio $G_{WD}^0\lr{\tau}/G_{DW}^0\lr{\tau}$ for free Wilson-Dirac and Domain Wall quarks.}
  \label{fig:jVjV_DW_vs_WD}
\end{figure*}

We also note that for lower temperatures and larger values of the chemical potential the current-current correlators become significantly noisier. In addition, their statistical distribution seems to develop heavy tails, so that contributions from outlier configurations become more and more important. These outlier configurations present a major challenge for calculating connected current-current correlators at low temperatures and high densities.

A comparison with the free quark results is shown on plots on the right in Fig.~\ref{fig:jVjV_WD}). For this as well as for all other calculations with free quarks we use the bare quark mass $a m = 0.01$, which corresponds to the optimal value of quark mass in the DW Dirac operator which, for interacting theory, reproduces the pion mass obtained with staggered fermions (see Fig.~\ref{fig:mpi_vs_mq}). This choice is dictated by the expected smallness of mass renormalization effects for DW fermions, which makes the bare and renormalized masses close to each other. Clearly, for free WD fermions mass renormalization is absent, and we can use the same value of bare mass as for free DW fermions. In any case, current-current correlators depend only weakly on the bare quark mass, and changing its value by $\pm 50 \%$ does not lead to any noticeable change in results for our lattice parameters.

The comparison with free-quark results shows that for all values of the chemical potential the relative difference between current-current correlators in interacting and non-interacting theories does not exceed $50\%$, which agrees with previous studies at zero chemical potential \cite{Kaczmarek:1012.4963}. At not very large densities, chemical potential moves current-current correlators closer to the free quark result. An interesting feature is that for $a \mu = 0.5$ this trend is reversed, and for these values of $\mu$ the current-current correlator around midpoint becomes smaller than the corresponding free-quark result.

\begin{figure}[h!tpb]
  \centering
  \includegraphics[angle=-90,width=0.45\textwidth]{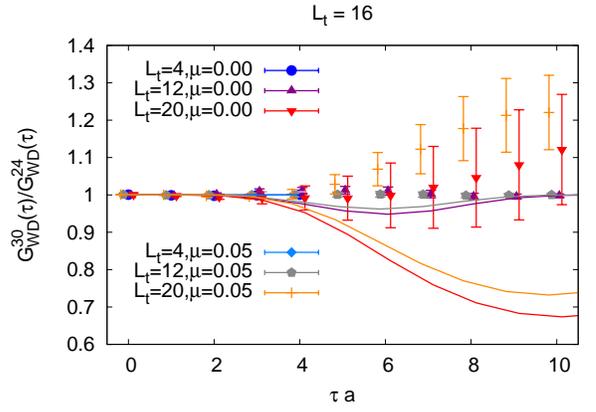}\\
  \caption{The ratio $G_{WD}\lr{\tau, L_s = 30}/G_{WD}\lr{\tau, L_s = 24}$ of current-current correlators $G_{WD}\lr{\tau}$ calculated on lattices with spatial sizes $L_s = 24$ and $L_s = 30$. Solid lines show the same relative difference calculated for free quarks. Data points are slightly shifted away from integer values of $\tau a$ in order to improve the readability of the plot.}
  \label{fig:jVjV_Ns24_vs_Ns30}
\end{figure}

In order to check how the chiral properties of lattice fermions might affect the electric conductivity, in Fig.~\ref{fig:jVjV_DW_vs_WD} we plot the ratios of connected contributions (\ref{current_current_connected}) to current-current correlators in (\ref{GreenKubo_conductivity}) calculated with Wilson-Dirac (WD) and with Domain-Wall (DW) quarks. Around mid-point, the results obtained with both WD and DW Dirac operators agree within statistical errors. For DW Dirac operators, the latter are noticeably larger due to smaller statistics (as calculations with DW quarks are more than an order of magnitude more expensive than with WD quarks). A salient feature of the current-current correlators for DW fermions is that they strongly deviate from the Wilson-Dirac result at short Euclidean time separations, where the contribution of five-dimensional bulk modes becomes important and is not completely cancelled by Pauli-Villars regulator fields.

\begin{figure*}[h!tpb]
  \centering
  \includegraphics[angle=-90,width=0.45\textwidth]{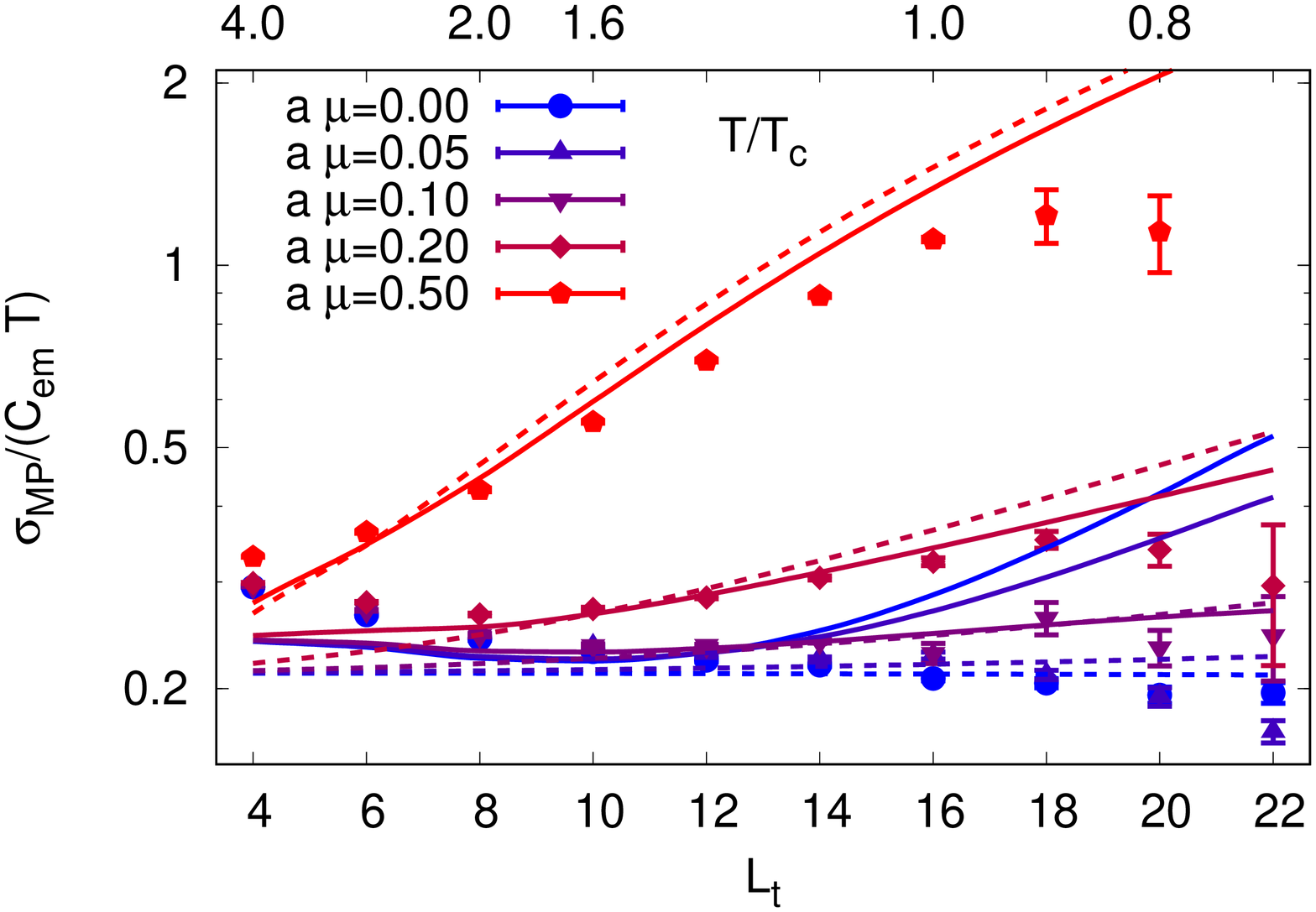}
  \includegraphics[angle=-90,width=0.45\textwidth]{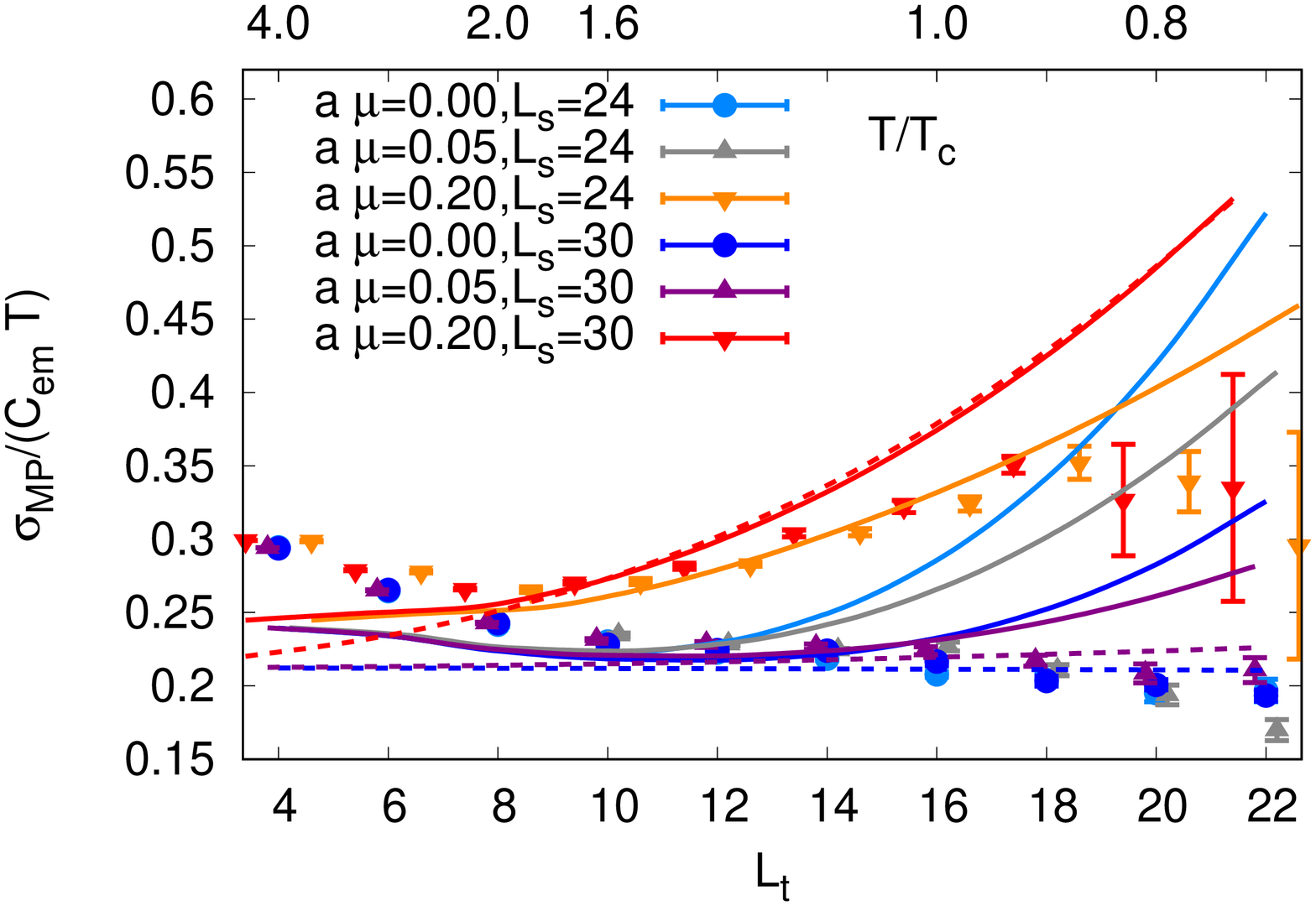}\\
  \caption{Midpoint estimate $\sigma_{MP}$ of the low-frequency electric conductivity obtained with Wilson-Dirac fermions as a function of temperature at different values of the chemical potential. \textbf{On the left:} for lattices with $L_s = 24$. \textbf{On the right:} comparison of results on lattices with $L_s = 24$ and $L_s = 30$. We switch from zero diquark source to nonzero $\lambda = 5 \cdot 10^{-4}$ at $L_t = 10$ for $L_s = 24$ and at $L_t=14$ for $L_s=30$. Solid lines are the results obtained with free quarks at the same temperature, lattice size and chemical potential. Dotted lines are the results for free continuum quarks at the same temperature and chemical potential.}
  \label{fig:conductivity_MP_summary}
\end{figure*}

In order to estimate possible finite-volume artifacts in our study, in Fig.~\ref{fig:jVjV_Ns24_vs_Ns30} we compare finite-density connected current-current correlators calculated on lattices with $L_s = 24$ and $L_s = 30$ using Wilson-Dirac fermions with $a \mu = 0.0$ and $a \mu = 0.05$ both in the full $SU\lr{2}$ lattice gauge theory and for free quarks. The deviations clearly grow towards lower temperatures and become quite significant for free quarks, but do not exceed $10 \%$ for the full gauge theory. An important observation is that in the full gauge theory deviations due to finite-volume effects appear to have opposite sign to those in the free quark case.

We now turn to the estimates of the low-frequency electric conductivity $\sigma_{MP}$ based on the mid-point values of current-current correlators, as defined in (\ref{MP_estimator_def}). As discussed in Section~\ref{sec:conductivity_preliminary} above, these estimates are completely model-independent and do not depend in any way on the method of performing numerical analytic continuation of Euclidean data. An analysis of current-current correlators for free quarks (see Appendix~\ref{apdx:finite_volume_effects}) suggests that the midpoint estimator is also somewhat less affected by finite-volume effects.

In Fig.~\ref{fig:conductivity_MP_summary} we show the dependence of the ratio $\sigma_{MP}/\lr{C_{em} T}$ on the inverse temperature $1/\lr{a T} \equiv L_t$ in lattice units, calculated using Wilson-Dirac fermions. Points with error bars correspond to lattice data in the full gauge theory, and solid lines are free quark results on the same lattices as well as in the infinite-volume and continuum limit. On the left plot, we present the data for $L_s=24$ and the full range of chemical potential values used in this work. On the right plot, we compare the data obtained on $L_s=24$ and $L_s=30$ lattices. For both plots we combine the data points obtained with zero diquark source $\lambda = 0$ at $L_t < 12$ and with $a \lambda = 5 \cdot 10^{-4}$ at $L_t \geq 12$. As we demonstrate in Fig.~\ref{fig:conductivity_MP_comparison} below, introducing a small diquark mass term has no noticeable effect on current-current correlators for all temperatures and chemical potentials which we consider.

\begin{figure*}[h!tpb]
  \centering
  \includegraphics[angle=-90,width=0.32\textwidth]{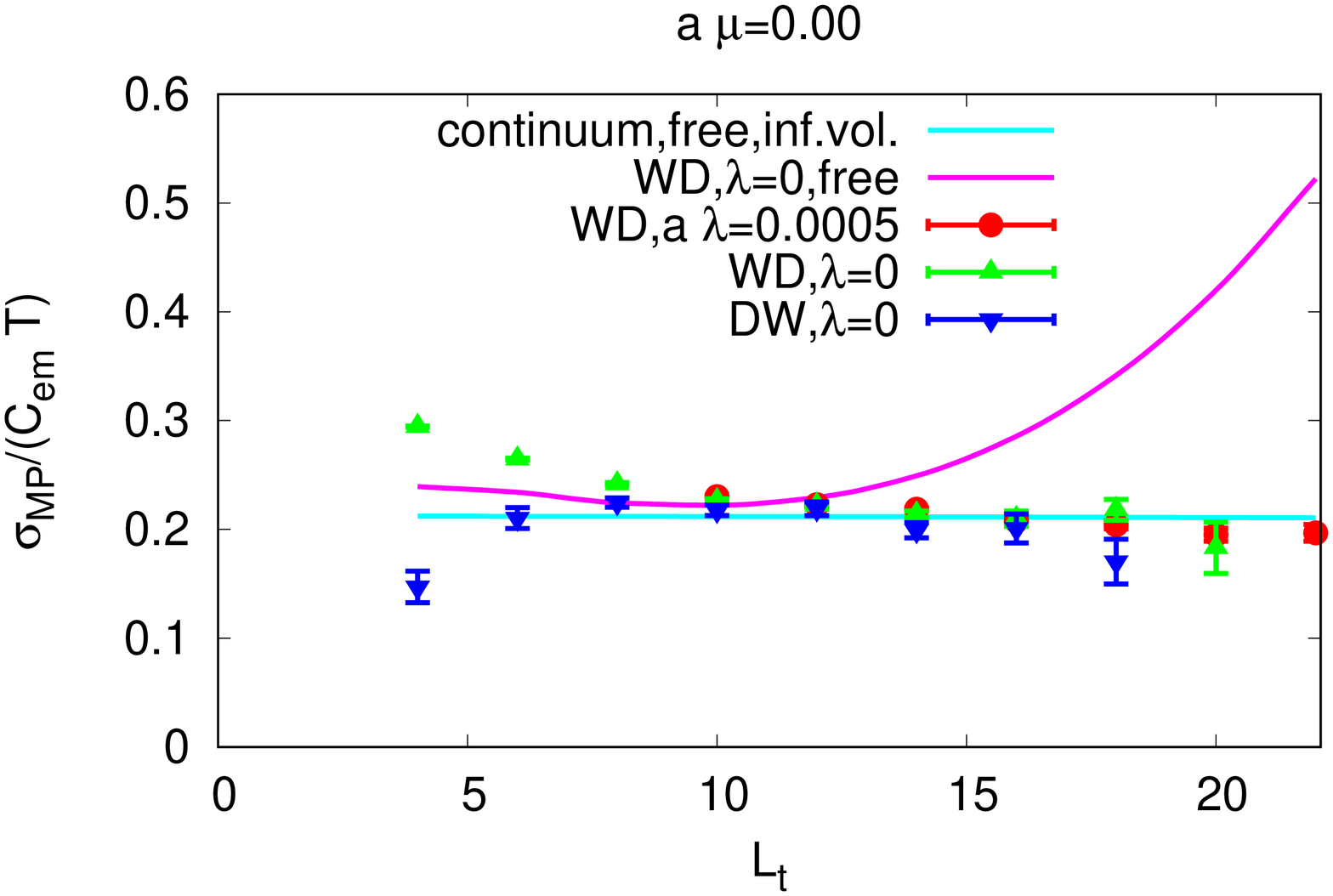}
  \includegraphics[angle=-90,width=0.32\textwidth]{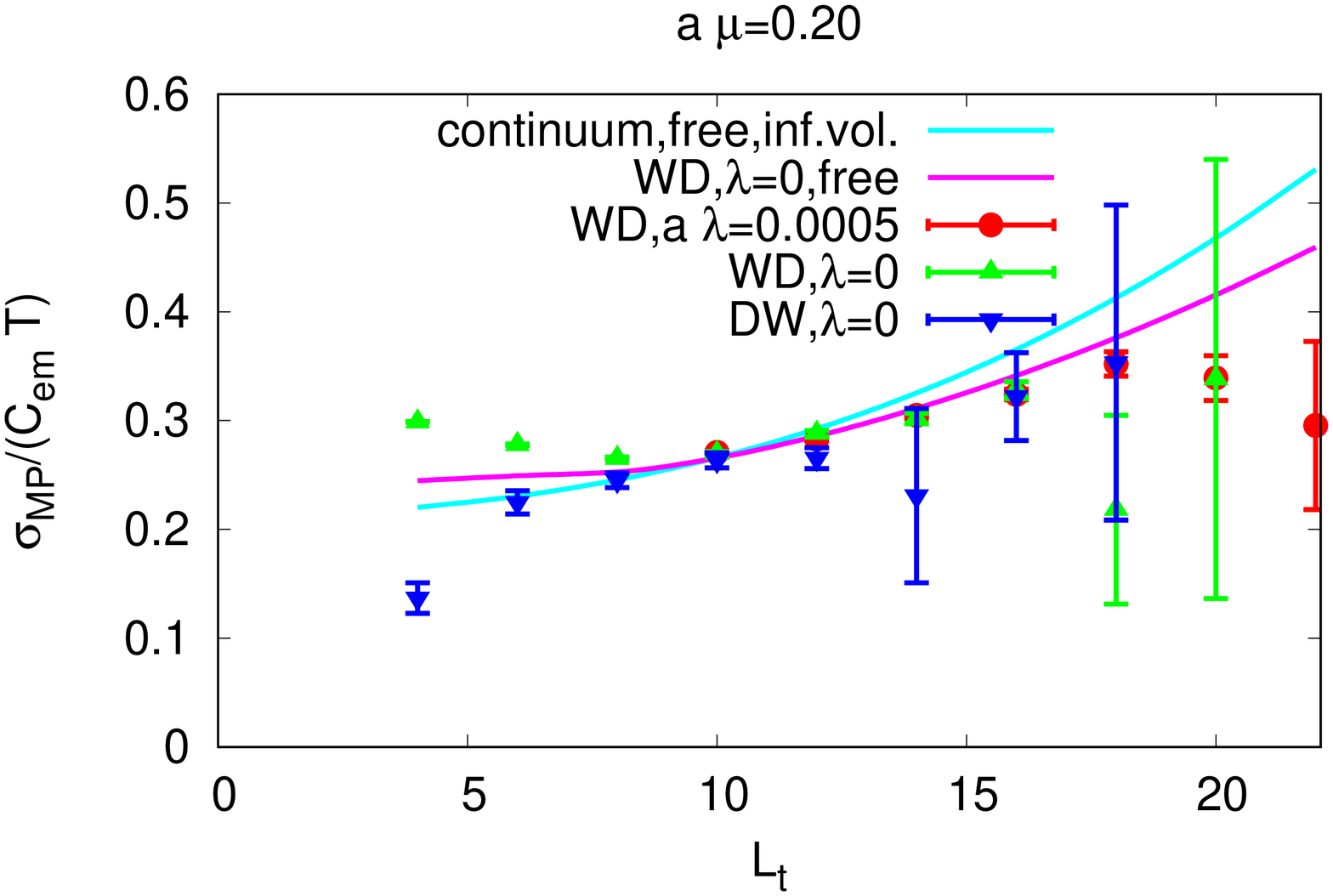}
  \includegraphics[angle=-90,width=0.32\textwidth]{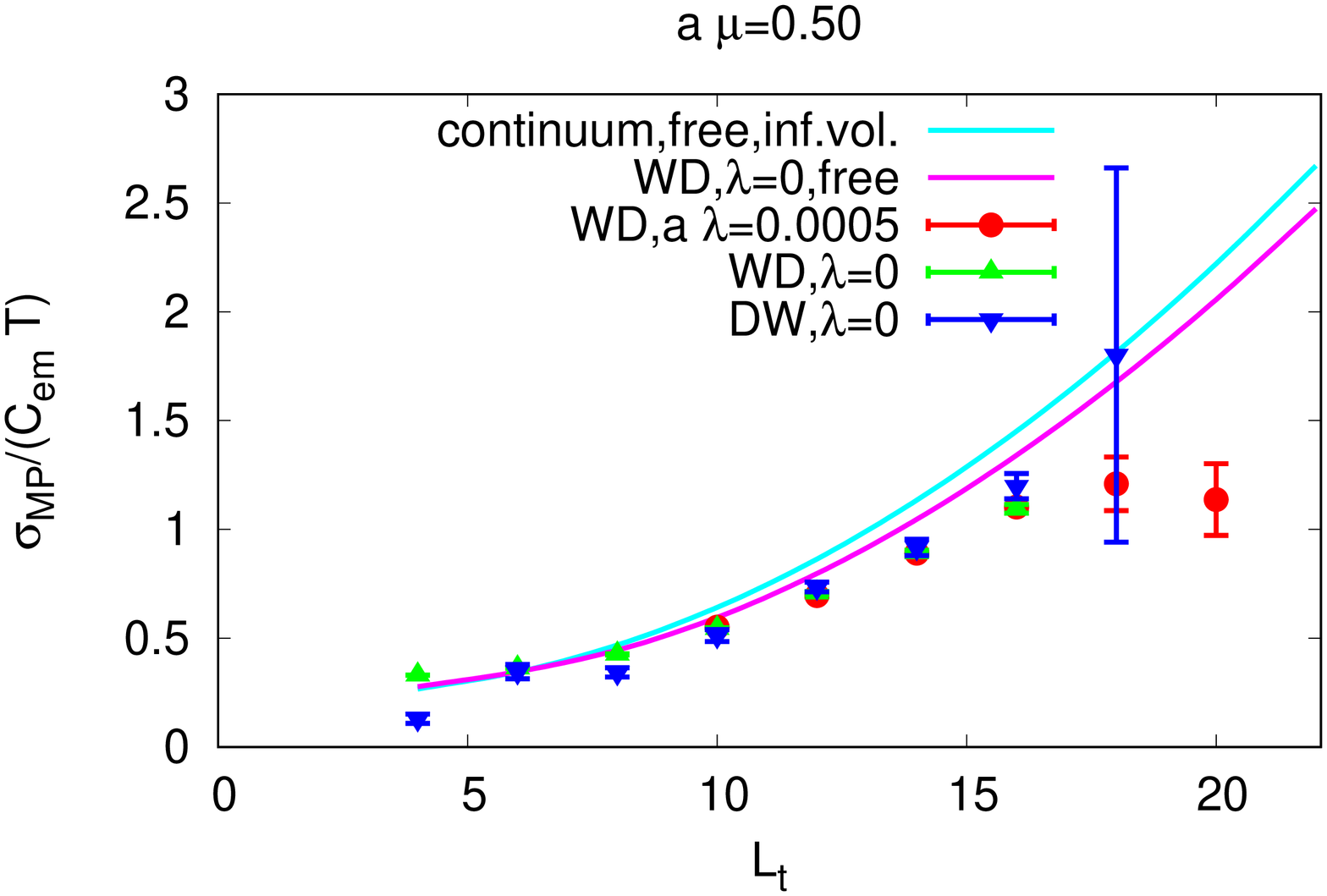}\\
  \caption{Midpoint estimates of the low-frequency conductivity for different lattice ensembles and different valence quark actions: Wilson-Dirac quarks with $\lambda = 0$ and $a \lambda=5 \cdot 10^{-4}$, Domain-Wall quarks with $\lambda=0$, free Wilson-Dirac quarks with $\lambda = 0$, and free quarks in the continuum and infinite-volume limits. For all lattice data $L_s = 24$.}
  \label{fig:conductivity_MP_comparison}
\end{figure*}

For small values of chemical potential the ratio $\sigma_{MP}/\lr{C_{em} T}$ in full gauge theory appears to be slowly decreasing towards lower temperatures. A comparison of the data for lattices with $L_s=24$ and $L_s=30$ suggests that finite-volume artifacts are comparable with statistical errors.

On the other hand, the corresponding free quark results grow towards lower temperatures, the faster the smaller is the volume. As a result, at low temperatures $\sigma_{MP}/\lr{C_{em} T}$ in the full gauge theory on $L_s = 30$ lattice is around $1.5$ times smaller than for free lattice quarks, in agreement with the expected drop of conductivity at low temperatures. From the right plot on Fig.~\ref{fig:conductivity_MP_summary} one can see that for $L_s=24$ lattice the difference appears to be even larger. On the other hand, the gauge theory results appear to be quite close to the values obtained for free continuum quarks.

As discussed in detail in Appendix~\ref{apdx:finite_volume_effects}, the growth of $\sigma_{MP}/T$ towards low temperatures for free quarks is a finite-volume artifact, and in the infinite-volume, continuum and massless limits $\sigma_{MP}/T$ at $\mu = 0$ is constant for free fermions: $\sigma_{MP}/T = \frac{N_c}{3 \pi} = 0.212$.

It is also interesting to note that for the two largest values of the chemical potential which we use, $a \mu = 0.2$ and especially $a \mu = 0.5$, the mid-point estimate $\sigma_{MP}/T$ appears to be closer to the free quark result than for the lower densities. This is probably due to the fact that large densities move the system closer to the high-energy regime of asymptotic freedom.

According to Fig.~\ref{fig:phase_diagram}, for $a \mu = 0.5$ and $L_t = 20,\, 22$ we should already be in the superconducting diquark condensation phase. Interestingly, superconductivity does not show up as a sharp increase in conductivity, here. Instead, the conductivity even falls slightly below the free quark result as we reach the diquark condensation phase with lowering the temperature at $a\mu =0.5$. Most likely, the dramatic changes expected in the transport-peak part of the electric conductivity are simply not captured by our frequency-smeared conductivity estimates, and are also to some extent compensated by the suppression of the higher-frequency conductivity at large $\mu$. A more detailed study of the electric conductivity in the superconducting phase would certainly be interesting, but is beyond the scope of this work.

\begin{figure*}[h!tpb]
  \centering
  \includegraphics[angle=-90,width=0.45\textwidth]{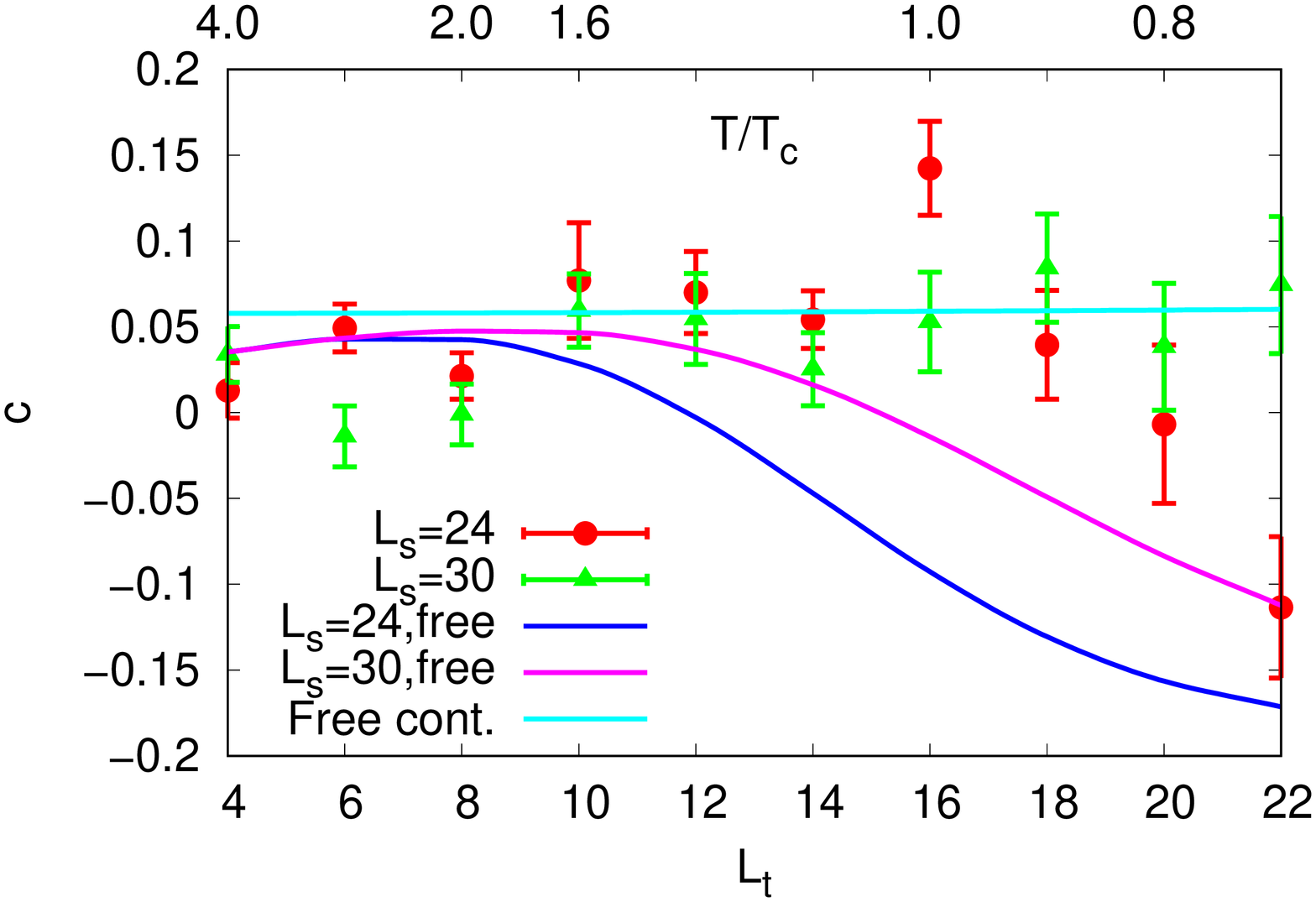}
  \includegraphics[angle=-90,width=0.45\textwidth]{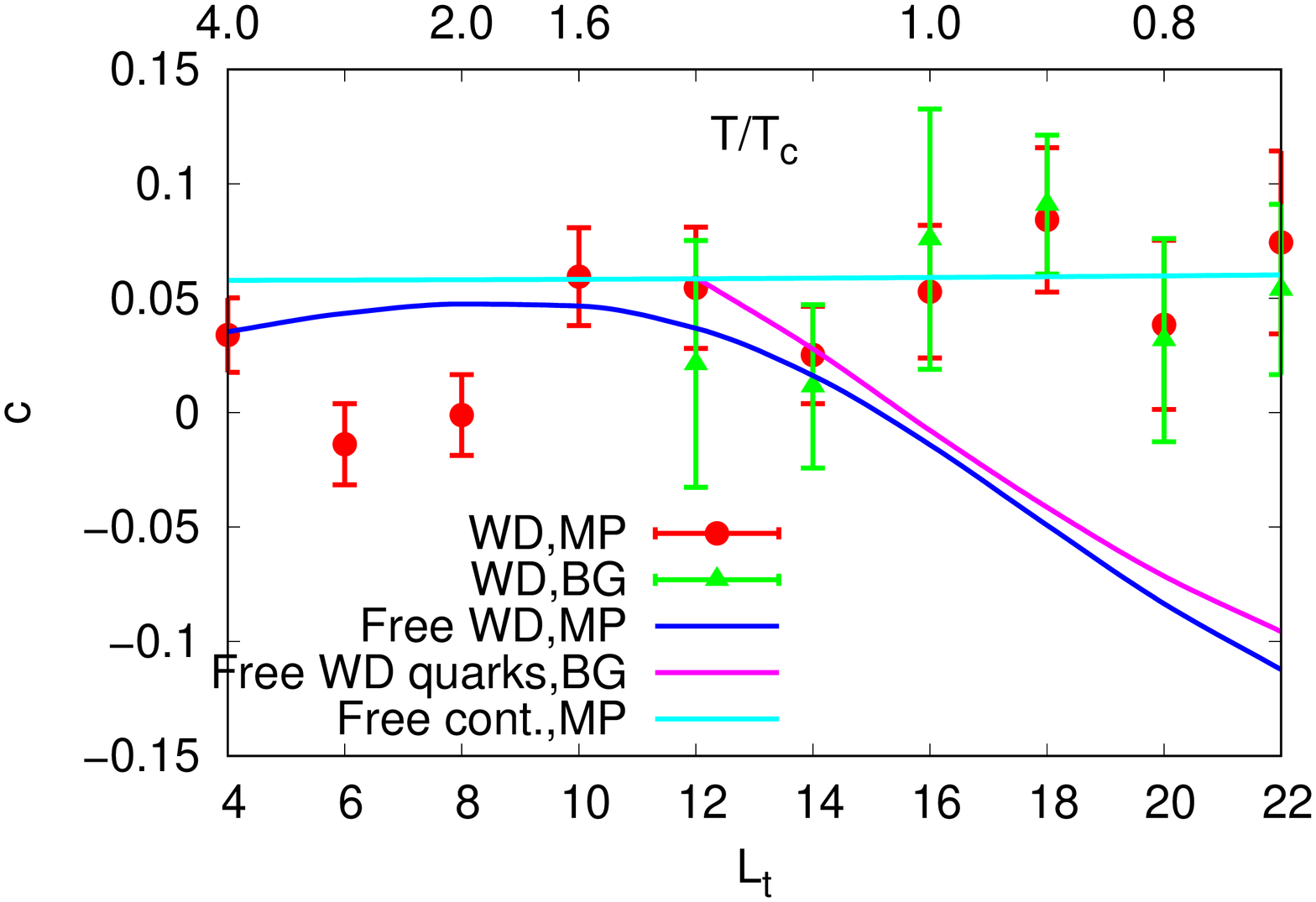}\\
  \caption{Numerical estimate of the coefficient $c$ in (\ref{sigma_vs_mu_parameterization}) from the finite difference between estimate of the low-frequency electric conductivity at $\mu = 0$ and $a \mu = 0.05$. \textbf{On the left:} using the mid-point estimate $\sigma_{MP}$ for $L_s = 24$ and $L_s = 30$. \textbf{On the right:} comparison of the results obtained using the mid-point and the Backus-Gilbert estimates for $L_s = 30$.}
  \label{fig:c_MP}
\end{figure*}

In order to further check the independence of our estimates on the choice of the fermionic action and diquark source term, in Fig.~\ref{fig:conductivity_MP_comparison} we compare the temperature dependence of the midpoint estimator $\sigma_{MP}/T$ for Domain Wall fermions and for Wilson-Dirac fermions with and without the diquark source term. One can see that all estimates agree within statistical errors.

Finally, we use our results for $\sigma_{MP}/T$ to estimate the expansion coefficient $c\lr{T}$ in (\ref{sigma_vs_mu_parameterization}), which characterizes the sensitivity of low-frequency electric conductivity to chemical potential. To this end we use the difference between $\sigma_{MP}$ at $\mu = 0$ and $a \mu = 0.05$. This value of $a \mu$ is a compromise between having a finite difference which is considerably larger than statistical errors, and being still in the small-$\mu$ QCD-like regime far from the diquark condensation phase. We thus approximate $c\lr{T}$ as
\begin{eqnarray}
\label{c_approx_def}
 c\lr{T}
 \approx
 \frac{
  \sigma_{MP}\lr{T, a\mu=0.05} - \sigma_{MP}\lr{T, \mu=0}
 }{
  \lr{a \mu L_t}^2 \, \sigma_{MP}\lr{T, \mu=0}
 } .
\end{eqnarray}
The resulting temperature dependence of $c\lr{T}$ for Wilson-Dirac fermions on lattices with $L_s = 30$ and $L_s = 24$ is illustrated on the left plot in Fig.~\ref{fig:c_MP}, along with reference results for free fermions on the lattice and in the continuum. While in the infinite-volume, continuum and massless limits the coefficient $c\lr{T}$ based on the midpoint estimate for free quarks is $c\lr{T} = \frac{15 - \pi^2}{9 \pi^2} = 0.0578$ and thus temperature-independent, the lattice data shows some temperature dependence, which is especially strong for free quarks and the full gauge theory on $L_s=24$ lattices. In particular, for sufficiently low temperatures the coefficient $c\lr{T}$ becomes negative. As discussed in Appendix~\ref{apdx:finite_volume_effects}, for free quarks this behavior is a finite-volume artifact, and in the large-volume limit the lattice free-quark estimate of $c\lr{T}$ becomes closer to the continuum value and depends weaker on the temperature.

A noticeable feature is that the temperature dependence of $c\lr{T}$ in the full gauge theory is quite different from the free quark result, especially in the vicinity of the chiral crossover, where $c\lr{T}$ takes its maximal value for both lattice sizes $L_s=24$ and $L_s=30$. The value of $c\lr{T}$ appears to be larger than the free quark result, with rather significant deviations between the data for $L_s=24$ and $L_s=30$ lattices. The temperature dependence of the data for $L_s=30$ appears to be weaker than for $L_s=24$. Both data sets, however, show a peak around crossover temperature - a pronounced one for $L_s=24$ and a small one for $L_s=30$. This suggests that the electric conductivity should be most sensitive to finite density in the crossover regime. Since this statement is based on the data obtained in the low-density QCD-like regime of $SU\lr{2}$ gauge theory, it should be also qualitatively correct for the full QCD. While our data still has quite large statistical and also systematic errors due to aforementioned finite-volume effects, a conservative estimate of the value of the coefficient $c\lr{T}$ around crossover temperature is $c\lr{T} = 0.10 \pm 0.07$, which is noticeably larger than the free quark result.

\subsection{Estimates of electric conductivity from the Backus-Gilbert method}
\label{subsec:numres_spectral}

In this Section we turn to the estimates of the electric conductivity based on the Backus-Gilbert method outlined in Section~\ref{sec:conductivity_preliminary}. We implement the resolution functions in the Backus-Gilbert transformation (\ref{BG_estimator_def}) on the discrete grid of frequency values $\omega = j \, T$ with $j = 0, 1, \ldots 2 L_t$. For each $L_t$ we tune the value of the Tikhonov regularization parameter $\Delta$ in the Backus-Gilbert method as follows: we perform the analysis with $\Delta = 1 \cdot 10^{-10}, 2 \cdot 10^{-10}, 5 \cdot 10^{-10}, 1 \cdot 10^{-9}, \ldots, 5 \cdot 10^{-7}$, starting from $\Delta = 1 \cdot 10^{-10}$, and choose the least value of $\Delta$ for which the Backus-Gilbert estimate $\sigma_{BG}\lr{\omega}$ of the electric conductivity on the lattice with $L_s=24$ is positive and its maximal relative error is less than $10 \%$ for all values of chemical potential. Such tuning yielded the following values: $\Delta = 1 \cdot 10^{-10}$ for $L_t = 12$, $\Delta = 1 \cdot 10^{-9}$ for $L_t=14$, $\Delta = 2 \cdot 10^{-9}$ for $L_t = 16$ and $\Delta = 2 \cdot 10^{-7}$ for $L_t = 18, 20, 22$. As illustrated in Fig.~\ref{fig:resolution_functions}, with these values of $\Delta$ the resolution functions $\delta_{BG}\lr{0, \omega }$ are still very close to resolution functions calculated with a very small reference value $\Delta = 10^{-12}$. They are noticeably narrower than the resolution function $\mathcal{N}^{-1} \, K\lr{a L_t/2, \omega}$ for the midpoint estimate (\ref{MP_estimator_def}).

\begin{figure*}[h!tpb]
  \centering
  \includegraphics[angle=-90,width=0.45\textwidth]{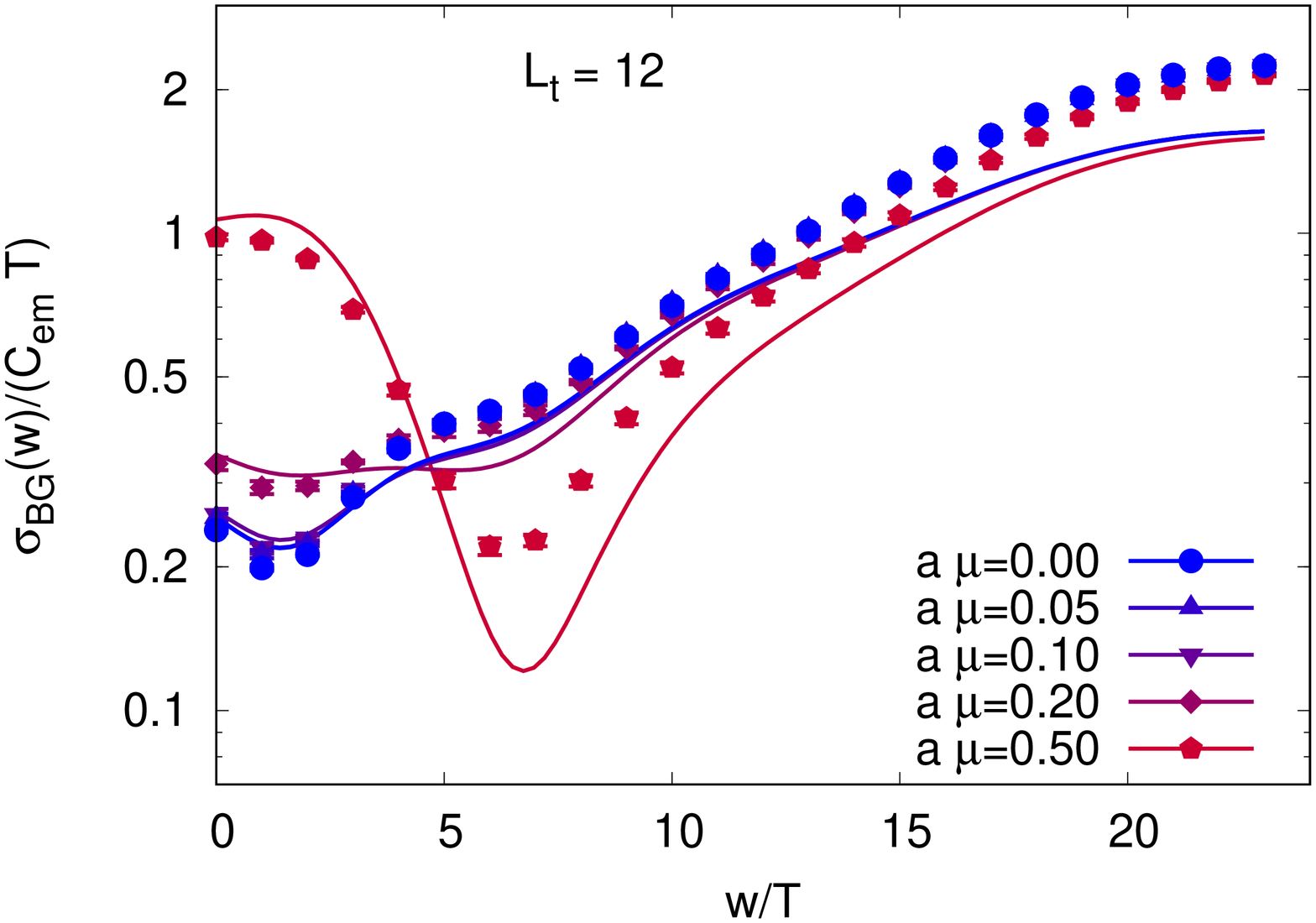}\includegraphics[angle=-90,width=0.45\textwidth]{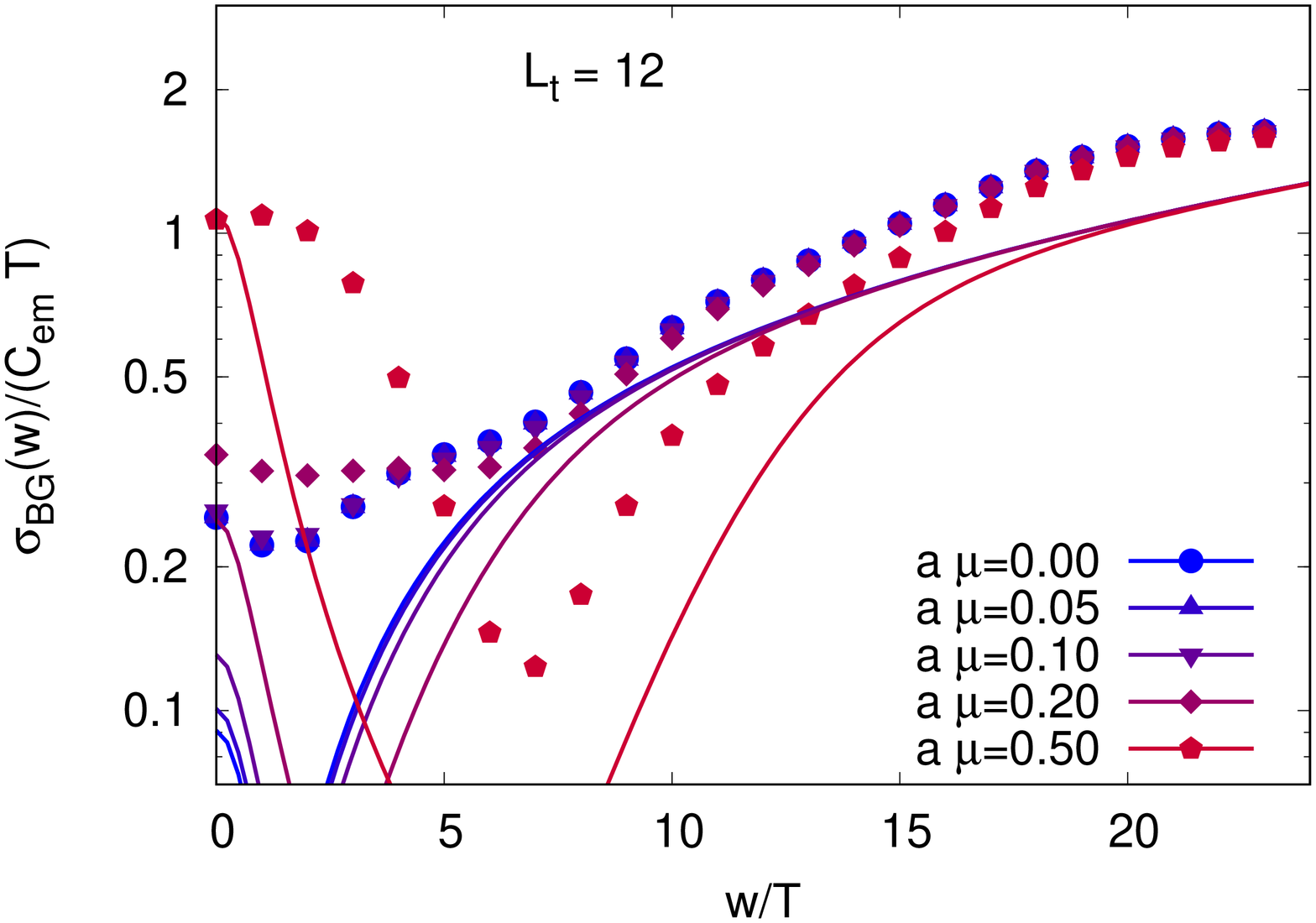}\\
  \includegraphics[angle=-90,width=0.45\textwidth]{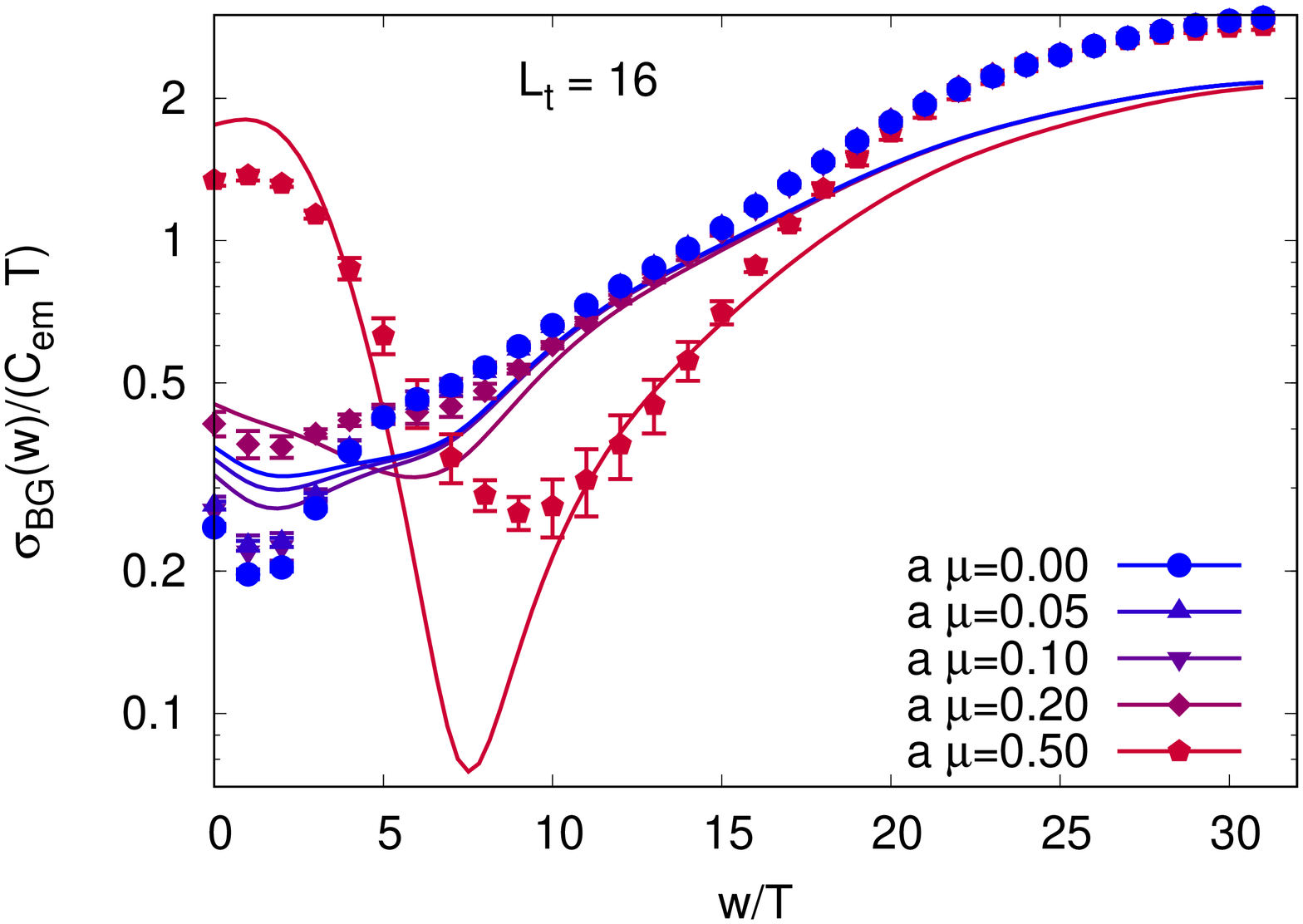}\includegraphics[angle=-90,width=0.45\textwidth]{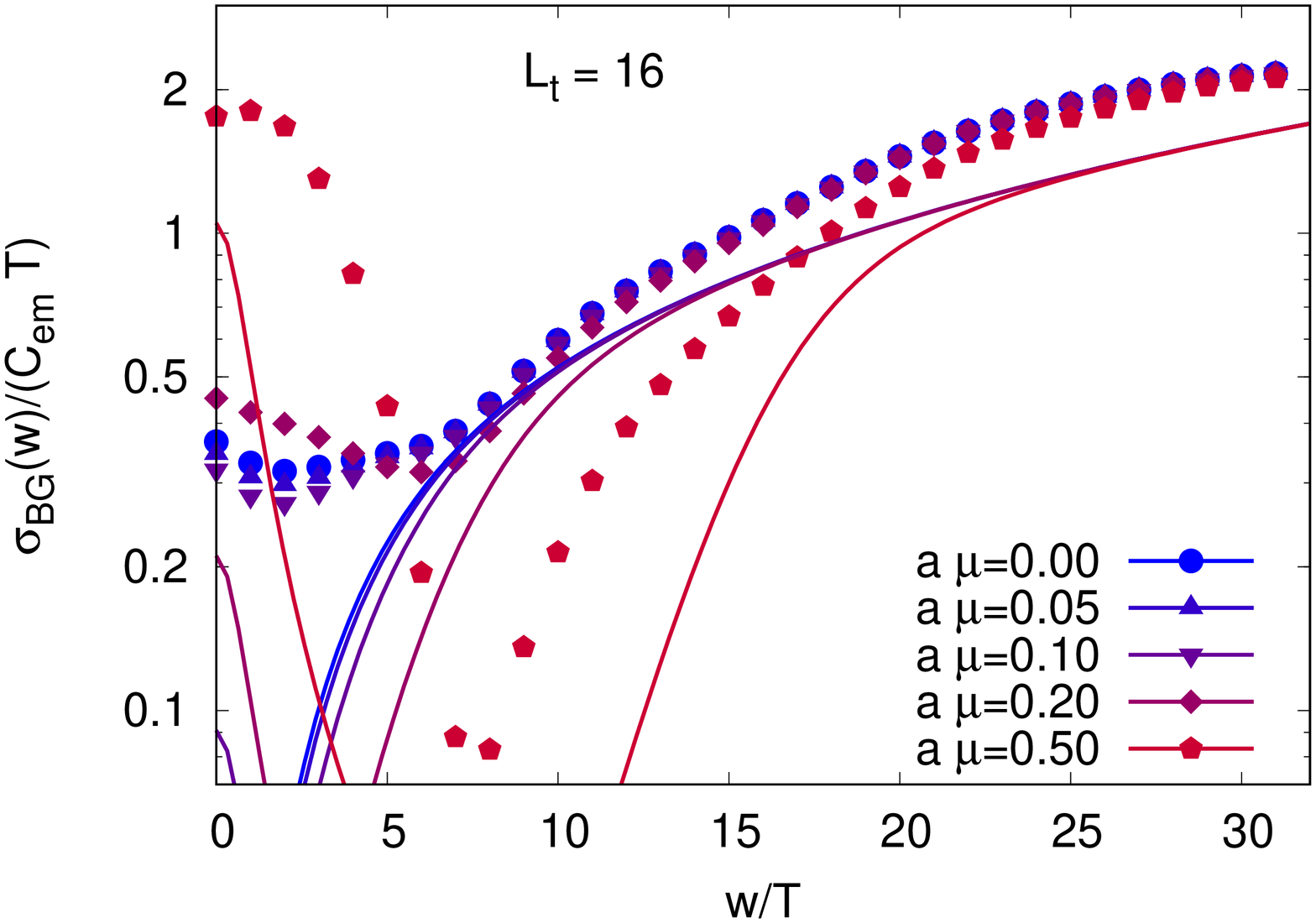}\\
  \includegraphics[angle=-90,width=0.45\textwidth]{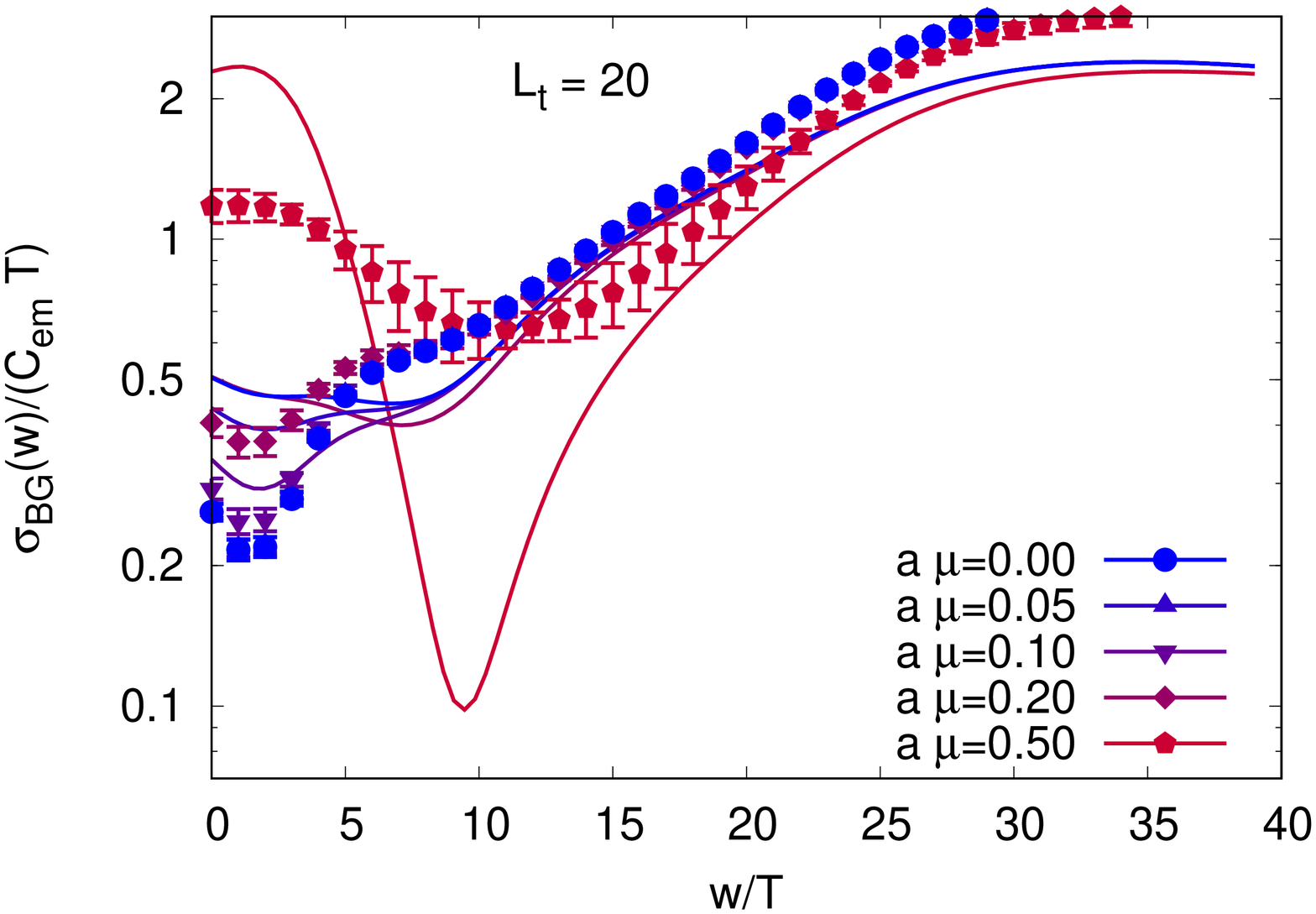}\includegraphics[angle=-90,width=0.45\textwidth]{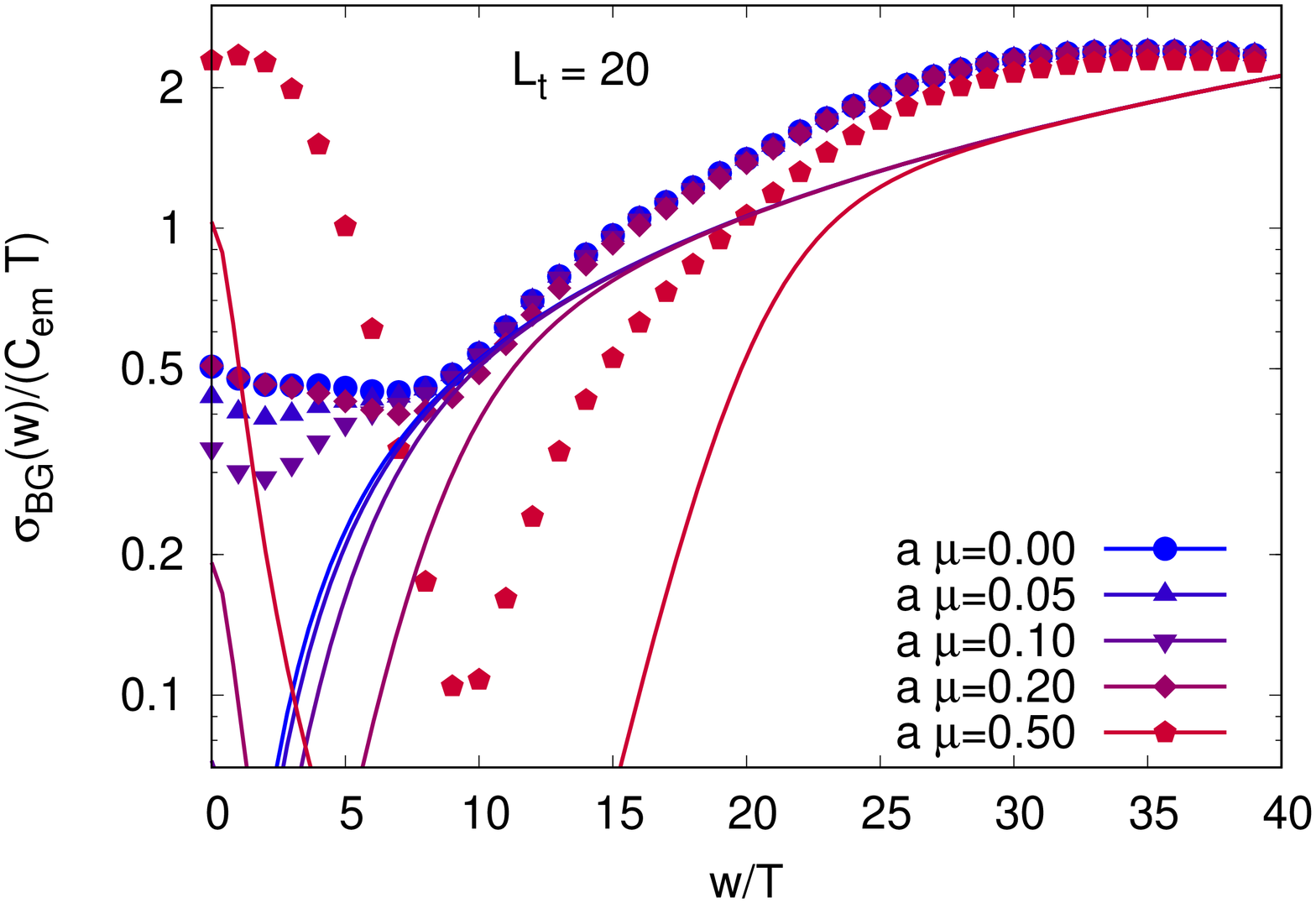}\\
  \caption{\textbf{On the left:} Frequency-smeared electric conductivity $\sigma_{BG}\lr{\omega}$ extracted from Euclidean correlation functions on lattices with $L_s = 24$ using the Backus-Gilbert method (points with error bars). Solid lines are the smeared electric conductivity obtained for free quarks using the same procedure. \textbf{On the right} we compare these free-fermion smeared conductivities (points) with analytic expression (\ref{free_quark_conductivity}) for the electric conductivity of continuum Dirac fermions (solid lines). To illustrate the magnitude of the ``transport peak'' term in (\ref{free_quark_conductivity}), the delta-function $\delta\lr{\omega}$ was replaced by a fixed-width Breit-Wigner distribution $\lr{\alpha/T}/\lr{1 + \lr{\omega/T}^2}$, where $\alpha$ is the $\delta$-function prefactor in (\ref{free_quark_conductivity}).}
  \label{fig:spectral}
\end{figure*}

In Fig.~\ref{fig:spectral} we show the Backus-Gilbert estimates of the electric conductivity $\sigma_{BG}\lr{\omega}$ for Wilson-Dirac fermions on lattices with $L_s=24$ at different values of chemical potential and different temperatures, and compare them with corresponding estimates for free quarks. To understand how the inherent smearing within the Backus-Gilbert method as well as lattice artifacts and finite-volume effects affect the frequency-dependent conductivity, in the plots on the right in Fig.~\ref{fig:spectral} we also compare the Backus-Gilbert estimates $\sigma_{BG}\lr{\omega}$ for free quarks with analytically calculated conductivities in the continuum theory. Analytic expressions for electric conductivity are summarized in Appendix~\ref{apdx:jvjv_free}. For illustrative purposes, in Fig.~\ref{fig:spectral} we have replaced the infinitely narrow transport peaks of free continuum quarks (the term proportional to the $\delta$-function in (\ref{free_quark_conductivity})) by Breit-Wigner distributions $\lr{\alpha/T}/\lr{1 + \lr{\omega/T}^2}$ of unit width, where $\alpha$ is the $\delta$-function prefactor in (\ref{free_quark_conductivity}).

\begin{figure*}[h!tpb]
  \centering
  \includegraphics[angle=-90,width=0.45\textwidth]{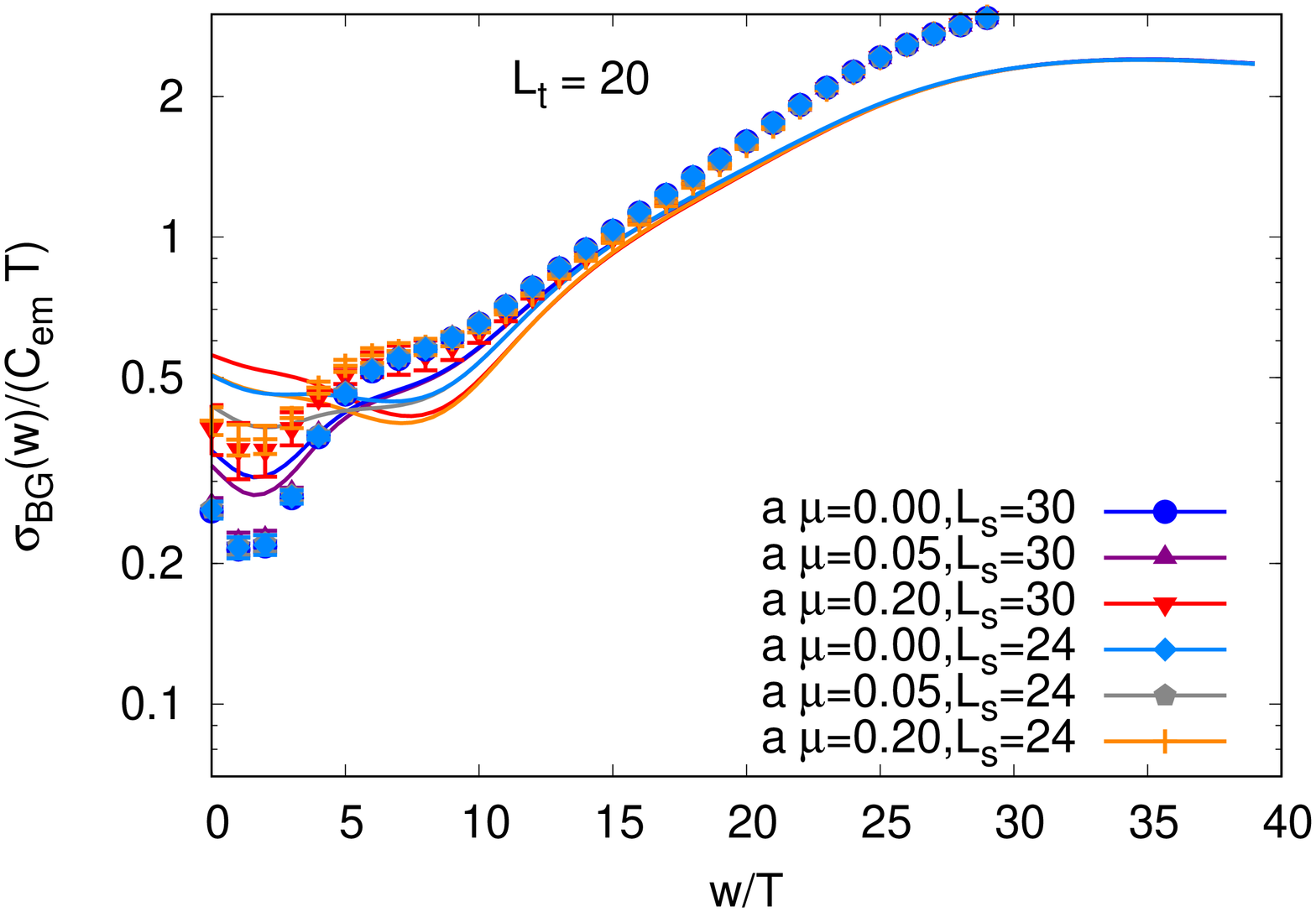}\includegraphics[angle=-90,width=0.45\textwidth]{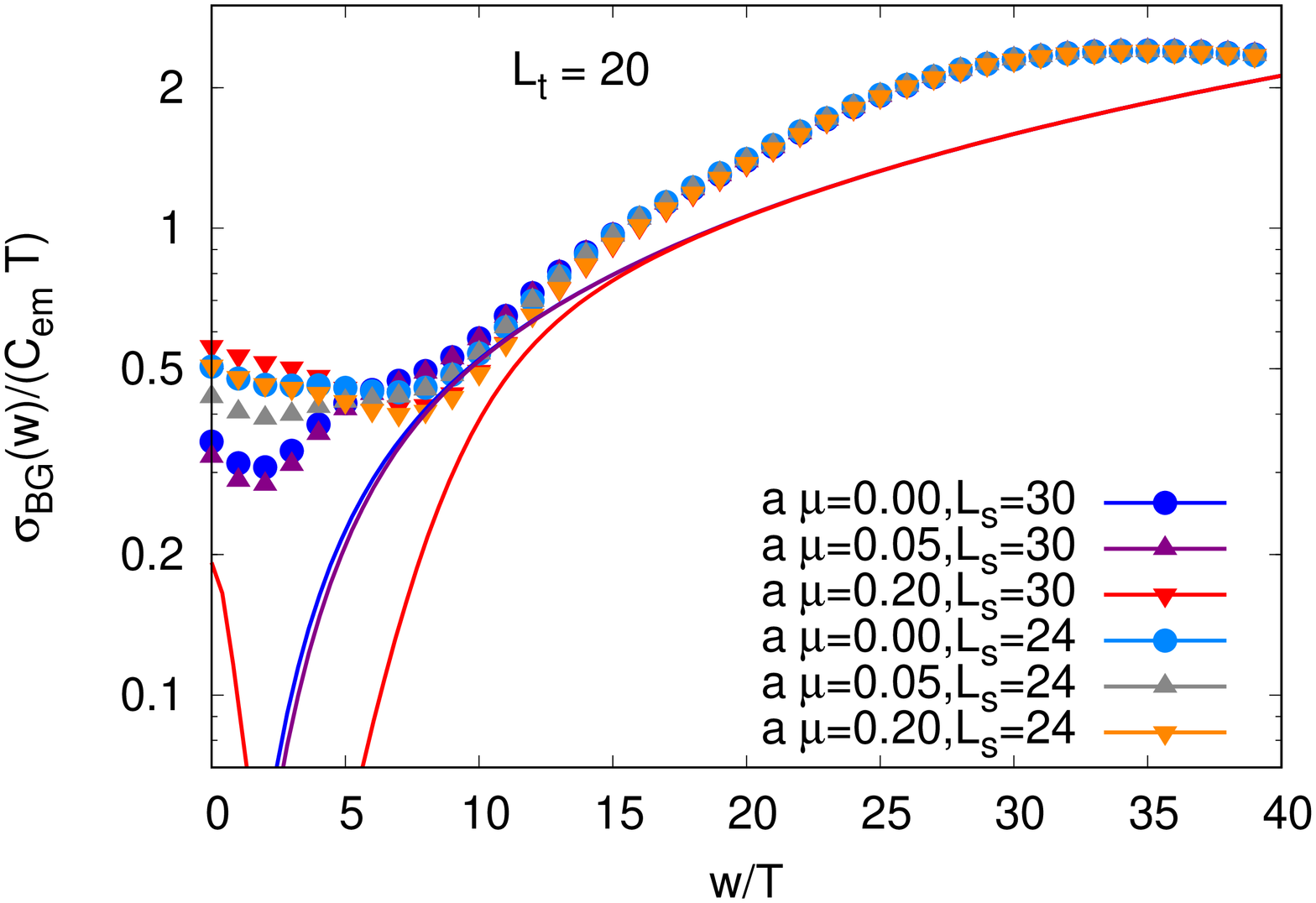}\\
  \caption{\textbf{On the left:} A comparison of frequency-smeared electric conductivities $\sigma_{BG}\lr{\omega}$ extracted from Euclidean correlation functions on lattices with $L_s=24$ and $L_s = 30$ using the Backus-Gilbert method (points with error bars). Solid lines are the frequency-smeared electric conductivities obtained for free quarks on the same lattices using the same procedure. \textbf{On the right} we compare these free-fermion conductivities (points) with analytic expression (\ref{free_quark_conductivity}) for the electric conductivity of continuum Dirac fermions (solid lines). The magnitude of the ``transport peak'' for continuum free fermions is illustrated in the same way as on Fig.~\ref{fig:spectral}.}
  \label{fig:spectral_Ns30}
\end{figure*}

Finite chemical potential affects the electric conductivity of free continuum quarks in two competing ways, which become especially evident in the zero-temperature limit (see equation~(\ref{free_quark_conductivity_zeroT})). On the one hand, the height of the transport peak at $\omega = 0$ grows approximately as $\mu^2$ (neglecting the small quark mass), i.e.~in proportion to the area of the Fermi surface. On the other hand, chemical potential makes the finite-frequency part of the electric conductivity vanish for $w < 2 \max\{m_q,\,\mu\}$. In the condensed-matter physics language, the transport peak and the finite-frequency part of the electric conductivity originate from intraband and interband transitions, respectively.

For low frequencies the Backus-Gilbert estimator $\sigma_{BG}\lr{\omega}$ receives contributions from both the transport peak and the finite-frequency electric conductivity. As a result, the strength of the transport peak is rather strongly over-estimated for low densities, as one can also see from the plots on the right in Fig.~\ref{fig:spectral}. For larger densities ($a \mu = 0.2$ and $a \mu = 0.5$) the finite-frequency part of electric conductivity is separated from the transport peak by a rather wide gap, wider than the width of the resolution functions in the Backus-Gilbert method. As a result, for large densities the Backus-Gilbert method captures the strength of the transport peak more precisely. As could be expected, the Backus-Gilbert estimates of electric conductivity most strongly deviate from the continuum results in the vicinity of the gap between the transport peak and the finite-frequency part of electric conductivity. This is a direct consequence of the smearing which removes sharp threshold effects and also smears out the transport peak. The deviation of lattice and continuum results in the high-frequency tails of conductivity is most likely a lattice artifact.

Comparing now the Backus-Gilbert estimates for the full gauge theory and for free quarks, we see that, as the temperature is decreased, the low-frequency electric conductivity becomes significantly smaller for the full gauge theory at all values of the chemical potential. On the other hand, for $\omega/T \sim 5$ (corresponding to $a \omega \approx 0.2 \ldots 0.4$), the data for the full gauge theory shows a kind of bump, where it becomes significantly larger than the free quark result. We associate this bump with a $\rho$-meson resonance, which becomes very wide due to smearing. At the largest value of the chemical potential $a \mu = 0.5$ this bump seems to vanish, but the gap between the transport peak and the finite-frequency part of electric conductivity is still much shallower than for free quarks. This suggests a strong broadening of the transport peak in full gauge theory.

\red{On Fig.~\ref{fig:spectral_Ns30} we also illustrate the lattice volume dependence of electric conductivity by comparing the frequency-smeared estimates of electric conductivity on lattices with spatial sizes $L_s=24$ and $L_s=30$ and temporal size $L_t=20$. For smaller $L_t$ (higher temperatures) finite volume effects are smaller. As already discussed in Subsection~\ref{subsec:numres_corrs} above, volume dependence is much stronger for free quarks than for the actual gauge theory data.}

\begin{figure*}[h!tpb]
  \centering
  \includegraphics[angle=-90,width=0.45\textwidth]{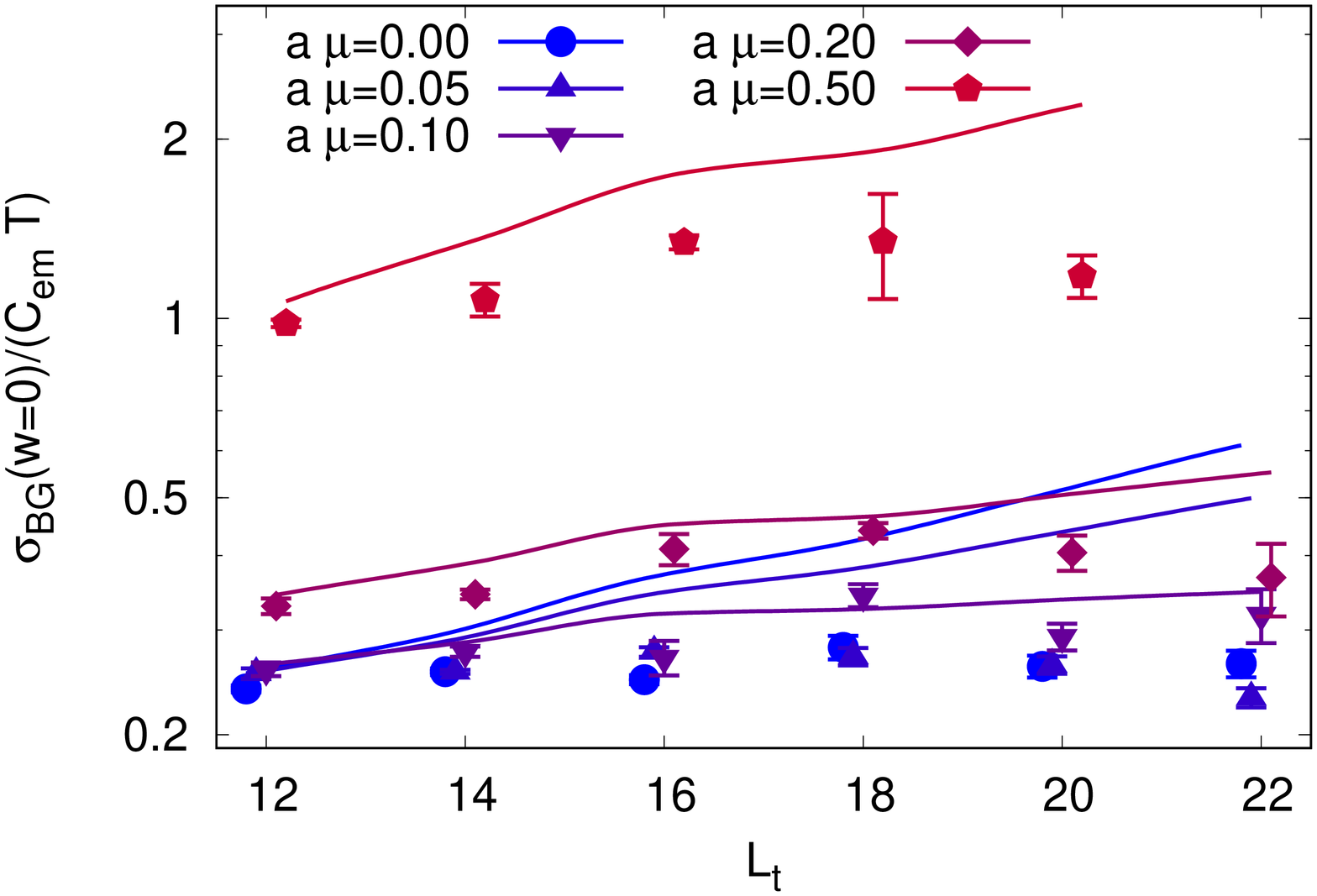}
  \includegraphics[angle=-90,width=0.45\textwidth]{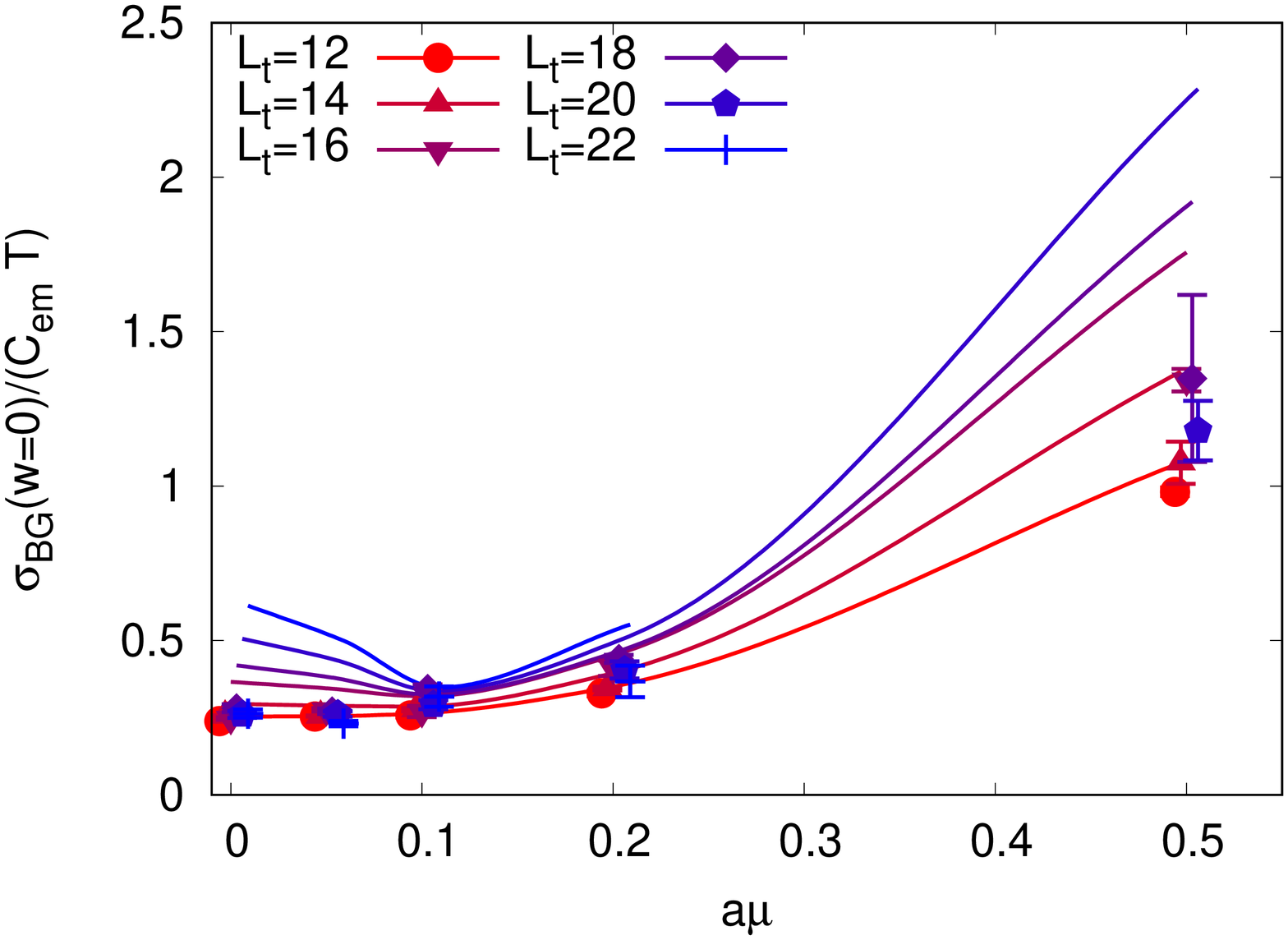}\\
  \caption{On the left: Backus-Gilbert estimate of the $\omega \rightarrow 0$ limit $\sigma_{BG}\lr{\omega=0 }/T$ of the electric conductivity as a function of the inverse temperature $L_t = 1/\lr{a T}$. On the right: $\sigma_{BG}\lr{\omega = 0}/T$ as a function of chemical potential at different $L_t$. For both plots $L_s = 24$. Data points are slightly shifted away from integer values of $L_t$ in order to improve the readability of the plot.}
  \label{fig:conductivity_BG}
\end{figure*}

We now use the zero-frequency limit of the smeared electric conductivity obtained with the Backus-Gilbert method to estimate the low-frequency conductivity as a function of temperature and chemical potential. As discussed in Section~\ref{sec:conductivity_preliminary} above and illustrated in Fig.~\ref{fig:resolution_functions}, the resolution of the Backus-Gilbert estimate is better than that of the midpoint estimate, which comes at the cost of the dependence on the regularization parameter $\Delta$. The results of these estimates are illustrated in Fig.~\ref{fig:conductivity_BG}, both as functions of temperature at fixed $\mu$ and as functions of $\mu$ at fixed temperature. Data points with error bars correspond to the full gauge theory, and solid lines correspond to Backus-Gilbert estimates for free quarks on the same lattices. The overall picture is consistent with the results from the midpoint estimators - for the full gauge theory the temperature dependence of the conductivity is much weaker than it is for free quarks. \red{Our estimate $\sigma\lr{T_c} \approx 0.25 \pm 0.02$ of electric conductivity in the vicinity of the crossover near $L_t=16$ is in a good agreement with other lattice QCD results \cite{Nikolaev:2008.12326}. }

While at high temperatures the conductivity is close to the free quark result, it differs by a factor of $2-3$ at low temperatures. For large densities the difference between the gauge theory results and the free quark results is somewhat more pronounced than for the midpoint estimators. In agreement with the midpoint estimates, in the presumably superconducting phase at $a \mu = 0.5$ and $L_t = 20,\, 22$ our estimate of low-frequency conductivity slightly drops.

\begin{figure}[h!tpb]
  \centering
  \includegraphics[angle=-90,width=0.45\textwidth]{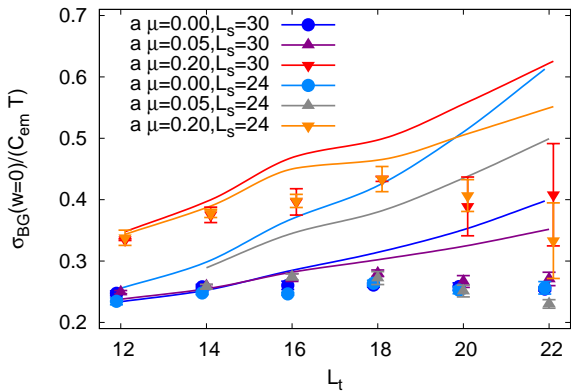}\\
  \caption{A comparison of Backus-Gilbert estimates of the $\omega \rightarrow 0$ limit $\sigma_{BG}\lr{\omega=0 }/T$ of the electric conductivity for lattice sizes $L_s=24$ and $L_s = 30$.}
  \label{fig:conductivity_BG_Ns30}
\end{figure}

\red{On Fig.~\ref{fig:conductivity_BG_Ns30} we also illustrate the effect of finite volume on the Backus-Gilbert estimate of low-frequency electric conductivity. Especially for $a \mu = 0.05$ finite-volume effects appear to be somewhat larger than for the midpoint estimate, but are still significantly smaller than for free quarks.}

Finally, we also use results from the Backus-Gilbert method to estimate the first nontrivial coefficient $c\lr{T}$ in the expansion (\ref{sigma_vs_mu_parameterization}) of electric conductivity in powers of $\mu/T$. We again use the finite-difference approximation (\ref{c_approx_def}), replacing $\sigma_{MP}$ with $\sigma_{BG}\lr{\omega = 0}$. The result is shown in Fig.~\ref{fig:c_MP} together with the result based on the midpoint estimate. Both results appear to be consistent with each other, within statistical errors, and they exhibit the largest deviations from the free quark result in the vicinity of the chiral crossover.

\section{Conclusions and discussion}
\label{sec:conclusions}

We have studied the low-frequency electric conductivity in finite-density $SU\lr{2}$ gauge theory with dynamical fermions at various temperatures across the chiral crossover, both within the phase with spontaneously broken chiral symmetry and around the transition to the diquark condensation phase. As a by-product of our study, we have also obtained new estimates of the phase boundaries of $SU\lr{2}$ gauge theory, as summarized in Fig.~\ref{fig:phase_diagram}. An interesting observation, which confirms the findings of \cite{Hands:1210.4496,Hands:1912.10975}, is that in $SU\lr{2}$ gauge theory the chiral crossover happens at rather low temperatures, $T_c/m_{\pi} \approx 0.37$. In contrast, in real QCD $T_c/m_{\pi} \gtrsim 1$.

We found that introducing finite density expectedly increases the electric conductivity. However, at low temperatures and small densities lattice data can show a very weak trend in the opposite direction due to finite-volume artifacts (see Fig.~\ref{fig:c_MP} and Appendix~\ref{apdx:jvjv_free}). For our largest chemical potential $a \mu = 0.5$ and low temperatures near the boundary of the diquark condensation phase, which is absent in real QCD, the conductivity can increase by a factor of about $5$ as compared to its zero-density value.

Our zero-density result $\sigma\lr{0}/T \approx 0.25 \pm 0.02$ at temperatures around $T_c$ \red{(see e.g. Fig.~\ref{fig:conductivity_BG})} is in agreement with the results obtained in full lattice QCD \cite{Aarts:1307.6763,Aarts:1412.6411,Meyer:1512.07249}. The decrease of the absolute value of $\sigma\lr{0}/T$ across the crossover in $SU\lr{2}$ gauge theory turns out to be not as significant as in full QCD. This is expectable, since for smaller $N_c$ the difference in the number of degrees of freedom between confinement ($O\lr{1}$) and deconfinement $O\lr{N_c^2}$ regimes is also smaller. However, in comparison with the free quark result the conductivity in the full gauge theory drops by around $50\%$ at $T/T_c \sim 0.8$, which is again in agreement with \cite{Aarts:1307.6763,Aarts:1412.6411,Meyer:1512.07249}. It is interesting that for all temperatures and densities which we have considered the conductivity is still much larger than the conductivity of a free pion gas, calculated in Appendix~\ref{apdx:jvjv_free}.

The result which should be most relevant for real QCD is our estimate of the first nontrivial coefficient $c\lr{T}$ in the expansion (\ref{sigma_vs_mu_parameterization}) of low-frequency electric conductivity in powers of chemical potential over temperature $\mu/T$. This result is obtained within the low-density QCD-like phase with spontaneously broken chiral symmetry and no diquark condensation. The maximal value of $c\lr{T}$ is $c\lr{T} \approx 0.10 \pm 0.07$ in the vicinity of the chiral crossover, that is, noticeably larger than the free quark result in the continuum. The temperature dependence of $c\lr{T}$ appears to be rather weak for the $L_s = 30$ data.

This estimate suggests that even for $T \approx T_c$ and $\mu/T \sim 1$ finite density cannot change the conductivity by more than $15 - 20 \%$, which validates zero-density calculations of the conductivity as being reasonably good approximations also at finite densities corresponding to values of the chemical potential as in the vicinity of the QCD critical endpoint. Away from the crossover, $c\lr{T}$ becomes closer to the free quark result.

The maximum of $c\lr{T}$ in the vicinity of the chiral crossover can be also explained by the following qualitative argument. As shown in Appendix~\ref{apdx:jvjv_free}, $c\lr{T}$ decreases towards higher temperatures for free quarks, but grows for free pion gas. Thus a maximum can be expected for intermediate temperatures between the regimes where each of these two approximations are valid.

We have also observed that, as the quark density increases and the temperature decreases, the contribution of disconnected fermionic diagrams to the current-current correlators becomes more significant. We cannot rule out that it can be as large as $\sim 10 \ldots 20 \%$ of the connected contributions for the largest value of the chemical potential $a \mu = 0.5$ which we have used. Therefore, disconnected contributions might potentially become as important as the connected ones in the quarkyonic phase.

\begin{acknowledgments}
The work of P.~B. was supported by the Heisenberg Fellowship from the German Research Foundation, project BU2626/3-1. D.~S. and L.~v.~S. were supported by the Helmholtz International Center (HIC) for FAIR. D.~S. also received funding from the European Union's Horizon 2020 research and innovation programme under grant agreement No.~871072, also known as CREMLINplus (Connecting Russian and European Measures for Large-scale Research Infrastructures).

We thank Lukas Holicki for sharing his OpenCL code on which we based the measurement code used in this work. We are grateful to Wolfgang~Cassing for a stimulating and informative discussion of the status of conductivity calculations in QCD, and Maksim Ulybyshev for his help with the Backus-Gilbert method.

This work was performed using the Cambridge Service for Data Driven Discovery (CSD3), part of which is operated by the University of Cambridge Research Computing on behalf of the STFC DiRAC HPC Facility (www.dirac.ac.uk). The DiRAC component of CSD3 was funded by BEIS capital funding via STFC capital grants ST/P002307/1 and ST/R002452/1 and STFC operations grant ST/R00689X/1. DiRAC is part of the National e-Infrastructure.

The simulations were also performed on the GPU cluster at the Institute for Theoretical Physics at Giessen University. Many thanks to Dominik Schweitzer for keeping this cluster alive during \#JLUoffline.
\end{acknowledgments}

%\bibliographystyle{apsrev4-1}
%\bibliography{Buividovich}

%merlin.mbs apsrev4-1.bst 2010-07-25 4.21a (PWD, AO, DPC) hacked
%Control: key (0)
%Control: author (72) initials jnrlst
%Control: editor formatted (1) identically to author
%Control: production of article title (-1) disabled
%Control: page (0) single
%Control: year (1) truncated
%Control: production of eprint (0) enabled
%

\appendix

\section{Current-current correlators and electric conductivity for free quarks and free pions}
\label{apdx:jvjv_free}

Using the standard tools of finite-temperature field theory, we obtain the following expression for the Euclidean current-current correlator of free quarks with $N_c$ colors:
\begin{eqnarray}
\label{jvjv_free_quarks}
 G_E^{q}\lr{\tau} =
 \frac{N_c}{12 \pi^2}
 \int\limits_{m}^{\infty} d\epsilon \,
 \frac{\lr{\epsilon^2 - m^2}^{\frac{3}{2}}}{\epsilon}
 \times \nonumber \\ \times
 \lr{
  \frac{1}{\cosh^2\lr{\frac{\epsilon-\mu}{2 T}}}
  +
  \frac{1}{\cosh^2\lr{\frac{\epsilon+\mu}{2 T}}}
 }
 + \nonumber \\ + \,\,
 \frac{N_c}{6 \pi^2}
 \int\limits_{m}^{\infty} d\epsilon \,
 \epsilon \sqrt{\epsilon^2 - m^2}
 \lr{2 + \frac{m^2}{\epsilon^2}}
 \times \nonumber \\ \times
 \frac{
  \cosh\lr{2 \epsilon \lr{\tau - \frac{1}{2 T}}}
 }{
  \cosh\lr{\frac{\epsilon-\mu}{2 T}}
  \cosh\lr{\frac{\epsilon+\mu}{2 T}}
 }
\end{eqnarray}
The corresponding electric conductivity in the Green-Kubo relation (\ref{GreenKubo_conductivity}) is
\begin{eqnarray}
\label{free_quark_conductivity}
 \sigma_q\lr{\omega}
 =
 \frac{\alpha_q N_c}{24 \pi \, T} \delta\lr{\omega}
 + \nonumber \\ +
 \frac{N_c}{24 \pi}
 \Re\lr{\omega^2 - 4 m^2}^{\frac{1}{2}}
 \lr{1 + \frac{2 m^2}{\omega^2}}
 \times \nonumber \\ \times
 \frac{
  \sinh\lr{\frac{\omega}{2 T}}
 }{
  \cosh\lr{\frac{\omega - 2 \mu}{4 T}}
  \cosh\lr{\frac{\omega + 2 \mu}{4 T}}
 } ,
\end{eqnarray}
where
\begin{eqnarray}
\label{free_quark_conductivity_const}
 \alpha_q =
 \int\limits_{m}^{\infty} d\epsilon \,
 \frac{\lr{\epsilon^2 - m^2}^{\frac{3}{2}}}{\epsilon}
 \times \nonumber \\ \times
 \lr{
  \frac{1}{\cosh^2\lr{\frac{\epsilon - \mu}{2 T}}}
  +
  \frac{1}{\cosh^2\lr{\frac{\epsilon + \mu}{2 T}}}
 } .
\end{eqnarray}

In the phase with spontaneously broken chiral symmetry, the conductivity is expected to be dominated by charged pions contributions. The leading-order contribution is just the conductivity of free massive charged scalar fields at finite chemical potential \cite{Fernandez:hep-ph/0512283}, with electric current defined as
\begin{eqnarray}
\label{charged_scalar_current_def}
 j_{\mu} = \frac{i}{2} \lr{\bar{\phi} \partial_{\mu} \phi - \lr{\partial_{\mu} \bar{\phi}} \phi } .
\end{eqnarray}
A straightforward calculation yields the following expression for the Euclidean correlator of spatial currents of a charged scalar field:
\begin{eqnarray}
\label{jvjv_free_scalar}
 G_E^{\pi}\lr{\tau}
 =
 \frac{1}{6 \pi^2} \int\limits_{m}^{\infty} d\epsilon \,
 \frac{\lr{\epsilon^2 - m^2}^{\frac{3}{2}}}{\epsilon}
 \times \nonumber \\ \times
 \left(
  \frac{1}{4 \sinh^2\lr{\frac{\epsilon + \mu}{2 T}}}
  +
  \frac{1}{4 \sinh^2\lr{\frac{\epsilon - \mu}{2 T}}}
  + \right. \nonumber \\ + \left.
 \frac{2 e^{\epsilon/T} \cosh\lr{2 \epsilon \lr{\tau - \frac{1}{2 T}}}}{\lr{e^{\lr{\epsilon + \mu}/T} - 1} \lr{{e^{\lr{\epsilon - \mu}/T} - 1}}}
 \right) .
\end{eqnarray}
The corresponding AC conductivity is
\begin{eqnarray}
\label{free_pion_conductivity}
 \sigma_{\pi}\lr{\omega} =
 \frac{\alpha_{\pi}}{48 \pi \, T} \delta\lr{\omega}
 + \nonumber \\ +
 \frac{1}{48 \pi}
 \frac{\Re \lr{\omega^2 - 4 m^2}^{\frac{3}{2}}}{\omega^2}
 \times \nonumber \\ \times
 \frac{
  \lr{e^{\omega/T} - 1}
 }{
  \lr{e^{\frac{\omega + 2\mu}{2 T}} - 1} \lr{e^{\frac{\omega - 2 \mu}{2 T}} - 1}
 } \, ,
\end{eqnarray}
where
\begin{eqnarray}
\label{free_pion_conductivity_const}
 \alpha_{\pi} = \int\limits_{m}^{\infty} d\epsilon \,
 \frac{\lr{\epsilon^2 - m^2}^{\frac{3}{2}}}{\epsilon}
 \times \nonumber \\ \times
 \lr{
  \frac{1}{\sinh^2\lr{\frac{\epsilon - \mu}{2 T}}}
  +
  \frac{1}{\sinh^2\lr{\frac{\epsilon + \mu}{2 T}}}
 } .
\end{eqnarray}
It is instructive also to consider the low-temperature limit of these expressions. In the limit $T \rightarrow 0$, free quark conductivity takes the form
\begin{eqnarray}
\label{free_quark_conductivity_zeroT}
 \lim\limits_{T \rightarrow 0} \sigma_q\lr{\omega}
 =
 \frac{N_c}{12 \sqrt{\pi}} \frac{\Re \lr{\mu^2 - m^2}^{\frac{3}{2}}}{\mu} \delta\lr{\omega}
 + \nonumber \\ +
 \frac{N_c}{24 \pi}
 \Re\lr{\omega^2 - 4 m^2}^{\frac{3}{2}}
 \lr{1 + \frac{2 m^2}{\omega^2}} \theta\lr{\omega - 2 \mu} ,
\end{eqnarray}
where $\theta\lr{x}$ is the Heaviside unit step function. In the same limit, the free pion conductivity takes the form
\begin{eqnarray}
\label{free_pion_conductivity_zeroT}
 \lim\limits_{T \rightarrow 0}\sigma_{\pi}\lr{\omega} =
 \frac{1}{48 \pi}
 \frac{\Re \lr{\omega^2 - 4 m^2}^{\frac{3}{2}}}{\omega^2}
 \theta\lr{\omega - 2 \mu} .
\end{eqnarray}
In contrast to the free fermion case, the term with the $\delta$-function vanishes in the limit of zero temperatures as $\frac{\sqrt{2 \pi m} \, T^{\frac{3}{2}}}{8 \pi} \delta\lr{\omega} \, e^{\lr{\mu - m}/T}$. Of course, for free bosons the conductivity is only defined for $\mu \leq m$.

It is instructive to compare the midpoint estimates of the low-frequency electric conductivity, which we use in Subsection~\ref{subsec:numres_corrs}, for free quarks and free pions. To this end we use the bare quark mass $m = 0.01$, and the pion mass $m_{\pi} = 0.158$, as determined in Section~\ref{sec:lattice_setup}. With chemical potential $a \mu = 0.05$, which is below the pion/diquark condensation threshold, and temperatures in the range $1/(a T) = 16 \ldots 22$, we find that the midpoint conductivity estimate is $5 \ldots 10$ times smaller for the pion gas than for free quarks. On the other hand, the midpoint estimate for the pion gas conductivity shows much stronger dependence on the chemical potential, as also noticed in \cite{Ghosh:1711.08257,Kaczmarek:1012.4963}. Also, for pion gas the coefficient $c\lr{T}$ in (\ref{sigma_vs_mu_parameterization}) grows with temperature, whereas for the free quark gas $c\lr{T}$ decreases with temperature.

\section{Efficient calculation of correlators of conserved currents for Wilson-Dirac and Domain Wall fermions}
\label{apdx:current_calculation}

Since for either the Wilson-Dirac or Domain Wall fermions the conserved current operator $j_{z,\mu} = \bar{\psi}_x \frac{\partial D_{xy}}{\partial \theta_{z,\mu}} \psi_y$ with the single-particle current operator
\begin{eqnarray}
\label{current_operator_def}
 \frac{\partial D_{xy}}{\partial \theta_{z,\mu}}
 = \nonumber \\ =
 i P_{\mu}^{+} U_{z,\mu} \delta_{x,z} \delta_{y,z+\hat{\mu}}
 -
 i P_{\mu}^{-} U^{\dag}_{z,\mu} \delta_{x,z+\hat{\mu}} \delta_{y,z} .
\end{eqnarray}
is localized on two lattice adjacent lattice sites $z$ and $z + \hat{\mu}$, a straightforward calculation of the connected part (\ref{current_current_connected}) of current-current correlators requires
\begin{eqnarray}
\label{no_inversions}
 2 \times 2 \times N_{d} \times N_c
\end{eqnarray}
inversions of the Dirac operator. In this expression $N_c$ is the number of independent source vector orientations in color space which is obviously equal to the number of colors, $N_d = 4$ is the number of independent source orientations in spinor space, the first factor of two comes from the necessity to have source vectors localized at two lattice sites $z$ and $z+\hat{\mu}$, and the second factor of two accounts for the inversions of both $\mathcal{D}$ and $\mathcal{D}^{\dag}$.

Especially for Domain Wall fermions these inversions become extremely costly due to the summation of five-dimensional vector current over the fifth dimension, which is necessary to obtain the conserved vector current and the correct form of the axial current.

Here we describe a small trick which was used in this work to halve the number of Dirac operator inversions. A straightforward idea is to try to diagonalize the single-particle current operator in the $2 N_c N_d$-dimensional linear space spanned on source vectors localized either at $z$ or $z + \hat{\mu}$ and having all possible colour and spin orientations. However, a simple check reveals that the matrix $\frac{\partial D_{x,y}}{\partial \theta_{z,\mu}}$ (with $x$, $y$ and the corresponding implicit spinor and color indices considered as matrix indices, and $z$ and $\mu$ as parameters) is nilpotent and cannot be diagonalized. A physical reason for this nilpotency is that the current operator $j_{x,\mu}$ moves electric charge from lattice site $x$ to site $x+\hat{\mu}$. After the first application of $j_{x,\mu}$ to some state there is no electric charge at site $x$, thus applying $j_{x,\mu}$ second time just produces zero. Instead of diagonalization, in this situation one should rather use Jordan decomposition. We have found that $\frac{\partial D_{x,y}}{\partial \theta_{z,\mu}}$ admits Jordan decomposition of the following form:
\begin{eqnarray}
\label{current_jordan_decomposition}
 \frac{\partial D_{\lr{x,\alpha,a};\lr{y,\beta,b}}}{\partial \theta_{z,\mu}}
 =
 \sum\limits_{A,\gamma=1,2}\sum\limits_{c=1}^{N_c}
 \psi^{\lr{A,\gamma,c}}_{x,\alpha,a} \bar{\chi}^{\lr{A,\gamma,c}}_{y,\beta,b} ,
 \nonumber \\
 \psi^{\lr{1,\gamma,c}}_{x,\alpha,a}
 =
 i \delta_{x,z} \phi^{\lr{+\gamma}}_{\alpha} \kappa^{\lr{c}}_a e^{i \theta_c} ,
 \nonumber \\
 \chi^{\lr{1,\gamma,c}}_{y,\beta,b}
 =
 \delta_{y,z+\hat{\mu}} \phi^{\lr{+\gamma}}_{\beta} \kappa^{\lr{c}}_b ,
 \nonumber \\
 \psi^{\lr{2,\gamma,c}}_{x,\alpha,a}
 =
 - i \delta_{x,z+\hat{\mu}} \phi^{\lr{-\gamma}}_{\alpha} \kappa^{\lr{c}}_a e^{-i \theta_c} ,
 \nonumber \\
 \chi^{\lr{2,\gamma,c}}_{y,\beta,b}
 =
 \delta_{y,z} \phi^{\lr{+\gamma}}_{\beta} \kappa^{\lr{c}}_b .
\end{eqnarray}
Here $\phi^{\lr{\pm \gamma}}_{\alpha}$ are orthonormal eigenspinors of the projection operators $P^{\pm}_{\mu}$, with
\begin{eqnarray}
\label{projection_eigenspinors_def}
 \lr{P^{\pm}_{\mu}}_{\alpha\beta}
 =
 \sum\limits_{\gamma=1,2}
     {\phi}^{\lr{\pm \gamma}}_{\alpha}
 \bar{\phi}^{\lr{\pm \gamma}}_{\beta} .
\end{eqnarray}
Similarly, orthonormal color vectors $\kappa^{\lr{c}}_a$ and phases $e^{i \theta_c}$ form an eigensystem of the link matrix $U_{z,\mu}$:
\begin{eqnarray}
\label{link_eigenvectors_def}
 \lr{U_{z,\mu}}_{ab}
 =
 \sum\limits_{c=1}^{N_c}
     {\kappa}^{\lr{c}}_{a}
     e^{i \theta_c}
 \bar{\kappa}^{\lr{c}}_{b} .
\end{eqnarray}
Omitting all matrix indices, the Jordan decomposition (\ref{current_jordan_decomposition}) can be compactly written as
\begin{eqnarray}
\label{current_eigenvalue_decomposition_compact}
\frac{\partial D}{\partial \theta_{z,\mu}}
=
\sum\limits_{A,\gamma,c} \psi^{\lr{A,\gamma,c}} \bar{\chi}^{\lr{A,\gamma,c}} .
\end{eqnarray}
Inserting this decomposition for one of the current operators in the current-current correlator (\ref{current_current_connected}), we obtain
\begin{eqnarray}
\label{jVjV_jordan}
 \sum\limits_{\vec{y}}
 \tr\lr{
  \frac{\partial D}{\partial \theta_{z,\mu}} D^{-1}
  \frac{\partial D}{\partial \theta_{y,\nu}} D^{-1}
 }
 = \nonumber \\ =
 \sum\limits_{A,\gamma,c}
 \bar{\chi}^{\lr{A,\gamma,c}} D^{-1}
 \lr{\sum\limits_{\vec{y}} \frac{\partial D}{\partial \theta_{y,\nu}}}
 D^{-1} \phi^{\lr{A,\gamma,c}} ,
\end{eqnarray}
where Dirac vectors $\chi$ and $\phi$ are constructed for the link $\lr{z,\mu}$ according to (\ref{current_jordan_decomposition}) and $\sum\limits_{\vec{y}}$ denotes summation over spatial lattice volume with time-like component $y_0$ fixed. The operator $\sum\limits_{\vec{y}} \frac{\partial D}{\partial \theta_{y,\nu}}$ is obviously a local lattice operator which can be applied to Dirac vectors in CPU time comparable with the application of the Dirac operator itself. Expression (\ref{jVjV_jordan}) suggests that the connected contribution to current-current correlators (\ref{current_current_connected}) can be calculated as follows. For each $A=1,2$, $\gamma=1,2$ and $c = 1 \ldots N_c$ we have to do two Dirac operator inversions, one to calculate $D^{-1} \phi^{\lr{A,\gamma,c}}$ and the other to calculate $\bar{\chi}^{\lr{A,\gamma,c}} D^{-1} = \lr{D^{\dag}}^{-1} \chi^{\lr{A,\gamma,c}}$. This amounts to $8 N_c$ Dirac operator inversions in total, to be compared with $16 N_c$ inversions which would be required for a more straightforward calculation.

The same trick can be applied to the calculation of disconnected current-current correlators of the form
\begin{eqnarray}
\sum\limits_{\vec{y}}
\vev{\tr\lr{\frac{\partial D}{\partial \theta_{z,\mu}} D^{-1}}
  \tr\lr{\frac{\partial D}{\partial \theta_{y,\nu}} D^{-1}
 }} .
\end{eqnarray}
In this case one can use the Jordan decomposition trick to calculate the trace $\tr\lr{\frac{\partial D}{\partial \theta_{z,\mu}} D^{-1}}$, and stochastic estimator techniques for the second trace.

\section{Current-current correlators in the presence of diquark sources}
\label{apdx:current_diquarks}

In the presence of diquark sources, the current-current connected correlator is different from (\ref{current_current_connected}) and takes somewhat more complicated form:
\begin{widetext}
\begin{eqnarray}
\label{current_current_connected_lambda}
 \vev{j_{x,\mu} j_{y,\nu}}_{conn} =
 \frac{\partial^2}{\partial \theta_{x,\mu} \partial \theta_{y,\nu}}
 \tr\ln\lr{D D^{\dag} + \lambda^2}
 = \nonumber \\ =
 \left. \re\tr\lr{
  D^{\dag} \lr{D D^{\dag} + \lambda^2}^{-1}
  \frac{\partial D}{\partial \theta_{x,\mu}}
  D^{\dag} \lr{D D^{\dag} + \lambda^2}^{-1}
  \frac{\partial D}{\partial \theta_{y,\nu}}
 } \right|_{\theta=0}
 + \nonumber \\ +
 \lambda^2 \left. \re\tr\lr{
  \lr{D^{\dag} D + \lambda^2}^{-1}
  \frac{\partial D}{\partial \theta_{x,\mu}}
  \frac{\partial D^{\dag}}{\partial \theta_{y,\nu}}
 } \right|_{\theta=0}
 + \nonumber \\ +
 \left. \re\tr\lr{
  D^{\dag} \lr{D D^{\dag} + \lambda^2}^{-1}
  \frac{\partial^2 D}{\partial \theta_{x,\mu} \partial \theta_{y,\nu}}
 } \right|_{\theta=0}
\end{eqnarray}
\end{widetext}
In the usual conductivity measurement setup, the last two terms in (\ref{current_current_connected_lambda}) are contact terms which only affect the time slice with $\tau = 0$. This time slice is anyway discarded in our analysis.

\section{Finite-volume and lattice artifacts for current-current correlators and electric conductivity for free quarks and free pions}
\label{apdx:finite_volume_effects}

In Section~\ref{sec:conductivity_results} we have seen that the $T$- and $\mu$-dependence of the electric conductivity on the lattice is quite different from the one in infinite-volume continuum theory. In this Appendix we quantify the finite-volume and lattice artifacts in electric conductivity for free Wilson-Dirac and Domain Wall quarks on the lattice and demonstrate that lattice results agree well with continuum theory in the large-volume limit. We consider both midpoint and Backus-Gilbert estimators. For estimates made with the Backus-Gilbert method, we use the same values of $\Delta$ as for the analysis of the real lattice data.

\begin{figure*}[h!tpb]
  \centering
  \includegraphics[angle=-90,width=0.45\textwidth]{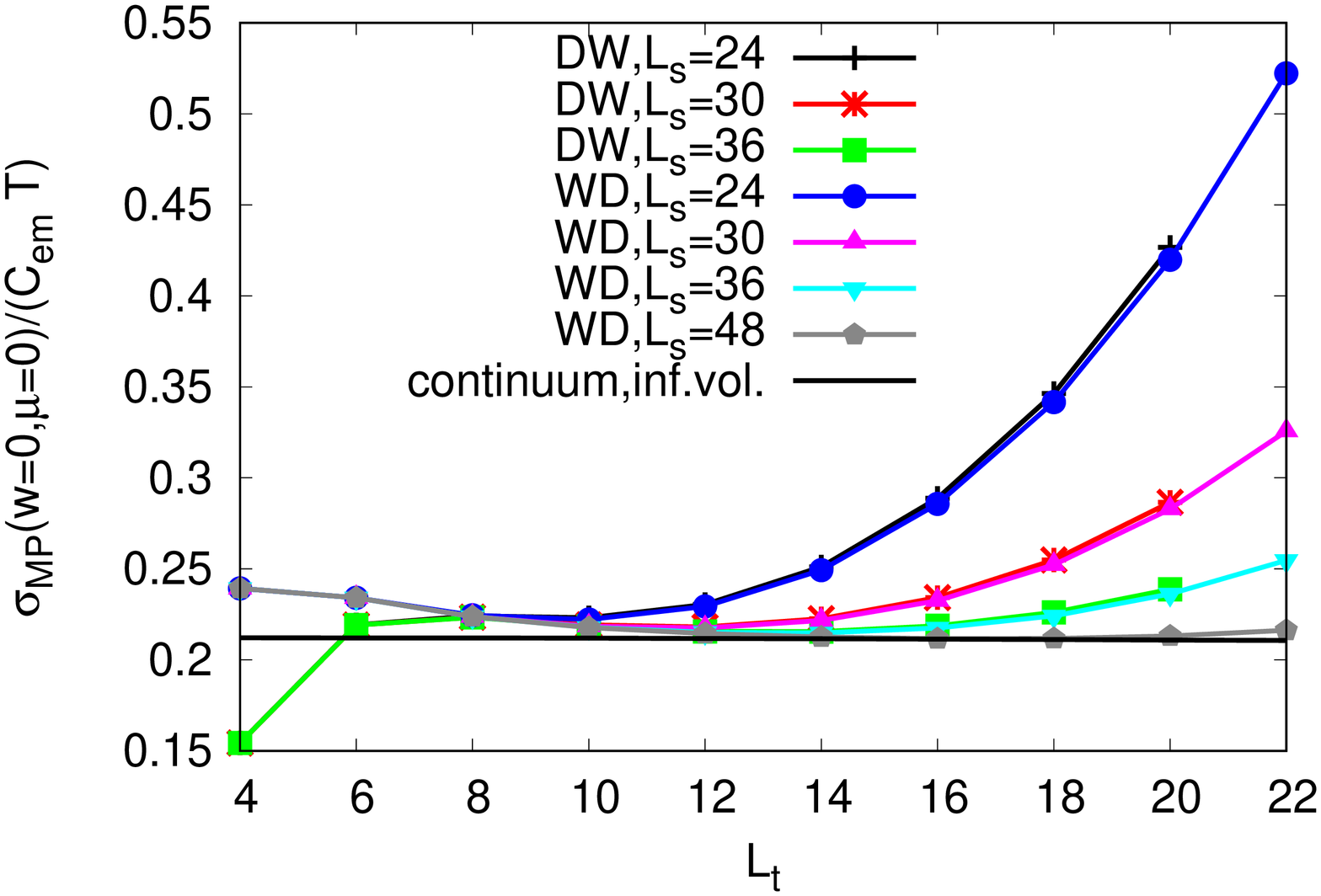}
  \includegraphics[angle=-90,width=0.45\textwidth]{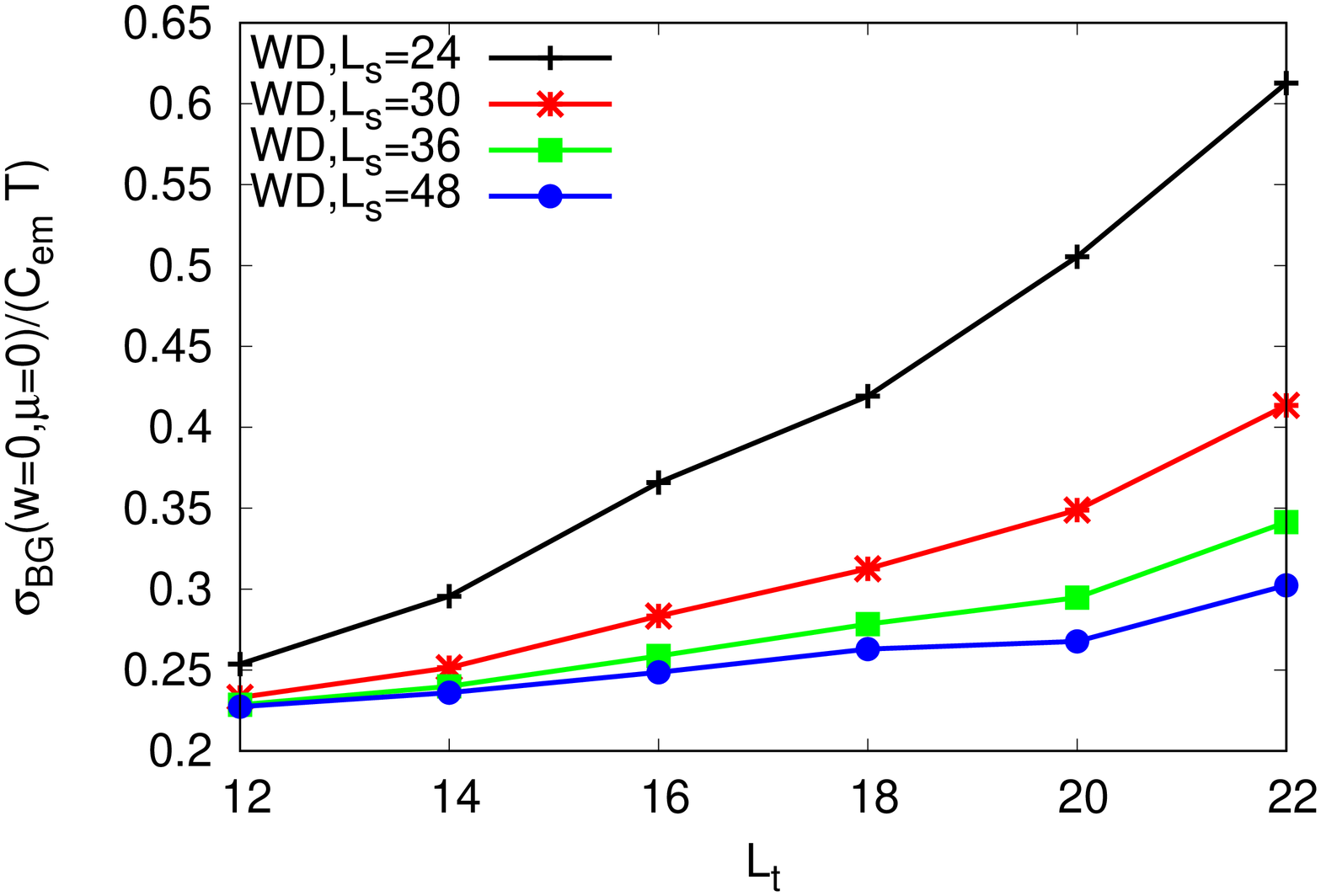}\\
  \caption{Temperature dependence of the low-frequency electric conductivity of free quarks at zero chemical potential and different lattice volumes. \textbf{On the left:} estimated from the correlator midpoint according to (\ref{MP_estimator_def}) compared with the free continuum result at the same bare quark mass. \textbf{On the right:} estimated using the Backus-Gilbert method.}
  \label{fig:free_finvol_s0}
\end{figure*}

In Fig.~\ref{fig:free_finvol_s0} we show the temperature dependence of zero-density electric conductivity, obtained with both the midpoint and the Backus-Gilbert estimators on lattices with different spatial volumes. As already discussed in Section~\ref{sec:conductivity_results}, for lattice size $L_s = 24$ the deviations from the infinite-volume limit are very large, and for $L_t = 22$ the lattice and the continuum results differ by a factor of $2.5$ for midpoint estimates. For Backus-Gilbert estimator the deviations are even larger. Only for twice larger lattice size these deviations reduce to few percents. However, we expect that for real gauge theory the correlation length is considerably smaller than for free quark gas, and hence finite-volume artifacts should be smaller. For higher temperatures, $L_t \lesssim 14$, where free quark approximation is not unreasonable, finite-volume effects are considerably smaller and do not exceed $20 \%$. A comparison of Wilson-Dirac and Domain Wall fermions suggests that all deviations are finite-volume rather than discretization artifacts, and discretization artifacts only become important for $L_t \lesssim 6$. They are much larger for Domain Wall fermions due to large contributions from bulk modes and Pauli-Villars regulator fields which compensate them.

\begin{figure*}[h!tpb]
  \centering
  \includegraphics[angle=-90,width=0.45\textwidth]{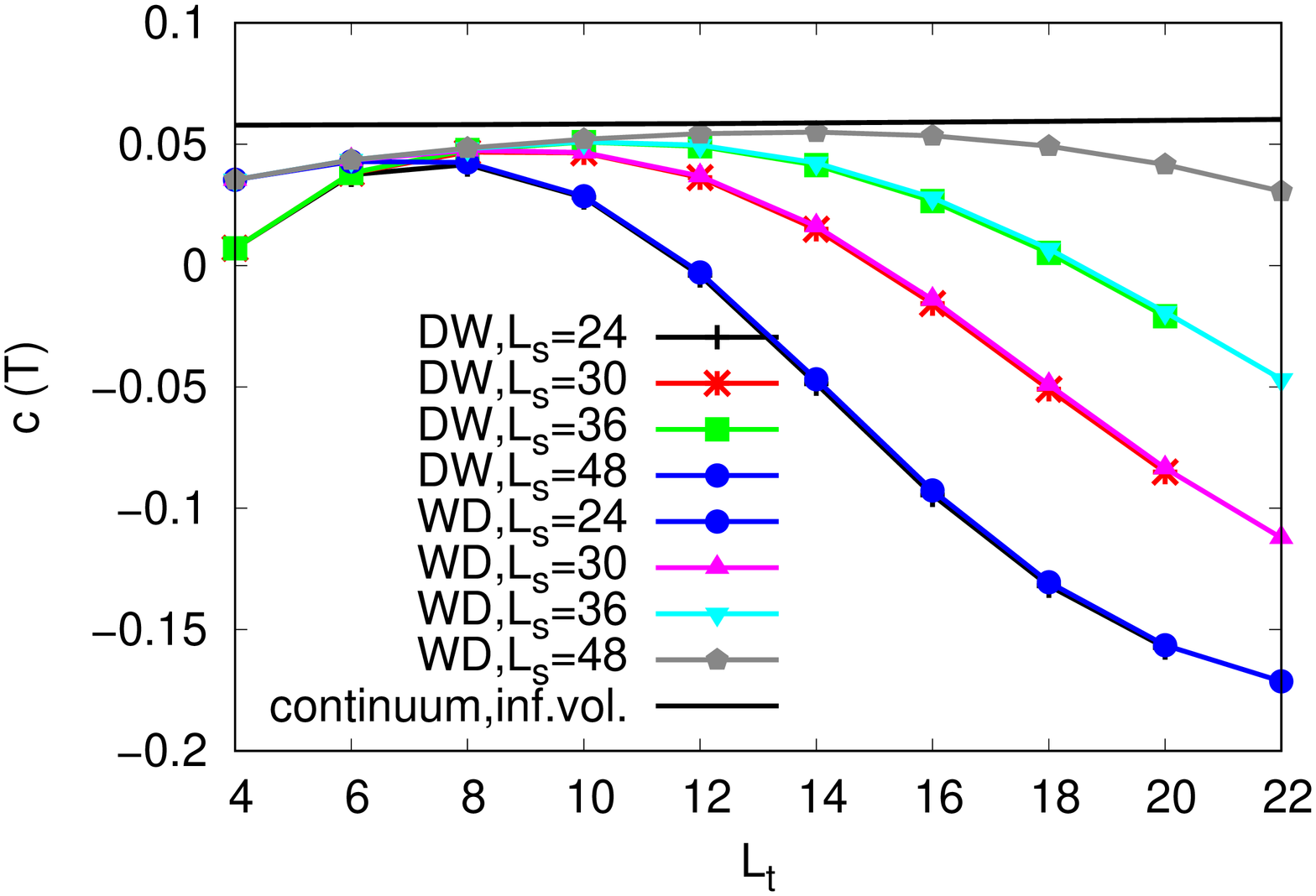}
  \includegraphics[angle=-90,width=0.45\textwidth]{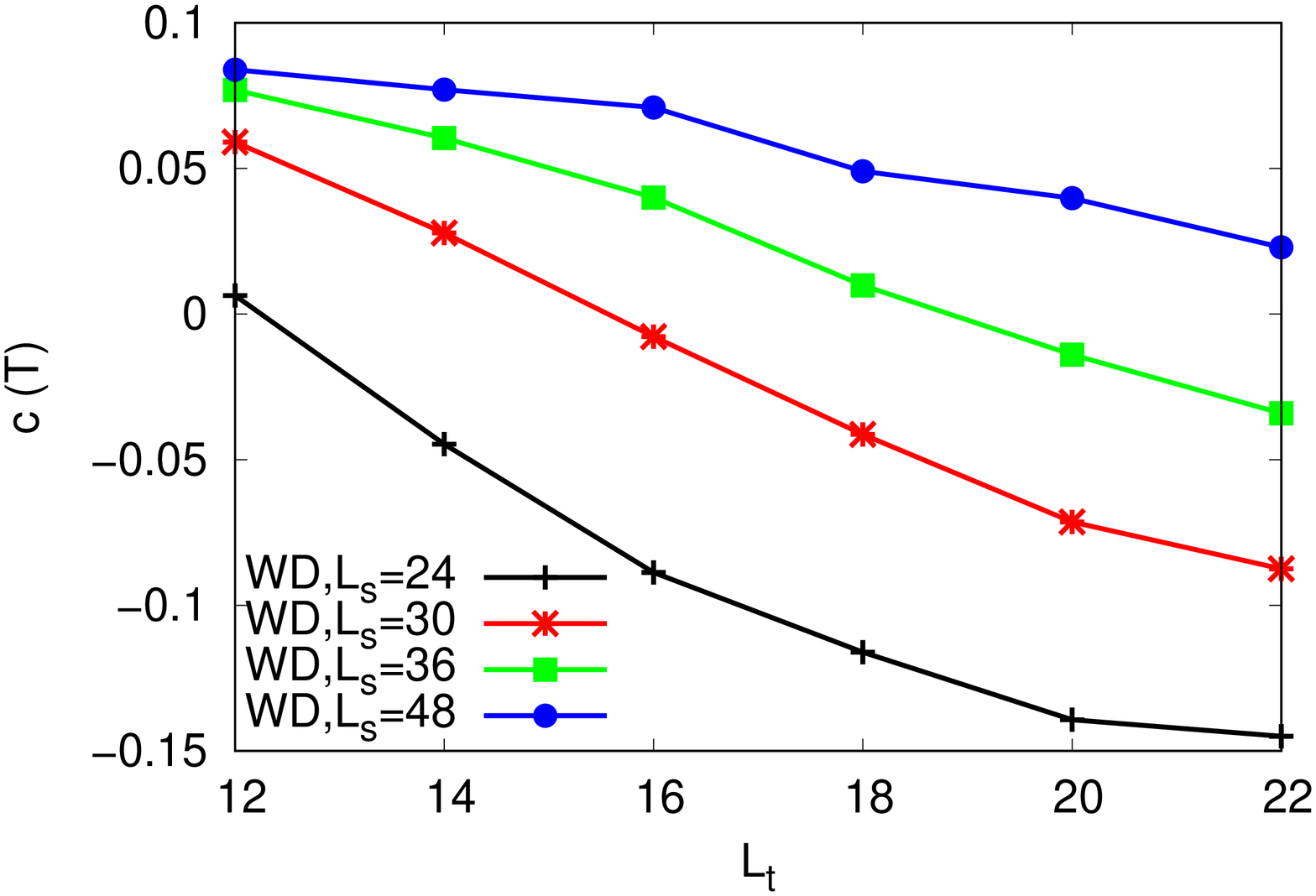}\\
  \caption{Temperature dependence of the second-order expansion coefficient $c\lr{T}$ of electric conductivity in powers of $\mu$ (\ref{sigma_vs_mu_parameterization}) at different lattice volumes. \textbf{On the left:} estimated from the correlator midpoint according to (\ref{MP_estimator_def}) compared with the free continuum result at the same bare quark mass. \textbf{On the right:} estimated using the Backus-Gilbert method.}
  \label{fig:free_finvol_c}
\end{figure*}

In Fig.~\ref{fig:free_finvol_c} we also illustrate the finite-volume effects in the estimates of the coefficient $c\lr{T}$ in the expansion (\ref{sigma_vs_mu_parameterization}), calculated from the finite difference between $a \mu = 0.05$ and $\mu = 0$. We use both the midpoint and Backus-Gilbert estimates. Again we see that $c\lr{T}$ quickly becomes negative towards lower temperatures, and becomes sufficiently close to the continuum value only for $L_s \gtrsim 48$. For estimates based on the Backus-Gilbert method the finite-volume artifacts are clearly larger than for the midpoint estimates. A comparison of Domain Wall and Wilson-Dirac fermions shows that these artifacts are indeed finite volume artifacts rather than discretization artifacts.

\end{document}